\documentclass[12pt]{scrartcl} 

\usepackage[authoryear]{natbib}
\usepackage{amsmath, amssymb, amsthm}
\usepackage{todonotes}
\usepackage[papersize={8.5in,11in},margin=1in]{geometry}  
\usepackage[doublespacing]{setspace}  
\usepackage{booktabs}
\usepackage[toc,page]{appendix}
\usepackage{graphicx}
\usepackage[ruled]{algorithm2e}

\SetKwInput{KwInput}{Input}
\SetKwInput{KwOutput}{Output}
\SetKwFor{ParFor}{for}{do in parallel}{end for}
\usepackage{svg}
\usepackage{url}
\usepackage{siunitx}
\usepackage{float}
\numberwithin{equation}{section}

\usepackage{chngcntr}
\usepackage{multirow}
\theoremstyle{plain}

\newtheorem{definition}{Definition}

\newtheorem{lemma}{Lemma}

\numberwithin{equation}{section}
\renewcommand{\v}[1]{\boldsymbol{#1}}
\newcommand{\blind}{0} 

\newcommand{\simiid}{\stackrel{\mathrm{iid}}{\sim}}

\newcommand{\bb}[1]{\mathbb{#1}}

\def\*#1{\mathbf{#1}}
\newcommand{\minus}{\scalebox{0.75}[1.0]{$-$}}

\usepackage{graphicx}
\usepackage{subcaption}
\usepackage{xr}

\usepackage{babel}

\newcommand{\im}{\mathrm{i}}

\title{Time-Varying Multi-Seasonal AR Models}
\if0\blind
{
\author{Ganna Fagerberg$^{a}$\thanks{Corresponding author: ganna.fagerberg@stat.su.se. 
\noindent $^a$Department of Statistics, Stockholm University. 
$^b$School of Business, University of New South Wales. $^c$Data Analytics Center for Resources and Environments (DARE).}, Mattias Villani$^{a}$ and Robert Kohn$^{b,c}$}
}\fi
\date{}

\begin{document}      

\maketitle

\begin{abstract}
We propose a seasonal AR model with time-varying parameter processes in both the regular and seasonal parameters. The model is parameterized to guarantee stability at every time point and can accommodate multiple seasonal periods. The time evolution is modeled by dynamic shrinkage processes to allow for long periods of essentially constant parameters, periods of rapid change, and abrupt jumps. A Gibbs sampler is developed with a particle Gibbs update step for the AR parameter trajectories. We show that the near-degeneracy of the model, caused by the dynamic shrinkage processes, makes it challenging to estimate the model by particle methods. To address this, a more robust, faster and accurate approximate sampler based on the extended Kalman filter is proposed. The model and the numerical effectiveness of the Gibbs sampler are investigated on simulated data. An application to more than a century of monthly US industrial production data shows interesting clear changes in seasonality over time, particularly during the Great Depression and the recent Covid-19 pandemic.
\end{abstract}

\section{Introduction}\label{sec:intro}

Autoregressive integrated moving average (ARIMA) models, originally popularized in the seminal book by \cite{boxjen@1970}, form a parsimonious and versatile class of time series models. ARIMA models have proven invaluable for analyzing time series data, particularly the subclass of AR models. The parsimony of ARIMA models is even more apparent when modeling seasonality with an elegant multiplicative structure that directly extends to multiple seasonal periods, an increasingly common feature in time series data. 

A drawback of the Box-Jenkins methodology is the requirement of stationarity, i.e.~that the process exhibits constant statistical properties over time, or can be differenced to stationarity. Time series data are often recorded over long study periods, and/or at high-frequency in rapidly changing environments. The statistical characteristics of such time series are likely to undergo substantial changes, which may occur both gradually and suddenly. Differencing the series to achieve stationarity can lead to over-differencing, resulting in misleading inferences \citep{granger1980introduction} and poor forecasts \citep{makridakis1997arma, smith1994forecasting}. Using data only from the most recent regime is wasteful since many correlations are expected to survive regime shifts, and prevents the model from assigning any probability to future regime changes.

A more flexible approach assumes local stationarity where the process is stationary over shorter time segments, or stationary in a small neighborhood around each time point, see e.g. \citet{dahlhaus2000likelihood} for a theoretical framework. One strand of literature explicitly models a small set of stationary regimes, with abrupt jumps between regimes, see e.g. the Markov switching type of models in \citet{hamilton1989new} or change-point models in \citet{chib1998estimation} and \citet{rosen2012adaptspec}, and dynamic mixture models in \citet{gerlach2000efficient}. The parameters in time-varying autoregressive (TVAR) models  \citep{prado2010time, wikle2014tvar, wood2011bayesian} instead evolve more gradually over time following a stochastic process, typically a random walk with Gaussian innovations. More recently, there has been progress in more realistic models for the parameter evolution that use local-global shrinkage priors \citep{polson2010shrink} to cater for smooth gradual changes, periods with no change, as well as abrupt jumps, see e.g. \citet{kalli2014time}, \citet{kowal2019dynamic}, \citet{cadonna2020triple} and \citet{knaus2023dynamic}.

Stability restrictions on the AR coefficients are rarely imposed in time-varying parameter AR models; \citet{everitt2013online}, \citet{Doucet2004smoothing} and \cite{wikle2014tvar} are exceptions with models for time-varying partial autocorrelations. A process with time-varying parameters may however easily drift into the non-stable regions for certain time periods leading to unwanted explosive forecast paths. The main reason for not imposing stability in time-varying models is most likely computational: the stability restrictions are recursive and non-linear, so the Kalman filter or the forward-filtering backward sampling (FFBS) algorithm \citep{carter1994gibbs, fruhwirth1994data} can therefore no longer be used for inference. 

Seasonality is a neglected aspect of time-varying models, and all of the aforementioned articles deseasonalize the data before the analysis. The regularities provided by seasonality are perhaps the most predictable pattern in a time series and there is much to gain from modeling time-varying seasonality jointly with other model parameters. The absence of time-varying seasonality within the ARIMA framework is probably due to the non-linearity resulting from the multiplicative seasonal structure, again precluding the use of the Kalman filter for inference. This is unfortunate, since the multiplicative seasonal AR representation in \cite{box2015time} directly extends to multiple seasonal patterns, which are an increasingly common feature in many datasets; for example, hourly measurements can exhibit daily, weekly and annual cycles \citep{xie2020forecasting, de2011forecasting}.

Our article makes the following contributions. First, a time-varying seasonal AR process is developed where both the regular and the seasonal AR parameters follow the dynamic shrinkage process priors in \citet{kowal2019dynamic}. The proposed model is shown to effectively track evolving patterns in parameter values over time, including jumps, while accurately identifying  periods of constant parameters. Second, the model is extended to allow for multiple seasonal periods. Third, we explicitly restrict the parameters to a stable process at every time point using the parameterization in \citet{Barndorff-Nielsen1973} and \citet{monahan1984note}, ensuring that the process is locally stable. Finally, Gibbs sampling algorithms are developed to handle the nonlinearities in the time-varying parameters stemming from both the multiplicative seasonality and stability restrictions.

We propose a fully Bayesian approach and develop a Gibbs sampling algorithm to deal with the non-linearity in the multiplicative seasonal structure and the stability restriction. The algorithm updates the regular and seasonal AR parameters using the particle Gibbs with ancestor sampling (PGAS, \citet{lindsten2014particle}) algorithm. PGAS is relatively time-consuming however, and may suffer from degeneracy due to the dynamic shrinkage prior, which encourages parameter paths to remain essentially constant over extended time periods. Therefore, we develop an alternative updating step using the forward-filtering backward sampling (FFBS) algorithm
\citep{carter1994gibbs, fruhwirth1994data}, with the filtering step performed by the extended Kalman filter \citep{thrun2002probabilistic} to handle the two types of non-linearities in the model. The extended Kalman filter is an approximate filter, but we demonstrate that it is quite accurate in seasonal AR models with stability restrictions. 

We explore the properties of the proposed model and posterior sampling algorithms in several simulation experiments, and show that the model compares favorably to several widely used time-varying parameter models. Finally, the model is applied to more than a century of monthly industrial production data in the US. We find that the strong seasonality is essentially wiped out during the Great Depression, and is also markedly different during the Covid-19 pandemic.

The proofs and additional results are available in the supplementary material to this article, referenced below with the prefix S, e.g. Figure \ref{fig:normalapprox2uniform} in Section \ref{app:t_transform}.

\section{Time-varying multi-seasonal AR processes}
Section \ref{subsec:static_ar} introduces the time-invariant, or static, multi-seasonal AR model (SAR) with stability constraints. The model is then extended in Section \ref{subsec: TV-SAR} to have time-varying parameters, using the dynamic shrinkage process (DSP) of \cite{kowal2019dynamic} to regularize the parameter evolution. Finally, Section \ref{subsec:ar_regr} recasts the full model as a nonlinear regression, providing a convenient structure for efficient inference. 

\subsection{Multi-seasonal AR processes with stability restrictions}\label{subsec:static_ar}
A general static AR model with one or more seasonal components can be written as \citep{wei2019time}
\begin{equation}
\label{eq: stat_AR}
   \phi(L) \prod_{j=1}^M\Phi_j(L^{s_j})(y_t-m) = \varepsilon_t, 
\end{equation}
where $m$ is the mean, $L$ is the lag operator with the property $L^k y_t = y_{t-k}$ and $\varepsilon_t \simiid \mathrm{N}(0,\sigma^2)$ is a white noise sequence. The regular AR polynomial of order $p$ is $\phi(L) = 1-\phi_{1} L^{1} -\phi_{2} L^{2} -\ldots - \phi_{p} L^{p}$. The $j$-th seasonal AR component with seasonal period $s_j$ and seasonal order $P_j$, the $j$th seasonal component is $\Phi_j(L^{s_j}) = 1-\Phi_{j1} L^{s_j} -\Phi_{j2} L^{2s_j} -\ldots - \Phi_{jP_j}$, a polynomial in $L^{s_j}$. Here $M$ stands for the total number of seasonal components. 

We denote this model by SAR$(p, \textbf{P})_\textbf{s}$, where $p$ is the order of the regular AR polynomial, $\v s = (s_1,\ldots,s_M)$ is the vector of seasonal periods, and $\*P = (P_1,\ldots,P_M)$ is the corresponding vector of seasonal orders. 

When $M=1$, the model in \eqref{eq: stat_AR} reduces to the standard single-seasonal AR model with one regular  and one seasonal polynomial with season $s$, SAR$(p, P)_s$: 
\begin{equation}\label{eq: stat_AR_traditional}
    \phi(L) \Phi(L^s)(y_t-m) = \varepsilon_t. 
\end{equation}
Models with $M>1$ allow for multiple seasonal periods. For example, a SAR model for hourly data with $M=2$ and $\v s=(24,168)$ captures both both daily ($s_1=24$) and monthly ($s_2=24\cdot 7 = 168$) seasonal cycles.

The  SAR$(p, \textbf{P})_\textbf{s}$ process in \eqref{eq: stat_AR} is stable if all lag polynomials $\phi(L)$ and  $\Phi_j(L^{s_j})$ have all their zeros outside the unit circle, i.e., $|\phi(z)|\neq 0$ and $|\Phi_j(z)|\neq 0$, for $|z| \leq 1$. A stable process has many desirable properties, e.g. non-explosive forecast paths, existence of a  spectral density, and well-behaved impulse response functions. A stable AR process is also stationary. The model in \eqref{eq: stat_AR} can be extended to include polynomials that apply regular differencing $\phi(L) = 1- L$ or seasonal differencing $\Phi_j(L^s_j) = 1- L^{s_j}$. Since the stability restrictions imply the same set of restrictions on the regular and seasonal AR parameters in each polynomial, we now restrict attention to the regular AR polynomial $\phi(L)=1 - \phi_1 L - \ldots - \phi_p L^p$. 

To enforce stability we use the reparametrisation in \citet{Barndorff-Nielsen1973} and \citet{monahan1984note} to map a set of unconstrained real-valued coefficients to the set of stable parameters, i.e.~parameters that determine a stable process. Let $\mathbb{S}^p \subset \mathbb{R}^p$ denote the region in parameter space where the $\mathrm{AR}(p)$ process is stable. The reparameterization is defined as follows.
\begin{definition}[stability parameterization]\label{def:monahan}
An $\operatorname{AR}(p)$ process can be restricted to be stable by a 1:1 and onto map
\begin{equation*}
\v\theta \rightarrow \v r \rightarrow \boldsymbol{\phi}.
\end{equation*}
 of unrestricted parameters $\v\theta = (\theta_1,\ldots,\theta_p)^\top \in \mathbb{R}^p$ via the partial autocorrelations $\v r = (r_1,\ldots,r_p)^\top \in \bb (\minus 1,1)^p$ to the stable AR parameters $\boldsymbol{\phi}=(\phi_1,\ldots,\phi_p)^\top \in \mathbb{S}^p$.
The mapping from $\v r$ to $\v\phi$ is given by setting $\phi_{1,1} = r_1$ followed by the recursion
\begin{equation*}
\phi_{k,j} = \phi_{k-1,j}-r_{k}\phi_{k-1,k-j},\;\text{ for } k=2,\ldots,p \text{ and } j=1,\ldots,k-1,
\end{equation*}
and finally returning $\v \phi = (\phi_{p,1},\ldots,\phi_{p,p})^\top$, where $\phi_{p,p}=r_p$.
\end{definition}

\begin{figure}
 \centering
 \includegraphics[scale=0.6]{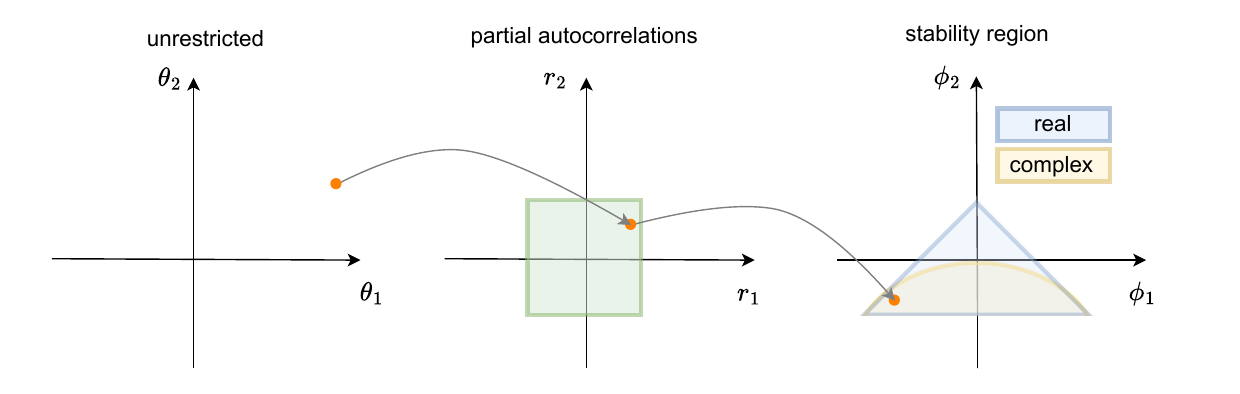}
    \caption{Illustrating the stability parametrization from an $\mathrm{AR}(2)$ process from the unrestricted $\v\theta = (\theta_1,\theta_2)^\top\in \bb R^2$ (left) via the partial autocorrelations $\v r = (r_1,r_2)^\top \in \bb (\minus 1,1)^2$ (middle) to the AR parameters $\boldsymbol{\phi}=(\phi_1,\phi_2)^\top$ (right); the regions with real (blue) and complex (beige) roots are also shown.}\label{fig:monahan}
\end{figure}
We use the $\v\theta \rightarrow \v r$ transformation suggested by \citet{monahan1984note}
\begin{equation}\label{eq:t_transf}
 r_k=\frac{\theta_{k}}{\sqrt{1+\theta_{k}^2}}\quad  \text{ for } k=1,\ldots,p.
\end{equation}
Figure~\ref{fig:monahan} illustrates the two mappings involved in the parametrization.

The unrestricted parameters $\v\theta = (\theta_1,\ldots,\theta_p)^\top$ in \eqref{eq:t_transf} are merely technical devices  and are not interpretable. The following lemma gives the prior on $\v\theta$ which implies a uniform distribution over the stable/stationary region for the AR parameters; its proof is in Section \ref{proofStationaryLemma}. A uniform distribution can be used as a baseline for more informative distributions.

\begin{lemma}\label{lem:uniform_t}
Let $\v\phi=\v g(\v\theta)$ be the mapping in Definition \ref{def:monahan}, and let $\theta_1,\ldots,\theta_p$ be independently distributed as
\begin{equation}\label{eq:t_odd_even}
\theta_k \sim
    \begin{cases}
       t\big(k+1,0,\frac{1}{\sqrt{k+1}}\big) & \text{ if } k \text{ is odd}\\
       t_\mathrm{skew}\big(\frac{k}{2},\frac{k+2}{2}, 0, \frac{1}{\sqrt{k+1}}\big) & \text{ if } k \text{ is even};
    \end{cases}
\end{equation}
then $\v\phi=\v g(\v\theta)$ is uniform over the stability region in $\v\phi$-space. 
Here $t(v, \mu, \sigma)$ is the usual $t$-distribution with $v$ degrees of freedom translated by $\mu$ and scaled by $\sigma$, and $t_\mathrm{skew}(a, b, \mu, \sigma)$ is the skew-\textit{t} extension in \cite{jones2003skew}.\newline 

\end{lemma}

\subsection{Locally stable time-varying multi-seasonal AR processes}\label{subsec: TV-SAR}

The time-varying, or dynamic, AR model with multiple seasonal components
\begin{equation}\label{eq:dynamic_AR}
   \phi_t(L) \prod_{j=1}^M\Phi_{jt}(L^{s_{j}})y_t = \varepsilon_t, 
\ \mathrm{N}(0,\sigma^2)
\end{equation}
extends the model in \eqref{eq: stat_AR} with time-varying regular and seasonal AR polynomials $\phi_{t}(L) = 1-\phi_{1t} L^{1} -\phi_{2t} L^{2} -\ldots - \phi_{pt} L^{p}$ and $\Phi_{jt}(L^{s_j}) = 1-\Phi_{j1t} L^{s_j} -\Phi_{j2t} L^{2s_j} -\ldots - \Phi_{jp_jt} L^{P_js_j}$, respectively. 

The errors $\varepsilon_t$ can be generalized to have time-varying noise variance; see Section \ref{subsec:update_sigma} and the empirical application in Section \ref{sec:applications}. 
We also assume that the process has a zero mean, but a non-zero time-varying mean $m_t$ can be included in the model. We denote this model by TVSAR$(p, \v P)_{\v s}$, where, as before, $\v s = (s_1,\ldots,s_M)$ is a vector with the $M$ seasonal periods and $\v P = (P_1,\ldots,P_M)$ are the corresponding seasonal lag orders.

To enforce local stability in the dynamic AR process, we employ the parameterization described in Definition~\ref{def:monahan} to each regular and seasonal AR polynomial in \eqref{eq:dynamic_AR}. As in the static case, we restrict our attention to the regular AR polynomial, noting that the same mapping applies to the seasonal polynomials as well. 

At each time period $t$, we map the unrestricted time-varying parameters $\v\theta_t = (\theta_{1t},\ldots, \theta_{pt})^\top$ to the time-varying AR parameters $\v\phi_{t}=(\phi_{1t},\ldots,\phi_{pt})^\top$. The unrestricted parameters $\v\theta_t$ evolve over time following the dynamic shrinkage process (DSP) prior in \citet{kowal2019dynamic}. The DSP is a time series extension of the widely used horseshoe prior \citep{carvalho2010horseshoe}, allowing the parameters to be essentially unchanged for long periods followed by large jumps or potentially persistent periods of rapid change. The process for the evolving TVSAR parameters in any given AR polynomial with $p$ lags is 
\begin{align}
\label{eq: evol_process}
   \v\phi_t &= \v g(\v\theta_t) \nonumber \\
    \theta_{kt} &= \theta_{k,t-1} +\nu_{kt}, \hspace{3.5cm} 
    \nu_{kt}  \overset{\mathrm{indep}}{\sim} \mathrm{N}\big(0, \exp(h_{kt})\big)  \nonumber \\ 
    h_{kt} &= \mu_k + \kappa_k (h_{k,t-1}-\mu_k)+\eta_{kt}, \hspace{1.0cm}   
    {\eta}_{kt} \overset{\mathrm{iid}}{\sim} {Z}(1/2,1/2, 0, 1), 
\end{align}
for $k=1,\ldots,p$, where $\v\phi_t = \v g(\v\theta_t)$ maps the unrestricted parameters $\v\theta_t \in \mathbb{R}^p$ to the stable parameters $\v\phi_t \in \mathbb{S}^p$. The global means, $\v\mu = (\mu_1,\ldots,\mu_p)^\top$, of log-volatilities, $\v h_t = (h_{1t},\ldots,h_{pt})^\top$,  control the overall degree of time-variation in $\v\theta_t$ and consequently in $\v \phi_t$. The parameter $\kappa_k$ regulates the persistence of the log-volatility process $h_{kt}$ for the $k$th parameter. The innovations to the log-volatility are drawn from the heavy-tailed $Z$-distribution \citep{barndorff1982normal} with zero location and unit scale parameters. It is straightforward to add a scale parameter $\sigma_\eta$ to $\eta_{kt} \sim {Z}(1/2,1/2, 0, \sigma_\eta)$, but it is unnecessary in our model. These innovations admit a scale-mixture representation with P\'olya-Gamma distributed mixing variables, yielding a conditionally Gaussian model amenable to efficient inference via data augmentation \citep{kowal2019dynamic}. 

A small global $\mu_{k}$ gives a prior with long periods of essentially constant evolution in $\theta_k$. When a positive local heavy-tailed $\eta_{kt}$ innovation occurs, it causes the log-volatility to temporarily spike and remain high for a duration governed by the persistence parameter $\kappa_{k}$. To see the connection to the original horseshoe prior in \citet{carvalho2010horseshoe}, consider the case with no persistence, $\kappa_k=0$. The variance of the innovations to the $k$th parameter is then $\exp(h_{kt}) = \tau_k^2 \lambda_{kt}^2$, where $\tau_k^2 = \exp(\mu_{k})$ is the global variance and $\lambda_{kt}^2=\exp(\eta_{kt})$ is the time-varying local variance. The implied prior on each $\lambda_{kt}$ is  $C^+(0,1)$, i.e.~a heavy tailed half-Cauchy as in the original horseshoe prior. \citet{kowal2019dynamic} demonstrate that the DSP prior with $0<\kappa<1$ gives additional shrinkage in a time period that follows a period of heavy shrinkage. For a comprehensive treatment of the DSP process, see \cite{kowal2019dynamic}.

Lemma ~\ref{lem:uniform_t} gives the prior that implies a uniform distribution over the stability region. 
The log-volatilities at time $t=0$ are $\v h_0 = \v\mu + \v\eta_0$ with the elements of $\v\eta_0$ independent $Z \sim  (1/2,1/2, 0, 1)$ a priori. The prior for each $\kappa_k \sim \mathrm{TN}(\kappa_0,\psi_0^2,-1,1)$ is a normal truncated to $(-1,1)$, giving an easily sampled truncated normal as conditional posterior. \citet{kowal2019dynamic} use a P\'olya-Gamma construction for the prior on the global means $\mu_k$ to obtain the original $C^+(0,1)$ horseshoe prior for $\tau_k = \exp(\mu_k/2)$. This choice for the horseshoe prior is questioned by \citet{piironen2017sparsity} and we use $\mu_k \sim \mathrm{N}(\mu_0,\sigma_0^2)$, implying a log normal prior for $\tau_k$.

The DSP prior governs the time-variation in the unrestricted parameters $\v \theta_t$. The nonlinear mapping from the $\v \theta_t$ to the AR coefficients $\v \phi_t$ is smooth and invertible, and preserves the qualitative properties of the DSP prior, i.e. the ability to generate long spells of essentially constant parameter, rapid changes and jumps; see Section \ref{app:prior_evol}.

\subsection{The multi-seasonal TV-AR model as a nonlinear regression}\label{subsec:ar_regr}

The multi-seasonal $\mathrm{TVSAR}(p,\mathbf{P})_{\mathbf{s}}$ model in \eqref{eq:dynamic_AR} can be written as a single AR polynomial model by multiplying out the regular and $M$ seasonal AR polynomials $\phi_{t}(L)\prod_{j=1}^M \Phi_{jt}(L^{s_j})=\tilde{\phi}_{t}(L)$ and moving all lags to the right-hand side to obtain the regression model
\begin{equation}\label{eq:dynamic_AR_multiplied_out}
    y_t = \v x_{t}^\top \tilde{\v\phi}_{t} + \varepsilon_t,\quad \varepsilon_t \overset{\mathrm{iid}}{\sim} \mathrm{N}(0,\sigma^2),
\end{equation}
where $\v x_{t}$ is a vector containing all lags $y_{t-k}$ for which the coefficients in $\tilde{\phi}_{t}(L)$ are non-zero and the vector $\tilde{\v\phi}_{t}$ collects the coefficients for those lags. The elements of $\tilde{\v\phi}_{t}$ are products of the AR parameters in the different polynomials. For example, consider a model with a regular component with one lag and one seasonal component with period $s$ and one lag 
\begin{equation}\label{eq: toy_ex}
    (1-\phi_{1t} L)(1-\Phi_{1t} L^s)y_t = \varepsilon_t.
\end{equation} 
Multiplying out the two polynomials on the left hand side of \eqref{eq: toy_ex} and rearranging, the model can be written
\begin{equation}\label{eq: toy_additive}
    y_t = \phi_{1t} y_{t-1}+\Phi_{1t} y_{t-s} -\phi_{1t} \Phi_{1t} y_{t-(1+s)} +\varepsilon_t, 
\end{equation}
so $\v x_{t} = (y_{t-1}, y_{t-s}, y_{t-(1+s)})^\top$ and $\tilde{\v\phi}_t = (\phi_{1t},\Phi_{1t}, \phi_{1t}\Phi_{1t})^\top$ in \eqref{eq:dynamic_AR_multiplied_out}. The three-dimensional vector of regression coefficients $\tilde{\v\phi}_t$ is a nonlinear transformation of the two unrestricted AR parameters at time $t$. A general TVSAR model with one regular and $M$ seasonal polynomials involves a nonlinear transformation with a total of $r = p+\sum_{j=1}^M P_j$ unrestricted time-varying AR parameters.

We use the conditional likelihood function of the seasonal AR model based on the regression formulation in \eqref{eq:dynamic_AR_multiplied_out}, conditional on $p_{\max}$ pre-sample values $y_{0},y_{-1},\ldots,y_{-p_{\max}+1}$ where $p_{\max}$ is the maximal lag order in $\v x_{t}$. An alternative approach is to use the exact likelihood \citep{hamilton2020time} to avoid the loss of initial data, or to infer the initial observations as a separate step in the algorithm. 

In summary, the seasonal AR model with dynamic shrinkage priors is 
\begin{align}
\label{eq: SAR_DSP}
   y_t &= \v x_t^\top  \tilde{\v g}(\v\theta_t) + \varepsilon_t,\hspace{2.75cm}  \varepsilon_t \overset{\mathrm{iid}}{\sim} \mathrm{N}(0,\sigma^2) \nonumber \\
    \v\theta_{t} &= \v\theta_{{t-1}} +\v\nu_{t}, \hspace{3.35cm} 
    \v\nu_{t}  \overset{\mathrm{indep}}{\sim} \mathrm{N}\big(\mathbf{0}, \mathrm{Diag}(\exp(\v h_{t}))\big)  \nonumber \\ 
    {\v h}_{t} &= \v \mu + \v \kappa ({\v h}_{{t-1}}-\v \mu)+{\v\eta}_t, \hspace{1.25cm}   
    {\eta}_{kt} \overset{\mathrm{iid}}{\sim} {Z}(1/2,1/2, 0, 1), 
\end{align}
where $\v\theta_t$ is a vector with all  $r = (p +\sum_{j=1}^M P_j)$ AR coefficients and $\tilde{\v\phi}_t = \tilde{\v g}(\v\theta_t)$ maps the coefficients in each AR polynomial to the stability region followed by polynomial multiplication leading to $\tilde{\v\phi}_t$ in \eqref{eq:dynamic_AR_multiplied_out}. The $r\times r$ matrix $\v \kappa$ is diagonal with $(\kappa_1,\ldots,\kappa_r)^\top$ on its main diagonal and  $\mathrm{Diag}(\exp(\v h_{t}))$ is a diagonal matrix having  $\exp(\v h_t) =(\exp(h_{1t}),\ldots,\exp(h_{rt}))^\top$ on its main diagonal.

\section{Bayesian inference for time-varying multi-seasonal AR}

The joint posterior distribution $p(\v\theta_{0:T},\v h_{0:T},\sigma_{1:T}, \v\mu,\v\kappa \vert y_{1:T})$ is sampled using a block Gibbs sampler, with each of the above parameter vectors as separate updating blocks; here the notation $\v\theta_{0:T}$ means the sequence  of all $r$ unrestricted AR parameters $\v\theta_t$ for $t=0,1,\ldots,T$. See Algorithm~\ref{alg:gibbs_algorithm} in the supplementary material for an overview. 
 
We now explain the updating step for $\v\theta_{0:T}$, leaving the details of the updating steps for $\v\mu$, $\v\kappa$ and $\v h_{0:T}$ to Section \ref{app:GibbsUpdating}. 

\subsection{Updating the parameter paths}\label{subsec:PGAS}

Conditional on all other parameters, the TVSAR model in \eqref{eq: SAR_DSP} can be expressed as the following state-space model with $\v\theta_t$ as the state vector
\begin{align}
&\v \theta_0 \sim \mathrm{Uniform}(\mathbb{S}) \label{eq:ss_prior}  \\
&\v\theta_{t} \vert \v\theta_{t-1} \sim \mathrm{N}(\v\theta_{{t-1}} ,\mathrm{Diag}(\exp(\v h_{t}))\big)\label{eq:ss_transition}  \\
&y_t \vert \v\theta_{t} \sim \mathrm{N} \big(\v x_t^\top \tilde{\v g}(\v\theta_t), \sigma^2\big), \label{eq:ss_measurement}
\end{align}
where $\mathrm{Uniform}(\mathbb{S})$ denotes the uniform distribution over the stability region for the implied $\v\phi_0$ in each AR polynomial. Note that the state transition model in \eqref{eq:ss_transition} is linear and heteroscedastic Gaussian, and the measurement model in \eqref{eq:ss_measurement} is nonlinear with Gaussian additive errors.

Our goal is to design a Gibbs updating step where that draws from the joint conditional posterior of the parameter paths $p(\v \theta_{0:T}|y_{1:T}, \cdot)$, using the dot $\cdot$ as a placeholder for all the other parameters in the model. Sampling is ideally performed by the following two steps:
\begin{enumerate}
\item compute the sequence of \emph{filtering distributions} $p(\v \theta_{t}|y_{1:t}, \cdot)$ for $t=0,\ldots,T$; \vspace{-0.2cm}
\item sample from the \emph{joint smoothing distribution} $p(\v \theta_{0:T}|y_{1:T}, \cdot)$ via backward sampling, starting from the final filtering distribution $p(\v \theta_T|y_{1:T}, \cdot)$ and iterate back in time.
\end{enumerate}
For linear Gaussian models, the filtering posteriors are Gaussian and can be computed recursively in closed form by the Kalman filter recursions. The backward sampling step then simulates from multivariate Gaussian distributions. This is the Forward Filtering and Backward Sampling (FFBS) algorithm developed by \citet{carter1994gibbs} and \citet{fruhwirth1994data}. 

Conditional on the log-volatilities $\v h_{0:T}$, the model in \eqref{eq:ss_prior}-\eqref{eq:ss_measurement} is linear Gaussian in the transition model, but non-linear Gaussian in the measurement equation. We use the Particle Gibbs with Ancestor Sampling (PGAS) algorithm in \citet{lindsten2014particle} to sample from the conditional posterior of parameter trajectory $\v \theta_{0:T}$. Section \ref{app:pgas} in the Supplementary material explains the PGAS updating step, and refer to \cite{lindsten2014particle} for further details. The PGAS algorithm can suffer from particle degeneracy in near-degenerate models \citep{lindsten2015particle}, for example when the innovations in the transition model have close to zero variance for some $t$. This often occurs with the TVSAR in \eqref{eq: SAR_DSP} since the dynamic shrinkage priors encourage parameters to remain constant for extended time periods.  

For near-degenerate TVSAR models we suggest an alternative updating step for $\v\theta_{0:T}$ based on an approximate sampling step that replaces the Kalman filter in the FFBS sampler with the extended Kalman filter (EKF). The EKF linearizes the nonlinear observation and/or state equations via first-order Taylor expansions around the prior propagated state at each time step; see \citet{thrun2002probabilistic} for an introduction to the EKF. The derivatives needed in the EKF are computed by automatic differentiation \citep{baydin2018automatic}, making it possible to write fast and general code for TVSAR with any number of seasonal periods. The algorithm uses the same backward simulation recursions as in the FFBS algorithm; we call this algorithm FFBSx, where x is a mnemonic for extended. The extended Kalman filter needs a Gaussian prior for the initial state. We therefore approximate the prior in 
Lemma~\ref{lem:uniform_t} when using FFBSx by the closest normal distribution with respect to the Hellinger metric; see Section \ref{normal_approx_uniform} for details.

The accuracy of the EKF algorithm depends on the degree of nonlinearity in the measurement and transition equations, and how much the state and measurement noise deviate from Gaussian distributions. The TVSAR is well suited for the EKF since the state transition remains linear and Gaussian and the non-linearity is only present in the observation equation, which still has additive Gaussian errors. Our empirical results in Sections \ref{sec:simulations} and \ref{sec:applications} show that the algorithm is fast, robust and accurate for the TVSAR.

\section{Simulation experiments}\label{sec:simulations}
We conduct several simulation experiments to evaluate the effectiveness and robustness of the multi-seasonal AR model, assess the performance of the various simulation algorithms, and compare the TSVAR model against a set of benchmark models. The experimental setups include data generating processes with single and multiple seasonalities, and also an experiment where the estimated model has redundant lags.

\subsection{Setup and benchmark models}\label{subsec:sim_setup}
We simulate $100$ time series from the data generating process for each experiment. The simulated time series consist of $T=1000$ observations, and we generate \num{10000} draws from the posterior distribution for each method, after a burn-in of \num{3000} iterations. The post burn-in draws are thinned by a factor of $10$ before computing the spectral densities. In Experiments 1 and 2, we fit the same number of lags as the number of lags in the data generating process. Experiment~3 in Section \ref{app:exp3} of the supplementary material shows that the dynamic shrinkage prior is able to shrink redundant lags to zero in the TVSAR when a model with too many lags is applied to the data. The empirical application in Section \ref{sec:applications} uses the log predictive scores to determine the optimal number of lags. 

Our proposed models and algorithms are compared with the following alternatives: a linear time-varying AR model with dynamic shrinkage priors \citep{kowal2019dynamic} without explicit stability restrictions implemented in the R package \texttt{dsp}, a kernel smoothing estimation method using the \texttt{tvReg} package in R  \citep{tvRegPackage} and the time-varying AR with the dynamic triple gamma shrinkage prior \citep{knaus2023dynamic} implemented as the function \texttt{shrinkDTVP} in the \texttt{shrinkTVP} package \citep{knaus2021shrinkage}. We also compare with the AdaptSpec method \citep{rosen2012adaptspec,BertolucciBayesSpec2021} that estimates a local spectral density using smoothing splines over time segments learned by the algorithm. We use default settings for all benchmark methods with the exception of AdaptSpec, where the default number of spline knots is increased to 40 instead of the default with seven spline knots and the maximum number of segments is increased from the default 10 to 20. Preliminary experiments show that this gives AdaptSpec a better chance to capture the highly multimodal spectral density in seasonal models.

All of the time-varying AR benchmark models are designed for non-seasonal data. To give a fairer comparison, we construct seasonally adapted versions of the benchmark models. For example, a TVSAR$(1,1)_{12}$ can be formulated as a regular AR with non-zero coefficient $\phi_1$ at lag $1$, $\Phi_1$ at lag $12$ and $-\phi_1 \Phi_1$ at lag $13$, see \eqref{eq: toy_additive}. We then fit benchmark models with $13$ lags and impose zero-restrictions on the coefficients for lags 2 to 11. Note, however, that the coefficient on lag 13 is estimated as an unrestricted coefficient in the benchmark models since they cannot handle the non-linear multiplicative structure. The time-domain benchmark models do not impose stability/stationarity restrictions, and we therefore discard posterior draws outside the stability region to ensure a well-defined spectral density. Discarding non-stable draws can be formalized as using a prior truncated to the stability region with rejection sampling after the MCMC, where the acceptance probability is zero for unstable draws and one for stable draws; see Appendix B in \citet{cogley2005drifts}. This however requires the whole trajectory from $t=1$ to $t=T$ to be accepted/rejected, which leads to too many rejected draws. We instead use a pragmatic approach for the benchmark models by accepting or rejecting individually for different time periods. The difficulty with rejection sampling of whole trajectories shows the advantage of formally imposing stability restrictions in the TVSAR model.

In all three experiments, we use iid priors for the static parameters: $\mu \sim \mathrm{N}(-15,3^2)$, $\kappa \sim \mathrm{TN}(0.5, 0.3^2,-1,1)$ and $\sigma^2 \sim \mathrm{Inv-}\chi^2(3, s^2)$, a scaled inverse chi-squared distribution, where $s^2$ is the residual variance from a linear AR model fitted to the time series. Note that $\mu$ is on the log-scale, so the implied prior on the global standard deviation $\tau=\exp(\mu/2)$ is log normal with $95\%$ equal tail interval $(2.92\cdot 10^{-5}, 0.01)$. This prior encourages parameters to remain constant, but, as we will see, still allows for sizeable jumps via the heavy-tailed log-volatility innovations, $\eta_t$.

\subsection{Performance metrics}\label{subsec:performance_metrics}
Since the set of compared models includes models in both the time and frequency domains, and models with both multiplicative and additive structures, we compare the methods based on their ability to estimate the time-varying spectral density. The spectral density of a seasonal AR process at time $t$ is \citep{wei2019time}
\begin{equation}
    \label{eq: spectral_density}
    f(t,\omega)= \frac{\sigma^2}{\pi}\frac{1}{|\phi_{pt}(e^{-\im \omega})|^2}\frac{1}{|\Phi_{Pt}(e^{-\im s\omega    })|^2  } \text{ for } \omega \in (0, \pi ) \text{ and } t=1,\dots,T,
\end{equation}
where $\phi_{pt}(L)$ and $\Phi_{Pt}(L^s)$ are the usual regular and seasonal AR lag polynomials at time $t$ and $\omega \in (0, \pi ]$ is the radial frequency. This expression extends to seasonal AR models with multiple seasonal periods in the obvious way. Following \citet{rosen2012adaptspec}, we use the Mean Square Error (MSE) of the estimated log spectral density over all time periods as the performance metric
\begin{equation}
     \label{eq: MSE}
     \mathrm{MSE}( \widehat{\log f}) =\frac{1}{T m}\sum_{t=1}^{T}\sum_{k=1}^{m}\big(\widehat{\log f(t,\omega_k)}-\log f(t,\omega_k)\big)^2,
 \end{equation}
where $\log f(t,\omega)$ is the true log spectral density at time $t$ and frequency $\omega$; $\widehat{\log f(t,\omega)}$ is its estimate, and $m=314$ is the total number of frequencies on a grid from $0.01$ to $\pi$. The posterior median of the log spectral density $\log f(t,\omega)$ is used as the point estimate in the Bayesian models since it is more robust to outliers in the MCMC.

\subsection{Experiment 1 - Single-seasonal TVSAR}\label{subsec:experiment1}
\begin{figure}
 \centering
 \includegraphics[width=0.8\textwidth]{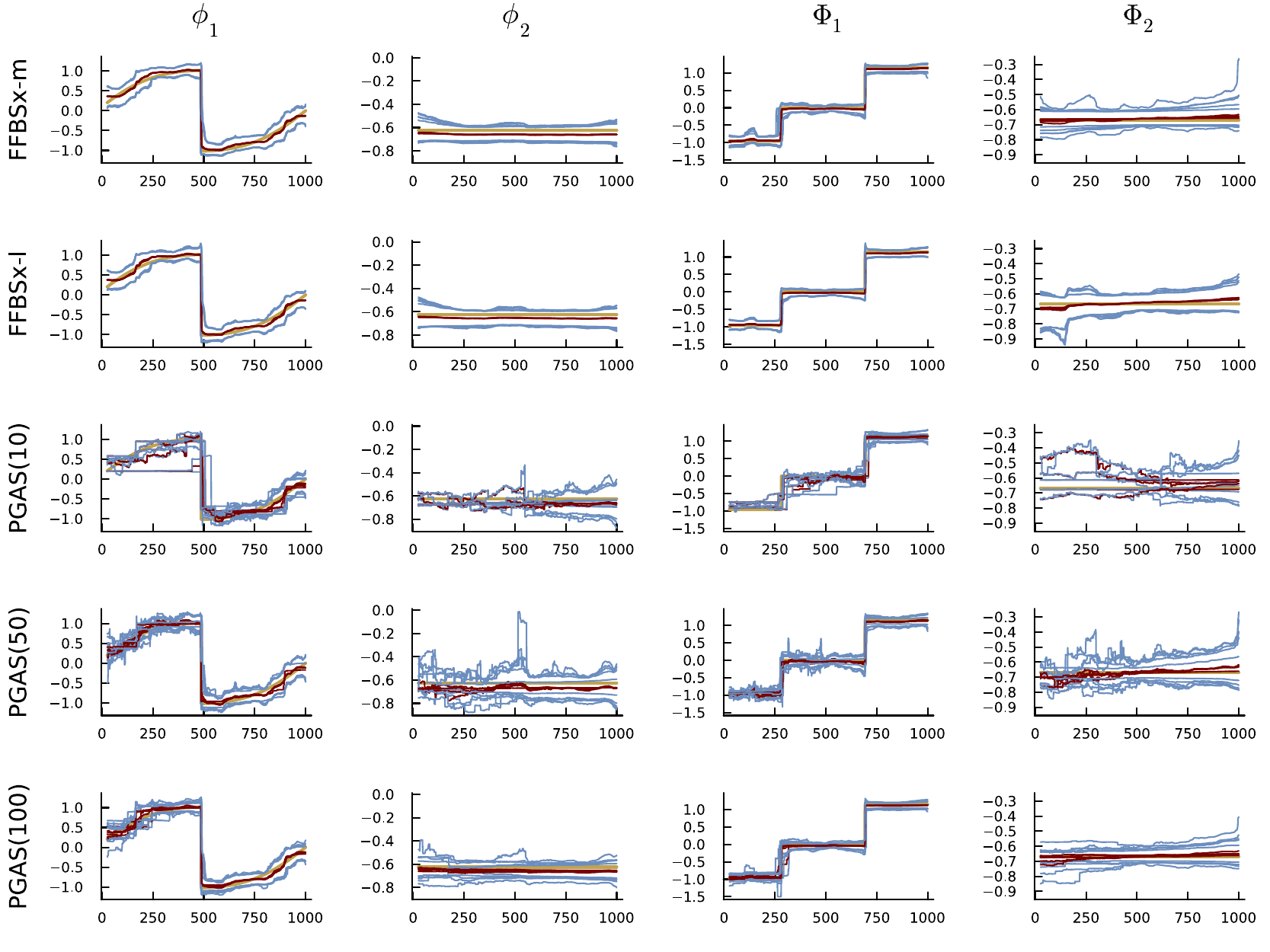}
    \caption{Experiment 1. Assessing MCMC convergence by re-estimating the model on the same dataset with five different initial values. The red and blue lines are posterior medians and $95\%$ equal tail credible bands over time for each of the five repeated runs. The beige line is the true parameter evolution in the $\phi$-parameterization.}\label{fig:exp1convergence}
\end{figure}

The first experiment simulates data from the time-varying TVSAR$(2,2)_{12}$
\begin{equation}\label{eq:exp1}
   (1-\phi_{1t} L -\phi_{2t} L^2)(1-\Phi_{1t} L^{12} -\Phi_{2t} L^{24})y_t = \varepsilon_t,
\end{equation}
where $\varepsilon_t \sim \mathrm{N}(0,1)$. 
The beige lines in Figure~\ref{fig:exp1convergence} show the evolution of the regular AR parameters $\phi_{kt}$ and the seasonal AR parameters $\Phi_{kt}$; for details, see equations \eqref{eq: Exp1_nonseas} and \eqref{eq: Exp1_seas} in Section \ref{app: exp1} of the supplementary material.

We first explore the numerical performance of the Gibbs sampler with a PGAS update step for different numbers of particles. As explained in Section \ref{subsec:PGAS}, particle Gibbs samplers like PGAS can struggle when the state transition model is nearly degenerate. Our simulation setup is a rather extreme case of near-degeneracy where some parameters are constant over extended segments or even constant throughout time. 

We compare PGAS with the more robust FFBSx algorithms based on the extended Kalman filter. The potential approximation errors from FFBSx can be attributed to the two non-linearities in the TVSAR: one from its the multiplicative structure and another from the stability restrictions. Therefore, we also apply the FFBSx algorithm to a TVSAR model without imposing the stability restrictions. The FFBSx applied to the TVSAR with stability restrictions is referred to as FFBSx-m (where the m stands for the Monahan transformation in \eqref{eq:t_transf}, while FFBSx-l refers to FFBSx applied to the TVSAR model without stability restrictions (l stands for linear).

Figure \ref{fig:exp1convergence} assesses the convergence of the algorithms by plotting the posterior medians and equal-tail $95\%$ credible intervals from five different runs on the same dataset, each with different initial values. The initial values for the global parameters are set so that $\mu_k=\mu$ and $\kappa_k = \kappa$ for all $k$, with $\mu$ and $\kappa$ drawn from their respective priors in each run of the algorithm. The initial values for the log-volatilities $h_{k,t}$ are set to the randomly generated $\mu$ for all $k$ and $t$. 
PGAS($N$) is the PGAS algorithm with $N$ particles. All PGAS algorithms use a proposal at time $t=0$ from an initial FFBSx-m run with 500 iterations. The results in Figure \ref{fig:exp1convergence} use a minimal offset of $10^{-16}$ (close to the machine precision in Julia) in the update for $\v h_{0:T}$; we return to this choice of offset below.

Figure \ref{fig:exp1convergence} shows that both the FFBSx and PGAS samplers accurately capture the evolving patterns of varying complexities and correctly identify constant parameters. However, the PGAS sampler struggles with the near-degeneracy in this example, requiring at least $100$ particles to achieve relatively stable results across different runs. In contrast, the FFBSx samplers are stable, while yielding a very similar posterior to PGAS(100). This is reassuring since FFBSx is an approximate algorithm. Moreover, the additional nonlinearity handled by FFBSx-m compared to single nonlinearity in FFBSx-l does not deteriorate stability and fit.

The effective sample size (ESS, \citep{chopin2020introduction}) of the PGAS algorithms can be very low for the earlier part of the time period, while the ESS for the FFBSx algorithms are larger and stable across time; see Table \ref{table:Experiment1_ESS} in Section \ref{app: exp1} for the ESS for the spectral density at some selected time periods and frequencies. The relatively low ESS comes from learning the global and local variances in the dynamic shrinkage process, a well known challenging issue with horseshoe-type of priors \citep{piironen2017sparsity}, an area which would benefit from more research.

The FFBSx samplers are substantially faster than the PGAS samplers. One thousand draws with FFBSx-m takes 25 seconds, compared to 23 seconds for FFBSx-l and 258 seconds with PGAS(100). The timings are for Julia code on a single core of an Apple M1 processor with 16 GB of RAM memory; the code is not optimized and timings should therefore be considered as rather loose upper bounds. The robustness, speed and accuracy of FFBSx makes it our preferred algorithm. When we refer to the TVSAR model below without mentioning the sampling algorithm, we mean the TVSAR with stability restrictions estimated with the FFBSx algorithm.

\begin{figure}
 \centering
 \includegraphics[width=0.3\textwidth]{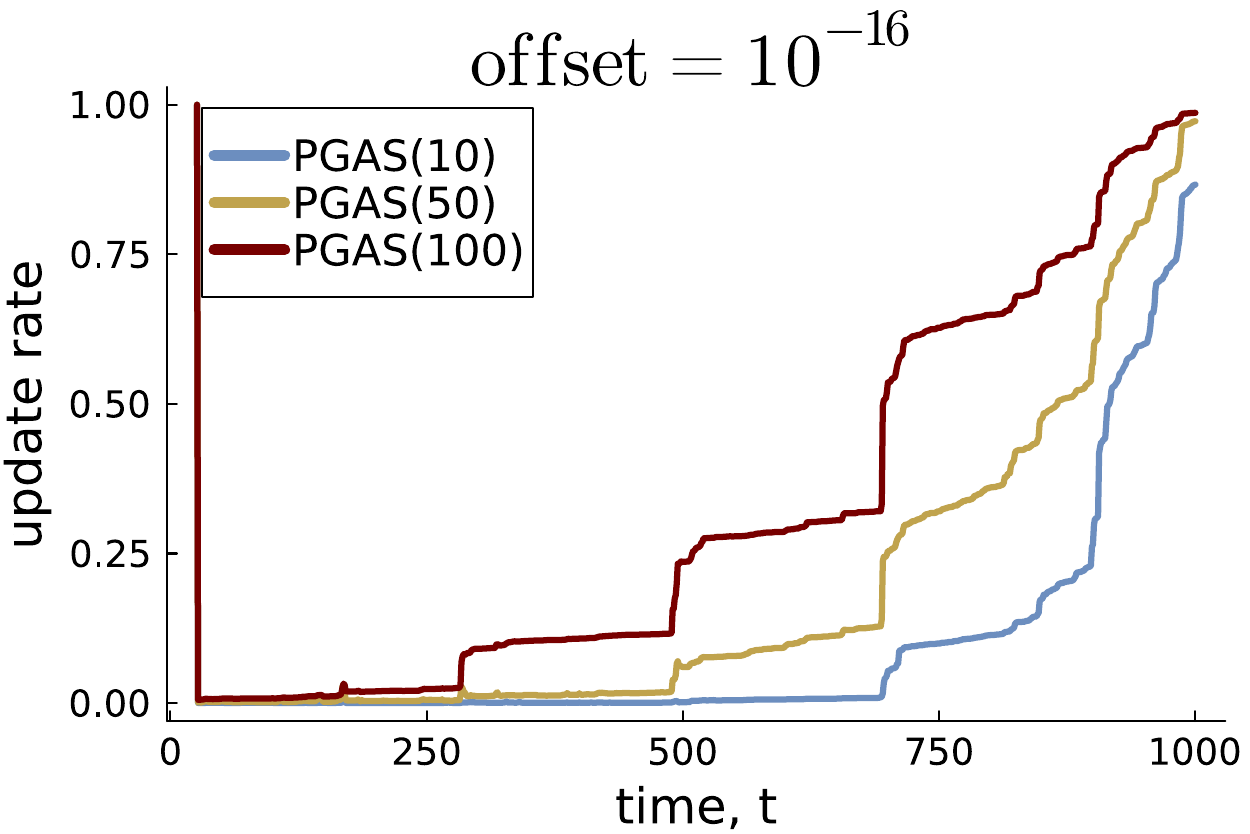}
 \includegraphics[width=0.3\textwidth]{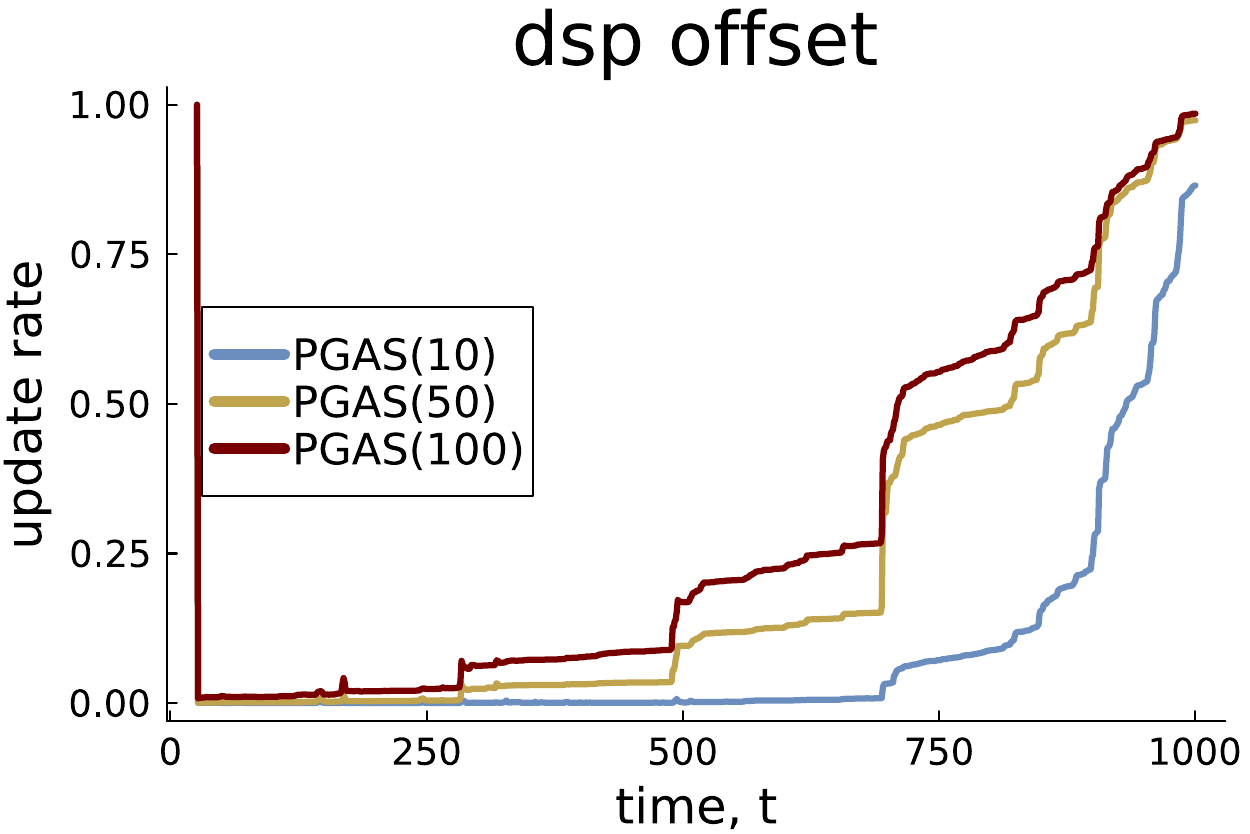}
 \includegraphics[width=0.3\textwidth]{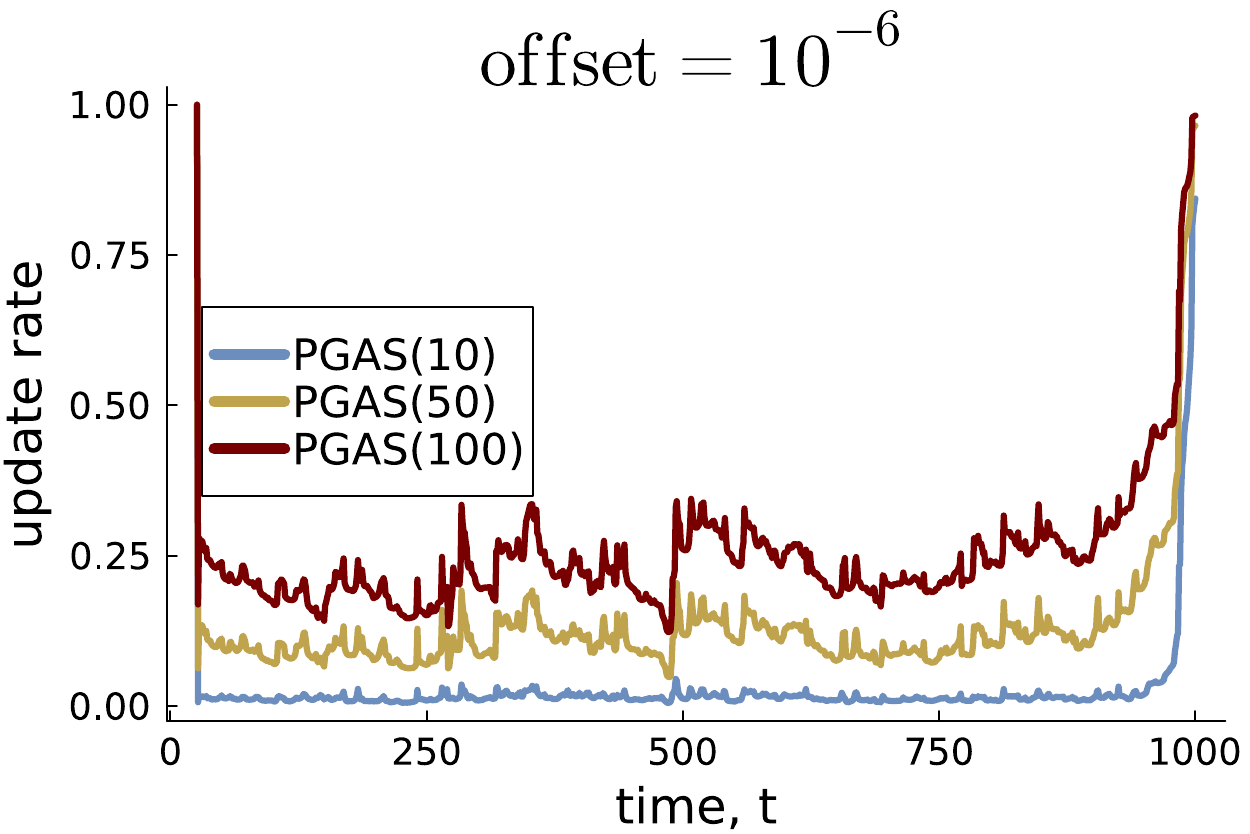}
\caption{Experiment 1. PGAS update rates for three different offsets in the log-volatility model for $\v h_t$ for different number of particles.}\label{fig:exp1updaterates}
\end{figure}

\begin{figure}
 \centering
 \includegraphics[width=1\textwidth]{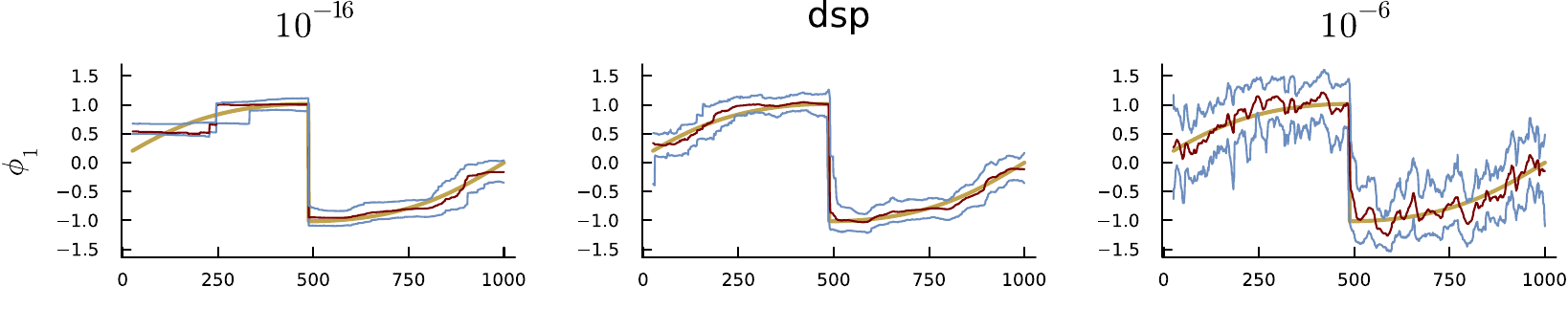}
\caption{Experiment 1. Posterior median and 95\% credible intervals for $\phi_{1t}$ from PGAS(100) from one run for the three different offsets in the log-volatility model for $\v h_t$. See Figure \ref{fig:exp1_postOffsets} of the supplement for the remaining parameters. }\label{fig:exp1_postOffsets_phi1}
\end{figure}

Figure~\ref{fig:exp1updaterates} plots the update rates for the TVSAR parameters for the PGAS algorithms for three different offsets in the update for $\v h_{1:T}$ in \eqref{logvolatility_offset}. Offsets are standard in stochastic volatility models \citep{kim1998stochastic} to avoid taking logs of zero, but also artificially inflates the inferred $h_t$, which can be important when using the DSP prior. The dsp offset in Figure~\ref{fig:exp1updaterates} uses the rule in \eqref{dsp_offset} from the \texttt{dsp} package \citep{kowal2019dynamic}. The update rate is the proportion of times the parameter changes over the iterations and is here computed on all \num{10000} draws before thinning. A larger offset makes the state transition less degenerate with higher update rates, but with more wiggly parameter evolutions, making it harder to detect constant parameters; see 
Figure~\ref{fig:exp1_postOffsets_phi1} for $\phi_{1t}$ and 
Figure~\ref{fig:exp1_postOffsets} for the remaining parameters. Importantly, the results in Figure \ref{fig:exp1convergence} are based on the smallest offset $10^{-16}$, showing that FFBSx works well even with essentially no offset. The effect of the near-degeneracy on PGAS can be less pronounced in other examples; see Experiment 2 below, and the results for TVSAR with one regular and one seasonal lag in Section~\ref{app: sar11}. 

\begin{figure}
 \centering
 \includegraphics[width=0.45\textwidth]{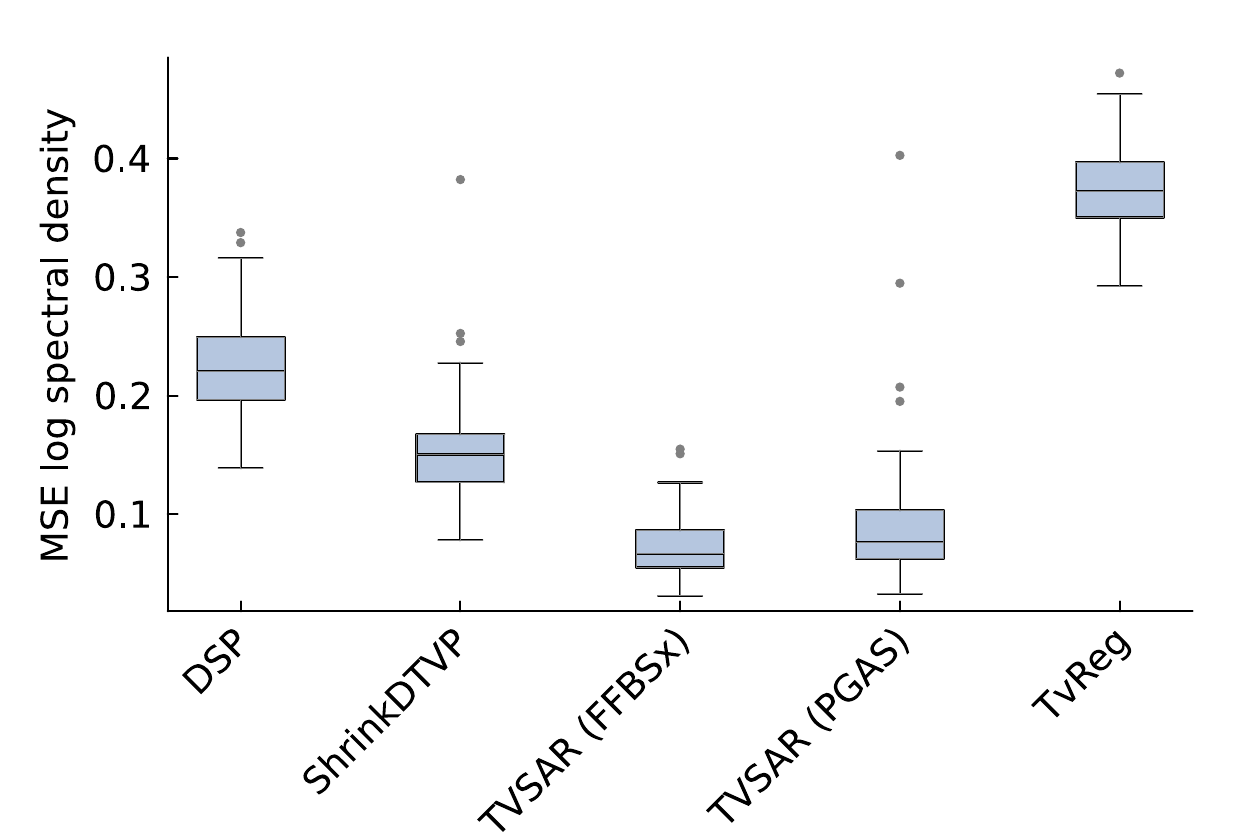} \includegraphics[width=0.45\textwidth]{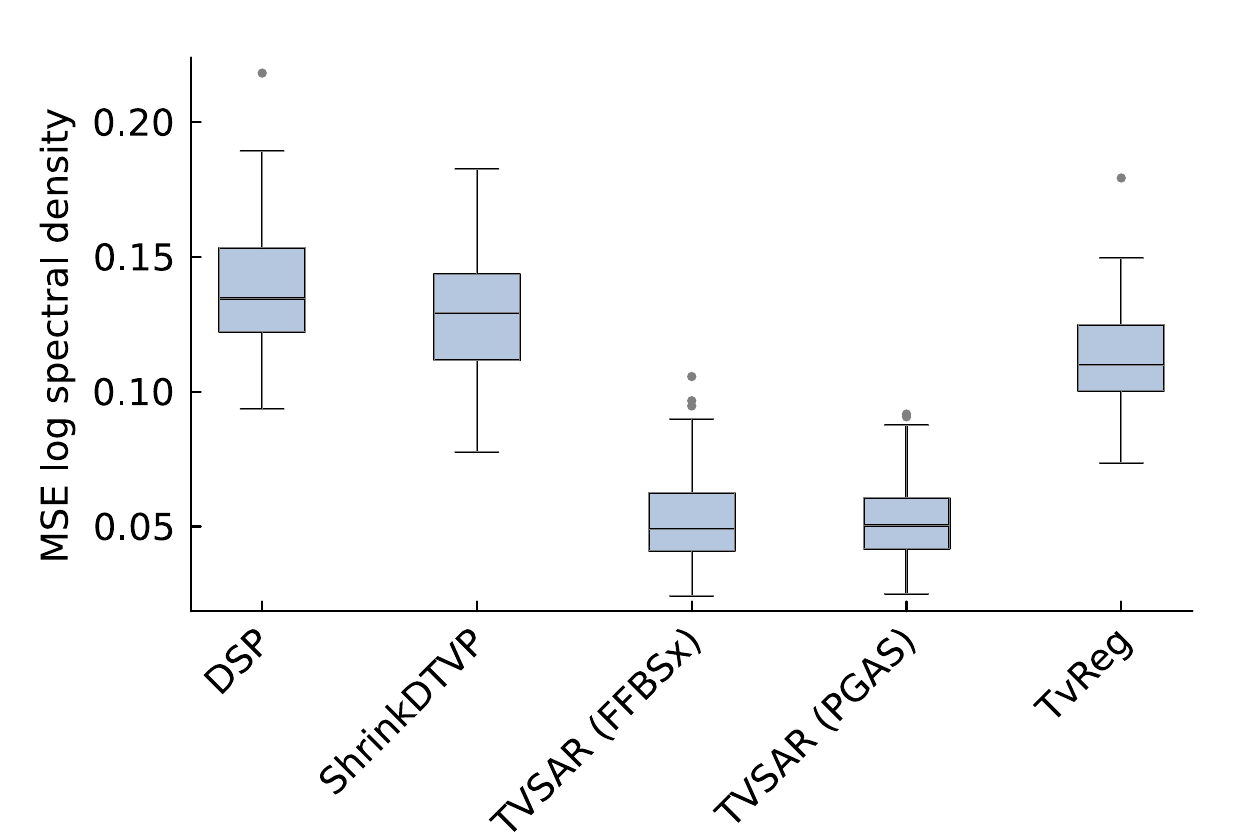}
    \caption{Box plots of the MSE for the log spectral density over time and frequency from 100 simulated datasets for Experiment 1 (left) and Experiment 2 (right).}\label{fig:exp1MSE}
\end{figure}   

The left graph of Figure \ref{fig:exp1MSE} displays box plots of the MSE metric in \eqref{eq: MSE} for the compared models over the $100$ simulated datasets. The median MSE for AdaptSpec is approximately $0.9$ with a couple of outliers around $1.5$, and is not shown in Figure \ref{fig:exp1MSE} to make the vertical scale more readable for comparing the other methods; AdaptSpec is not specifically designed for seasonal data and therefore struggles here. The TVSAR model estimated with the FFBSx algorithm performs best among the compared models, with a consistently low MSE across the datasets. The performance of the TVSAR model without stability restrictions is similar and is not shown. The linear DSP model outperforms TvReg with a sizeable margin, but the dynamic triple Gamma shrinkage prior (ShrinkDTVP) is the best among the benchmark methods. 

\begin{figure}
 \centering
 \includegraphics[width=0.9\textwidth]{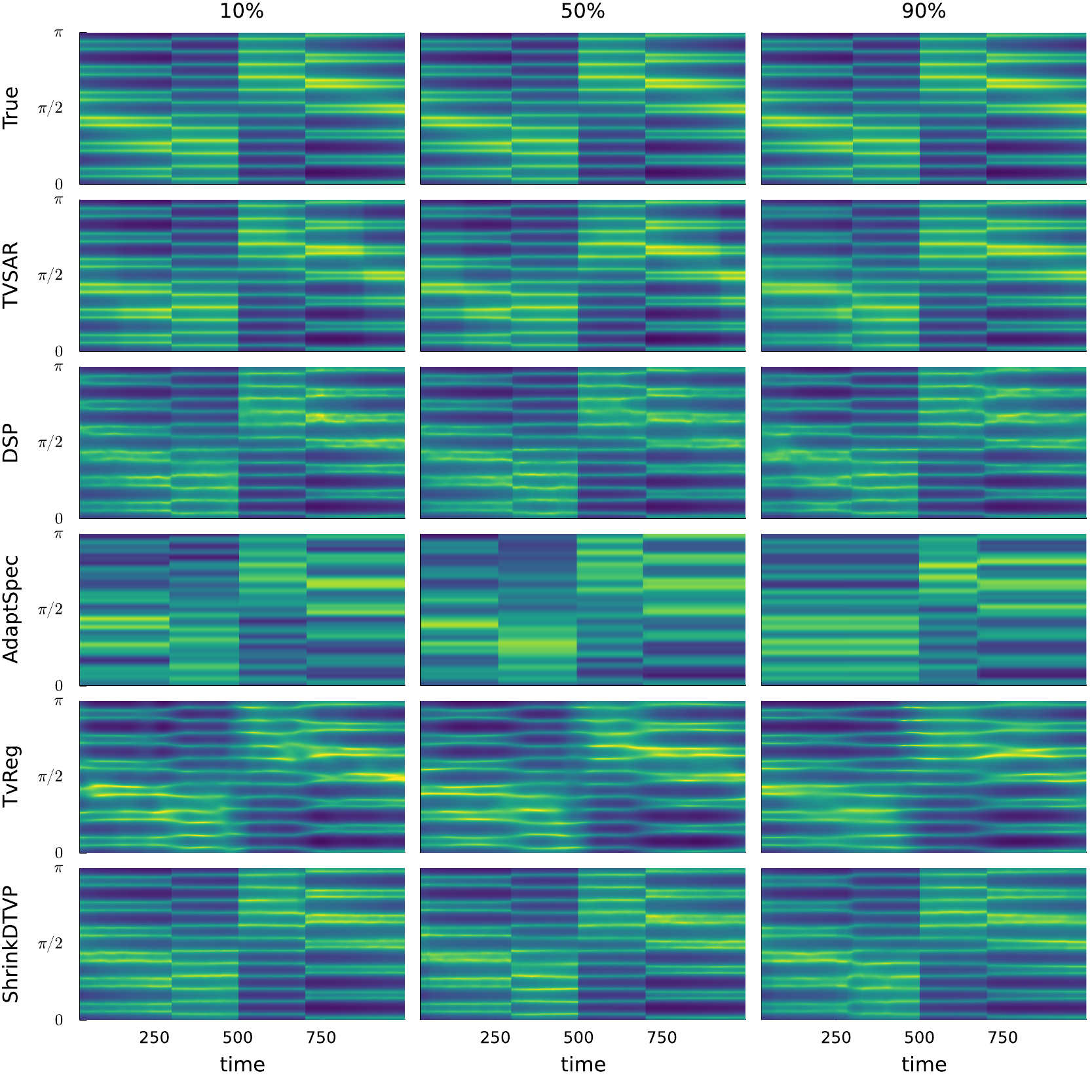}
    \caption{Experiment 1. Heatmaps of the estimated log spectral density over time for the different methods. The columns correspond to different datasets chosen from the percentiles of each model's MSE distribution, to show the performance of each model when it performs well ($10\%$), average ($50\%$) and poorly ($90\%$).}\label{fig:exp1heatmaps}
\end{figure}

Figure~\ref{fig:exp1heatmaps} displays heatmaps of the estimated time-varying log spectral density for the different models. Following \citet{rosen2012adaptspec}, we plot the posterior median of the time varying log spectral densities for three different datasets, corresponding to the $10\%$, $50\%$, and $90\%$ percentiles of the MSE values for each method. This show how each method performs when it performs well ($10\%$), average ($50\%$), and poorly ($90\%$). Note that the columns of Figure~\ref{fig:exp1heatmaps} are therefore for potentially different datasets for the different methods. See Figures~\ref{fig:exp1spectralovertime} and \ref{fig:exp1spectraloverfreq} in the supplementary material for an alternative view where  the fitted spectral densities are plotted at some selected times and frequencies. Our TVSAR model (with stability restrictions, estimated by FFBSx) gives less wiggly fits than DSP, while TvReg oversmooths.  ShrinkDTVP performs relatively well in capturing the jumps, but is also more wiggly than the TVSAR model.

\subsection{Experiment 2 - Multi-seasonal TVSAR}\label{subsec:experiment2}
\begin{figure}
 \centering
 \includegraphics[width=0.9\textwidth]{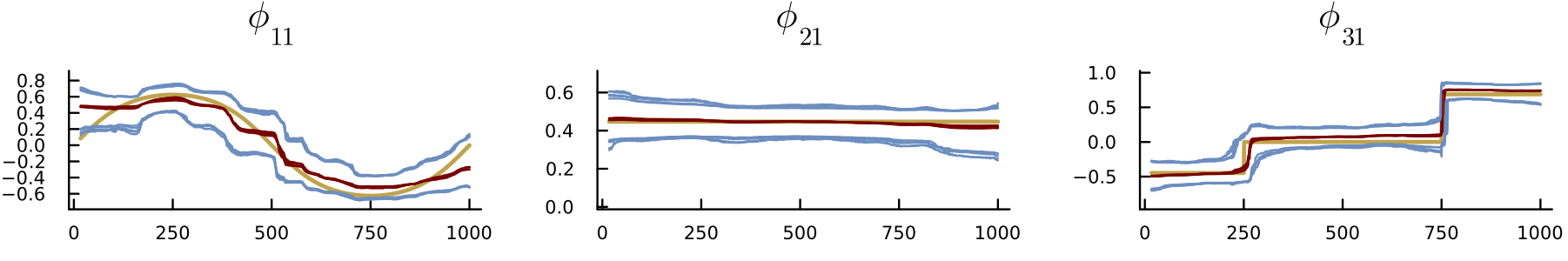}
    \caption{Experiment 2. Assessing MCMC convergence by re-estimating the model using the FFBS-x algorithm on the same dataset with five different initial values. The red and blue lines are posterior medians and $95\%$ equal tail credible bands over time for each of the five repeated runs. The beige line is the true parameter evolution in the $\phi$-parameterization. See also Figure \ref{fig:exp2convergence} for PGAS results with different number of particles.}\label{fig:exp2convergence_FFBSx}
\end{figure}

The second experiment simulates data from a multi-seasonal TVSAR with two seasonal periods $\v s=(4,12)$, non-seasonal order $p=1$ and seasonal orders $\v P=(1, 1)$ (i.e., one lag for each seasonal period). This corresponds to the TVSAR$(p=1, \v P = [1,1])_{[4,12]}$ model
\begin{equation}\label{eq: exp2}
   (1-\phi_{11t} L) (1-\Phi_{11t} L^4)(1-\Phi_{21t} L^{12})y_t = \varepsilon_t, 
\end{equation}
with $\varepsilon_t \sim \mathrm{N}(0,1)$. The time evolution of the parameters are plotted as beige  lines in Figure \ref{fig:exp2convergence_FFBSx}; see \eqref{eq: Exp2_nonseas} and \label{eq: Exp2_seas2} in Section \ref{app: exp2} of the supplementary material for details.

Figure \ref{fig:exp2convergence_FFBSx} shows that the five runs of FFBSx-m are nearly identical and very close to the results from PGAS(100) in Figure \ref{fig:exp2convergence}, showing that the extended Kalman filter is very accurate, even with both nonlinearities present in the model. Figures \ref{fig:exp2convergence} and \ref{fig:exp2updaterates} also show that the posterior from the PGAS algorithm across different initial values is stabler compared to Experiment~1, but at least $N=50$ particles are necessary. 

The right panel of Figure \ref{fig:exp1MSE} displays box plots of the MSE for the log spectral density over 100 simulated datasets for Experiment 2. The TVSAR model performs best, regardless of the sampling algorithm. The median MSE of AdaptSpec is approximately $0.33$ (not shown in the figure). In contrast to Experiment 1, TvReg is now slightly better than DSP and ShrinkDTVP, which here perform similarly. The improved performance of TvReg is likely due to the smoother parameter trajectories in Experiment 2, which TvReg is known to handle well \citep{tvRegPaper}. The results in Section \ref{app: exp2} in the supplementary material shows that the properties of the fitted spectral densities are qualitatively similar to the corresponding results in Experiment 1. 

\section{Application to US industrial production}\label{sec:applications}

\begin{figure}
 \centering
 \includegraphics[width=0.4\textwidth]{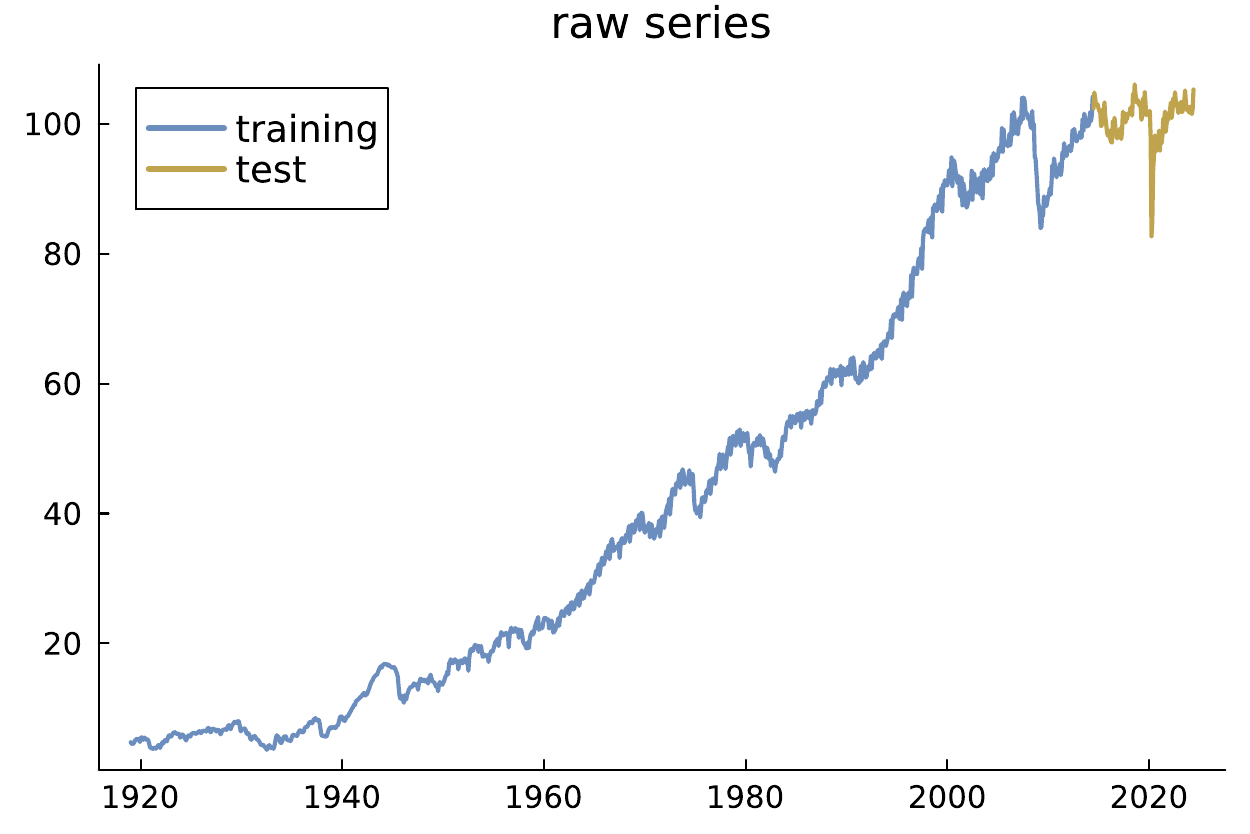}
  \includegraphics[width=0.4\textwidth]{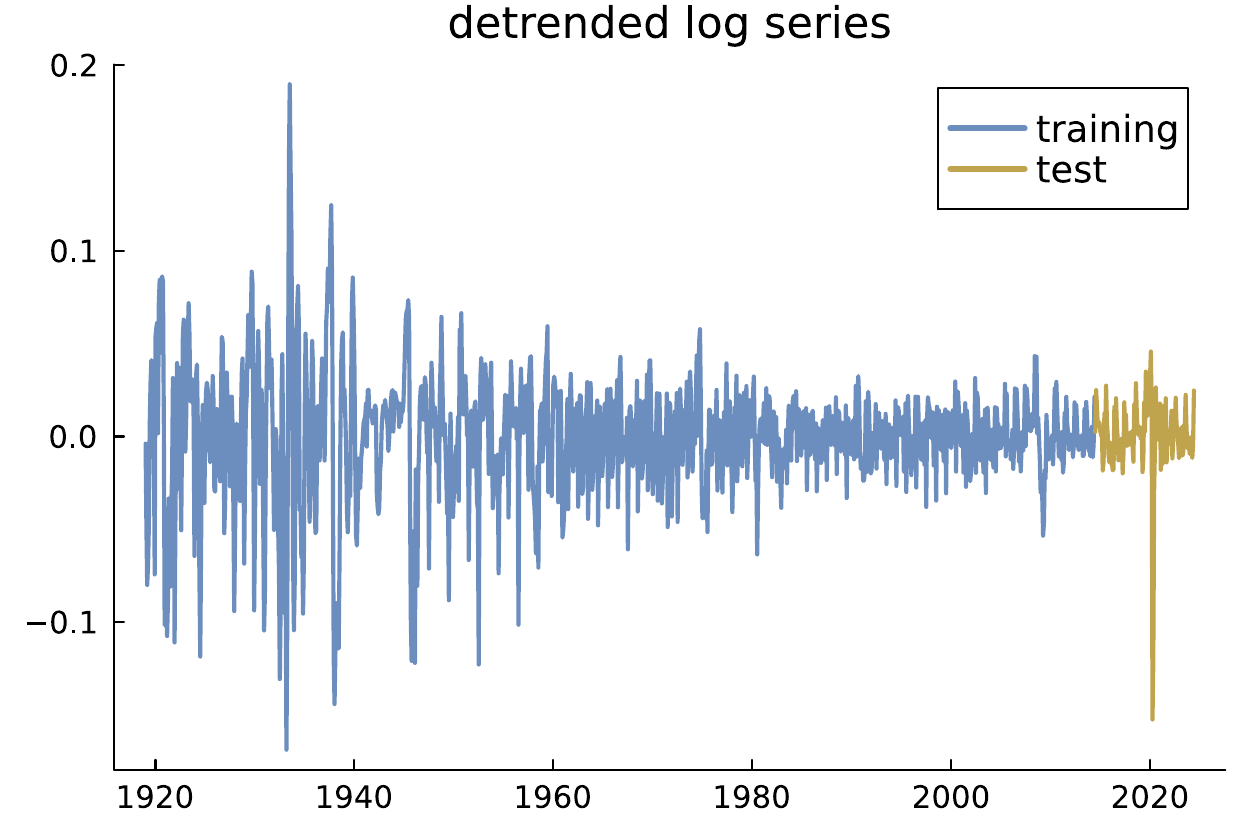}
    \caption{US industrial production time during Jan 1919 - June 2024. The test period used for model comparison, July 2014 - June 2024, is plotted in beige.}\label{fig:usip_data}
\end{figure}

We analyze monthly industrial production in the US between Jan 1919 - June 2024 from the FRED database (\url{https://fred.stlouisfed.org/series/INDPRO}); the left panel in Figure~\ref{fig:usip_data} shows the data, with the test data used for model selection in beige. To focus on the time-varying AR parameters, we log and detrend the data prior to the analysis by subtracting a local mean estimated with the \texttt{TvReg} package. The right panel of Figure \ref{fig:usip_data} shows the detrended data. Following \citet{kowal2019dynamic}, we do not remove the time-varying variance of the data. Instead, we model the variance by a stochastic volatility component for $\varepsilon_t$, where the log-volatility $g_t = \log\sigma_t^2$ is modeled as an AR(1) process
\begin{equation}\label{eq:SVmodel}
g_t = \mu_g + \kappa_g(g_{t-1} - \mu_g) + \zeta_t, \quad \zeta_t \overset{\mathrm{iid}}{\sim}\mathrm{N}(0,\sigma^2_\zeta).
\end{equation}
The parameter path $g_{1:T}$, along with the global mean, $\mu_g$, the innovation variance,$\sigma^2_\zeta$, and the persistence of the log-volatility process, $\kappa_g$, are updated in separate Gibbs steps. 

Given the monthly frequency of the data, we assume a seasonal period of $s = 12$, corresponding to an annual cycle. To determine a suitable number of regular lags $p$ and seasonal lags $P$, Table \ref{table:usip_lps} reports the out-of-sample log predictive score (LPS) for TVSAR models with different number of lags over the test period July 2014 - June 2024, with a total of $120$ monthly observations. The most recent posterior distribution available at each time point is used when computing the predictive distribution; see Section \ref{app: lps} of the supplementary material for details on the LPS estimates. Note that this is a difficult period to forecast due to two massive outliers during the Covid-19 pandemic, which are here handled in-sample by the stochastic volatility part of the model, but are of course hard to forecast ahead of time. The TVSAR$(1,3)_{12}$ model has the highest LPS, but the more parsimonious TVSAR$(1,2)_{12}$ is within one LPS unit of the TVSAR$(1,3)_{12}$ model and will therefore be used here for illustration. 

\begin{table}
\begin{center}
\begin{tabular}{c | c c c}
      & P = 1  &  P = 2 &  P = 3  \\
      \cline{1-4}
p = 1 & 361.73 & \textit{365.99} & \textbf{366.82} \\
p = 2 & 360.85 & 362.08 & 364.89 \\
p = 3 & 357.71 & 365.93 & 366.14  \\ 
\end{tabular}
\caption{Log predictive scores (LPS) for $1$-step-ahead forecasts over the 120 months test period July 2014 - June 2024 in the US industrial production data. Averages across three separate runs. The model with largest LPS is in bold font and the most parsimonious model within 1 LPS of the model with highest LPS is shown in italic.}
\label{table:usip_lps}
\end{center}
\end{table}

\begin{figure}
 \centering
 \includegraphics[width=0.8\textwidth]{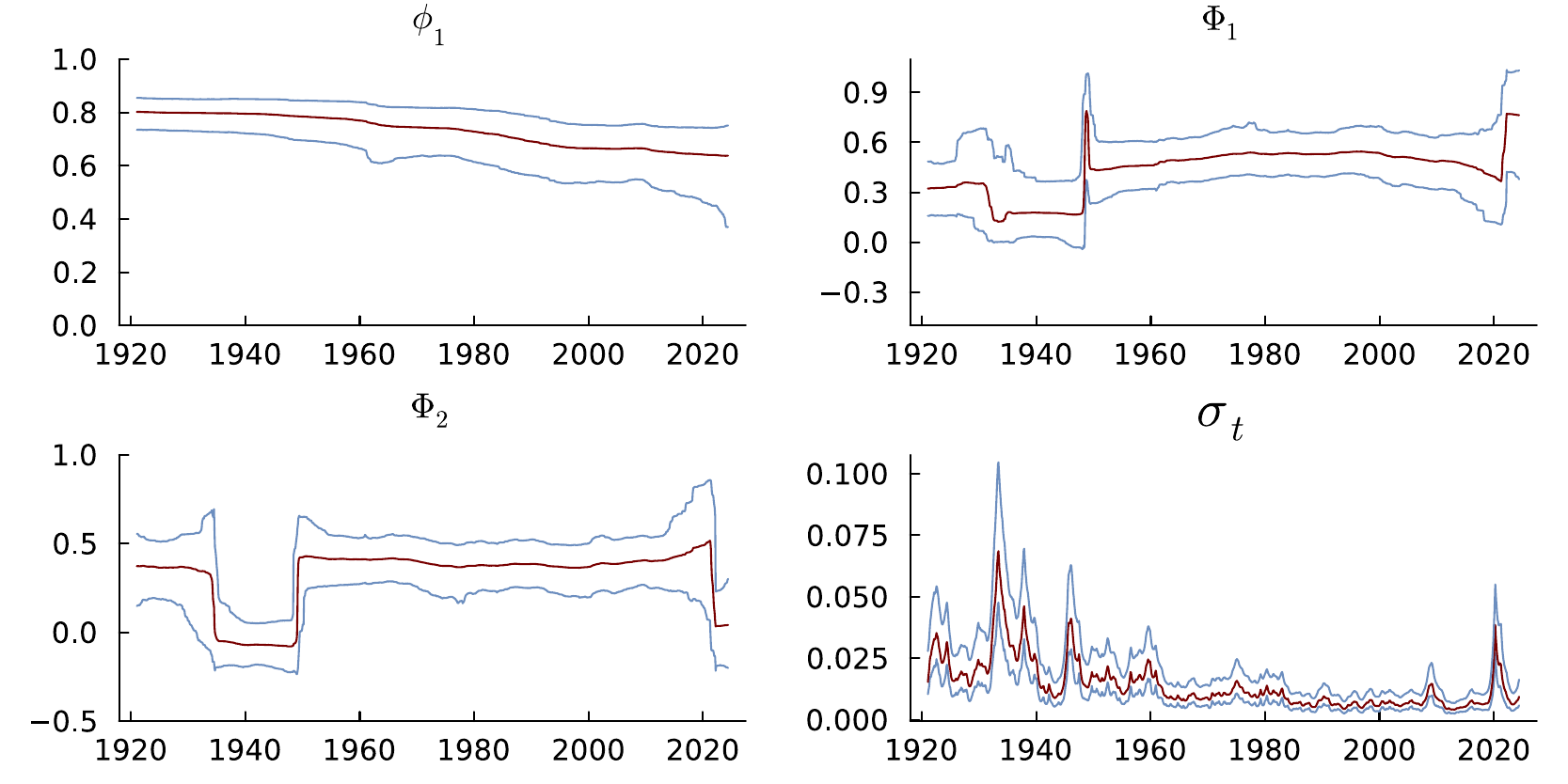}
    \caption{US industrial production. The posterior median and $95\%$ credible intervals from the TVSAR$(1,2)_{12}$ model with dynamic shrinkage prior and stochastic volatility, obtained with the FFBSx algorithm using an offset of $10^{-16}$.}\label{fig:usip_SAR12_ffbsx}
\end{figure}

Figure~\ref{fig:usip_SAR12_ffbsx} shows the posterior medians and $95\%$ credible intervals for the AR coefficients in the TVSAR$(1,2)_{12}$ model from \num{10000} post burn-in draws obtained with the FFBSx algorithm. The regular AR coefficient in Figure~\ref{fig:usip_SAR12_ffbsx} is stable over the century with only a mild downward trend. The two seasonal AR coefficients show substantial variation over time and jump to new levels during the Great Depression and WW2 period and the recent Covid-19 pandemic, while remaining very stable in the period in between these events. The posterior for the error standard deviation $\sigma_t$ in the lower right panel of Figure~\ref{fig:usip_SAR12_ffbsx} shows that there is a great deal of time-varying volatility in the data. Figure~\ref{fig:usip_SAR12_ffbsx_threeruns} shows that the FFBSx sampler has converged by plotting the posterior from three runs with different initial values drawn from the prior. PGAS with 100 particles gives a similar posterior to FFBSx; see Figure~\ref{fig:usip_SAR12_pgas100}.

Figure~\ref{fig:usip_SAR12_Gaussian_ffbsx} in the supplementary material displays the posterior results from a TVSAR$(1,2)_{12}$ model but with a traditional parameter evolution with homoscedastic Gaussian innovations. The results are qualitatively similar to the results from the dynamic shrinkage prior in 
Figure~\ref{fig:usip_SAR12_ffbsx}, but the parameter paths are much more wiggly in stable periods and oversmooth whenever there are jumps, a consequence of the constant parameter innovation variance. 

\begin{figure}
 \centering
 \includegraphics[width=0.9\textwidth]{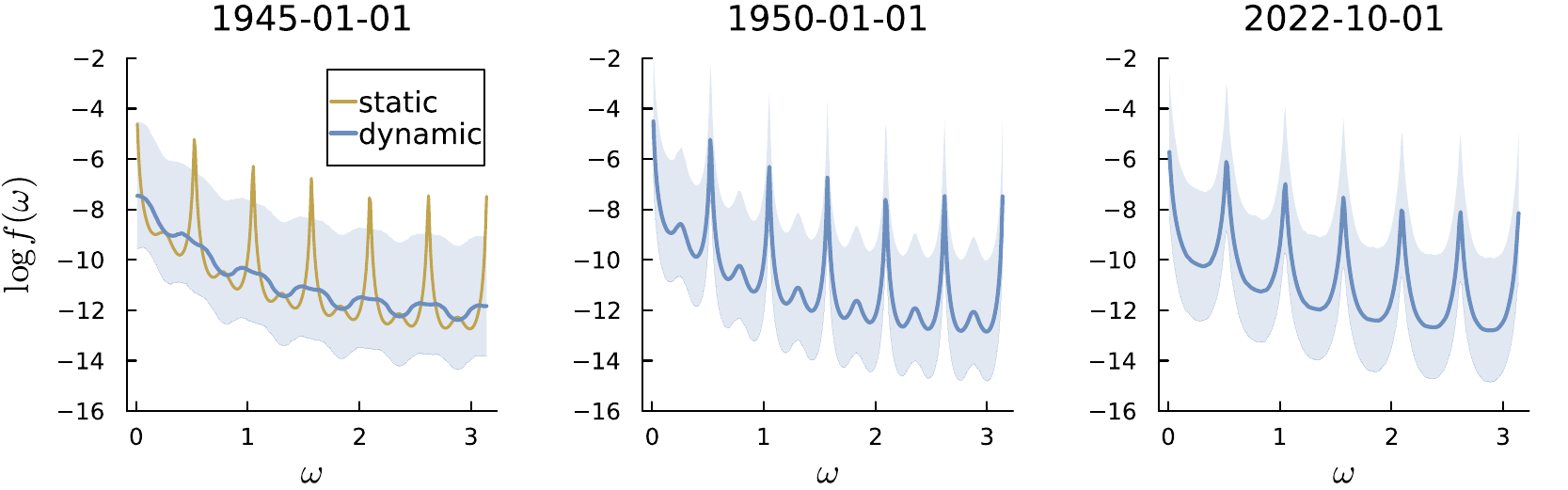}
    \caption{US industrial production. Fitted spectral density at three different time periods from TVSAR(1,2) model (blue line is posterior median and shaded regions are 95\% credible intervals) and static parameters (beige line is posterior median). Both models have time-varying error variance. An interactive graph is available at https://observablehq.com/@mattiasvillani/tvsar-spectral-usip.}\label{fig:usip_SAR12_spectral_snapshots}
\end{figure}

Figures~\ref{fig:usip_SAR12_spectral_snapshots} and \ref{fig:usip_SAR12_spectrograms} plot time evolution of the log spectral density over time. The spectral density during the Great Depression and WW2 period has substantially less seasonal variation compared to other time periods, and the Covid-19 pandemic also stands out. Fitting a $SAR(1,2)_{12}$ with constant parameters gives posterior median estimates $\hat\phi_1=0.759$, $\hat\Phi_1 = 0.539$ and $\hat\Phi_2 = 0.356$ and a fitted spectral density (beige line in Figure \ref{fig:usip_SAR12_spectral_snapshots}) dominated by the long stable period 1950-2020. 

\section{Conclusions}\label{sec:conclusions}

Our article proposes a time-varying multi-seasonal AR model with
the parameters in the non-seasonal and the seasonal
polynomials following the dynamic shrinkage process priors in 
\citet{kowal2019dynamic}. We show empirically that the model can capture extended periods of constant parameters, periods of rapid changes and abrupt jumps, as well as shrinking redundant lags to zero. In every time period the freely evolving parameters are mapped to the AR parameter stability region to ensure that parameters do not drift into explosive regions. Our framework can be used with any of the recently developed global-local shrinkage priors, for example the prior processes in \citet{kalli2014time},  \citet{cadonna2020triple} and \citet{knaus2023dynamic}. The model allows for multiple seasonal periods, which is a common feature in modern datasets where data is often observed hourly or at even high frequencies. The seasonalities are modelled with the parsimonious multiplicative structure in \citet{box2015time}. 

Our Gibbs sampling algorithm samples from the joint posterior distribution of all the model parameters. The multiplicative model structure and the restriction to a stable process makes the model nonlinear and the update step for the time-varying AR parameters challenging. We propose the FFBSx sampler based on the extended Kalman filter. We show that this approximate sampler is accurate by comparing its posterior with PGAS \citep{lindsten2014particle}, a state-of-the-art particle MCMC sampler which is known to be simulation consistent. The FFBSx sampler is fast, accurate and robust to the near-degeneracy in the state transition caused by the dynamic shrinkage process prior. The gradient in the FFBSx algorithm is efficiently computed by automatic differentiation, making it easy to write general computer code for any number of seasonal AR polynomials. 

The model and inference method are explored in three simulated examples with different numbers of seasonalities, and shown to compare well to several benchmarks. An application to more than a century of monthly US industrial production data detects interesting and large changes in seasonality during the Great Depression and the recent Covid-19 pandemic, separated by a long period of remarkably constant seasonality.

There are several interesting directions for future research. First, the Gibbs sampling of $\v h_{0:T}$, $\v\mu$ and $\v\kappa$ can benefit from better mixing, for example using the interweaving strategy in \citet{yu2011center,kastner2014ancillarity}. We have experimented with partial interweaving by sampling $\v\mu$ in both centered and non-centered parameterizations, but with limited success. Full interweaving is challenging since the $T$ P\'olya-Gamma variables needs to be sampled efficiently in the non-centered parameterization.

Second, as illustrated in Section \ref{app: nearnonstable}, the FFBSx sampler for the TVSAR model with stability restrictions can give a distorted posterior distribution when the data generating process is close to the non-stable region for extended periods. We are currently exploring other FFBS posterior samplers for the TVSAR model based on several recently proposed Kalman-based approximate filters in the signal processing literature \citep{skoglund2015extended, skoglund2019iterative, garcia2015posterior} to get a more accurate posterior approximation in the near non-stable setting. See Section \ref{app: nearnonstable} for an example where the unscented Kalman filter \citep{julier2004unscented} improves on the FFBSx sampler based on the extended Kalman filter. We are also exploring other functions from the unrestricted parameters to the partial autocorrelations than the one used in this article, to improve the performance of the posterior samplers.

Third, several extensions of the TVSAR model should be developed. An extension of the model to the ARMA case with the exact likelihood function is already in progress. Given the good performance of the dynamic triple gamma prior as a linear benchmark model, it would also be interesting to compare the performance of the DSP and triple gamma priors in the TVSAR model. Extensions to multivariate seasonal vector AR and ARMA models are possible using the multivariate stability and invertibility restrictions in \citet{ansley1986note}, but care is needed to handle the much larger number of parameters. 

Finally, we also plan to explore the forecasting performance of the proposed model in a separate article, along with the potential interplay between time-varying AR parameters and a time-varying mean of the process. 

\section*{Acknowledgments}
The authors thank the associate editor and two anonymous reviewers for many constructive comments that improved the manuscript.

\section*{Funding}
Mattias Villani was partially funded by the Swedish Research Council under Grant 2020-02846. The computations were enabled by resources provided by the National Academic Infrastructure for Supercomputing in Sweden (NAISS), partially funded by the Swedish Research Council through grant agreement no. 2022-06725.

\section*{Disclosure Statement}
The authors report there are no competing interests to declare.

\singlespacing 

\bibliographystyle{apalike}
\bibliography{localarma_ref}

@article{ansley1986note,
  title={A note on reparameterizing a vector autoregressive moving average model to enforce stationarity},
  author={Ansley, Craig F and Kohn, Robert},
  journal={Journal of Statistical Computation and Simulation},
  volume={24},
  number={2},
  pages={99--106},
  year={1986},
  publisher={Taylor \& Francis}
}

@Article{Barndorff-Nielsen1973,
  author    = {Barndorff-Nielsen, O and Schou, G},
  title     = {On the parametrization of autoregressive models by partial autocorrelations},
  journal   = {Journal of Multivariate Analysis},
  year      = {1973},
  volume    = {3},
  number    = {4},
  publisher = {Elsevier},
}

@article{barndorff1982normal,
  title={Normal variance-mean mixtures and z distributions},
  author={Barndorff-Nielsen, Ole and Kent, John and S{\o}rensen, Michael},
  journal={International Statistical Review},
  pages={145--159},
  year={1982},
  publisher={JSTOR}
}

@misc{BertolucciBayesSpec2021,
  author = {Bertolacci, M.},
  title = {{BayesSpec}},
  year = {2021},
  publisher = {GitHub},
  journal = {GitHub repository},
  howpublished = {\url{https://github.com/mbertolacci/BayesSpec.git}}
}

@article{cadonna2020triple,
  title={Triple the gamma—A unifying shrinkage prior for variance and variable selection in sparse state space and {TVP} models},
  author={Cadonna, Annalisa and Fr{\"u}hwirth-Schnatter, Sylvia and Knaus, Peter},
  journal={Econometrics},
  volume={8},
  number={2},
  pages={20},
  year={2020},
  publisher={MDPI}
}

@article{carvalho2010horseshoe,
  title={The horseshoe estimator for sparse signals},
  author={Carvalho, Carlos M and Polson, Nicholas G and Scott, James G},
  journal={Biometrika},
  volume={97},
  number={2},
  pages={465--480},
  year={2010},
  publisher={Oxford University Press}
}

@article{chib1998estimation,
  title={Estimation and comparison of multiple change-point models},
  author={Chib, Siddhartha},
  journal={Journal of econometrics},
  volume={86},
  number={2},
  pages={221--241},
  year={1998},
  publisher={Elsevier}
}

@article{dahlhaus2000likelihood,
  title={A likelihood approximation for locally stationary processes},
  author={Dahlhaus, Rainer},
  journal={The Annals of Statistics},
  volume={28},
  number={6},
  pages={1762--1794},
  year={2000},
  publisher={Institute of Mathematical Statistics}
}

@article{everitt2013online,
  title={Online {Bayesian} inference in some time-frequency representations of non-stationary processes},
  author={Everitt, Richard Geoffrey and Andrieu, Christophe and Davy, Manuel},
  journal={IEEE transactions on signal processing},
  volume={61},
  number={22},
  pages={5755--5766},
  year={2013},
  publisher={IEEE}
}

@article{granger1980introduction,
  title={An introduction to long-memory time series models and fractional differencing},
  author={Granger, Clive WJ and Joyeux, Roselyne},
  journal={Journal of time series analysis},
  volume={1},
  number={1},
  pages={15--29},
  year={1980},
  publisher={Wiley Online Library}
}

@book{hamilton2020time,
  title={Time series analysis},
  author={Hamilton, James D},
  year={1994},
  publisher={Princeton university press}
}

@article{hamilton1989new,
  title={A new approach to the economic analysis of nonstationary time series and the business cycle},
  author={Hamilton, James D},
  journal={Econometrica},
  pages={357--384},
  year={1989},
  publisher={JSTOR}
}

@article{kalli2014time,
  title={Time-varying sparsity in dynamic regression models},
  author={Kalli, Maria and Griffin, Jim E},
  journal={Journal of Econometrics},
  volume={178},
  number={2},
  pages={779--793},
  year={2014},
  publisher={Elsevier}
}

@article{kowal2019dynamic,
  title={Dynamic shrinkage processes},
  author={Kowal, Daniel R and Matteson, David S and Ruppert, David},
  journal={Journal of the Royal Statistical Society: Series B (Statistical Methodology)},
  volume={81},
  number={4},
  pages={781--804},
  year={2019},
  publisher={Wiley Online Library}
}

@article{monahan1984note,
  title={A note on enforcing stationarity in autoregressive-moving average models},
  author={Monahan, John F},
  journal={Biometrika},
  volume={71},
  number={2},
  pages={403--404},
  year={1984},
  publisher={Oxford University Press}
}

@article{jones1987beta,
 ISSN = {00359254, 14679876},
 URL = {http://www.jstor.org/stable/2347544},
 author = {M. C. Jones},
 journal = {Journal of the Royal Statistical Society. Series C (Applied Statistics)},
 number = {2},
 pages = {134--138},
 publisher = {[Wiley, Royal Statistical Society]},
 title = {Randomly Choosing Parameters from the Stationarity and Invertibility Region of Autoregressive-Moving Average Models},
 urldate = {2023-02-16},
 volume = {36},
 year = {1987}
}

@article{smith1994forecasting,
  title={Forecasting costs incurred from unit differencing fractionally integrated processes},
  author={Smith, Jeremy and Yadav, Sanjay},
  journal={International Journal of Forecasting},
  volume={10},
  number={4},
  pages={507--514},
  year={1994},
  publisher={Elsevier}
}

@article{wikle2014tvar,
author = {Yang, Wen-Hsi and Holan, Scott and Wikle, Christopher},
year = {2014},
month = {08},
pages = {},
title = {Bayesian Lattice Filters for Time-Varying Autoregression and Time-Frequency Analysis},
volume = {11},
journal = {Bayesian Analysis},
doi = {10.1214/15-BA978}
}

@article{knaus2021shrinkage,
  title={Shrinkage in the time-varying parameter model framework using the {R} package {shrinkTVP}},
  author={Knaus, Peter and Bitto-Nemling, Angela and Cadonna, Annalisa and Fr{\"u}hwirth-Schnatter, Sylvia},
  journal={Journal of Statistical Software},
  volume={100},
  number={13},
  year={2021},
  publisher={University of California at Los Angeles}
}

@article{Doucet2004smoothing,
author = {Simon J Godsill and Arnaud Doucet and Mike West},
title = {Monte {Carlo} Smoothing for Nonlinear Time Series},
journal = {Journal of the American Statistical Association},
volume = {99},
number = {465},
pages = {156-168},
year  = {2004},
publisher = {Taylor & Francis},
doi = {10.1198/016214504000000151},

URL = { 
    
        https://doi.org/10.1198/016214504000000151
    
    

},
eprint = { 
    
        https://doi.org/10.1198/016214504000000151
    
    

}

}

@article{lindsten2013backward,
  title={Backward simulation methods for {Monte Carlo} statistical inference},
  author={Lindsten, Fredrik and Sch{\"o}n, Thomas B },
  journal={Foundations and Trends in Machine Learning},
  volume={6},
  number={1},
  pages={1--143},
  year={2013},
  publisher={Now Publishers, Inc.}
}

@article{lindsten2014particle,
  title={Particle {Gibbs} with ancestor sampling},
  author={Lindsten, Fredrik and Jordan, Michael I and Schon, Thomas B},
  journal={Journal of Machine Learning Research},
  volume={15},
  pages={2145--2184},
  year={2014},
  publisher={MICROTOME PUBL}
}

@book{boxjen@1970,
  author = {Box, George.E.P. and Jenkins, Gwilym M.},
  publisher = {San Francisco: Holden-Day},
  title = {Time Series Analysis: Forecasting and Control},
  edition = {First},
  year = 1970
}

@book{wei2019time,
  title={Time Series Analysis Univariate and Multivariate Methods},
  author={Wei, W.W.S.},
  isbn={9780134995366},
  lccn={2017059693},
  series={Pearson Modern Classics for Advanced Statistics Series},
  url={https://books.google.se/books?id=_v-xswEACAAJ},
  year={2019},
  publisher={Pearson Education}
}

@TechReport{tvRegPaper,
    title = {tvReg: Time-Varying Coefficients Linear Regression for
      Single and Multi-Equations in R},
    author = {Isabel Casas and Ruben Fernandez-Casal},
    year = {2019},
    note = {R package version 0.5.9},
    institution = {SSRN},
    url =
      {https://papers.ssrn.com/sol3/papers.cfm?abstract_id=3363526},
  }

@Manual{tvRegPackage,
    title = {tvReg: Time-Varying Coefficients Linear Regression for
      Single and Multi-Equations},
    author = {Isabel Casas and Ruben Fernandez-Casal},
    year = {2023},
    note = {R package version 0.5.9},
    url = {https://CRAN.R-project.org/package=tvReg},
  }

@article{rosen2012adaptspec,
  title={AdaptSPEC: Adaptive spectral estimation for nonstationary time series},
  author={Rosen, Ori and Wood, Sally and Stoffer, David S},
  journal={Journal of the American Statistical Association},
  volume={107},
  number={500},
  pages={1575--1589},
  year={2012},
  publisher={Taylor \& Francis}
}

@article{rue2001fast,
  title={Fast sampling of {Gaussian Markov} random fields},
  author={Rue, H{\aa}vard},
  journal={Journal of the Royal Statistical Society: Series B (Statistical Methodology)},
  volume={63},
  number={2},
  pages={325--338},
  year={2001},
  publisher={Wiley Online Library}
}

@article{Andriue@ParticleGibbs2010,
title = {Particle {Markov chain Monte Carlo} methods},
author = {Andrieu, Christophe and Doucet, Arnaud and Holenstein, Roman},
year = {2010},
journal = {Journal of the Royal Statistical Society Series B},
volume = {72},
number = {3},
pages = {269-342},
url = {https://EconPapers.repec.org/RePEc:bla:jorssb:v:72:y:2010:i:3:p:269-342}
}

@book{prado2010time,
  title={Time series: modeling, computation, and inference},
  author={Prado, Raquel and West, Mike},
  year={2010},
  publisher={Chapman and Hall/CRC}
}

@article{wood2011bayesian,
  title={Bayesian mixtures of autoregressive models},
  author={Wood, Sally and Rosen, Ori and Kohn, Robert},
  journal={Journal of Computational and Graphical Statistics},
  volume={20},
  number={1},
  pages={174--195},
  year={2011},
  publisher={Taylor \& Francis}
}

@inproceedings{xie2020forecasting,
  title={Forecasting with multiple seasonality},
  author={Xie, Tianyang and Ding, Jie},
  booktitle={2020 IEEE International Conference on Big Data (Big Data)},
  pages={240--245},
  year={2020},
  organization={IEEE}
}

@article{de2011forecasting,
  title={Forecasting time series with complex seasonal patterns using exponential smoothing},
  author={De Livera, Alysha M and Hyndman, Rob J and Snyder, Ralph D},
  journal={Journal of the American Statistical Association},
  volume={106},
  number={496},
  pages={1513--1527},
  year={2011},
  publisher={Taylor \& Francis}
}

@article{fruhwirth1994data,
  title={Data augmentation and dynamic linear models},
  author={Fr{\"u}hwirth-Schnatter, Sylvia},
  journal={Journal of time series analysis},
  volume={15},
  number={2},
  pages={183--202},
  year={1994},
  publisher={Wiley Online Library}
}

@article{carter1994gibbs,
  title={On {Gibbs} sampling for state space models},
  author={Carter, Chris K and Kohn, Robert},
  journal={Biometrika},
  volume={81},
  number={3},
  pages={541--553},
  year={1994},
  publisher={Oxford University Press}
}

@article{polson2013bayesian,
  title={Bayesian inference for logistic models using {P{\'o}lya}--Gamma latent variables},
  author={Polson, Nicholas G and Scott, James G and Windle, Jesse},
  journal={Journal of the American Statistical Association},
  volume={108},
  number={504},
  pages={1339--1349},
  year={2013},
  publisher={Taylor \& Francis}
}

@article{lindsten2015particle,
  title={Particle ancestor sampling for near-degenerate or intractable state transition models},
  author={Lindsten, Fredrik and Bunch, Pete and Singh, Sumeetpal S and Sch{\"o}n, Thomas B},
  journal={arXiv preprint arXiv:1505.06356},
  year={2015}
}

@book{thrun2002probabilistic,
  title={Probabilistic robotics},
  author={Thrun, S and Burgard, W and Fox, D},
  year={2005},
  publisher={MA: MIT Press}
}

@article{kastner2014ancillarity,
  title={Ancillarity-sufficiency interweaving strategy ({ASIS}) for boosting {MCMC} estimation of stochastic volatility models},
  author={Kastner, Gregor and Fr{\"u}hwirth-Schnatter, Sylvia},
  journal={Computational Statistics \& Data Analysis},
  volume={76},
  pages={408--423},
  year={2014},
  publisher={Elsevier}
}

@article{jones2003skew,
  title={A skew extension of the t-distribution, with applications},
  author={Jones, M Chris and Faddy, MJ1959820},
  journal={Journal of the Royal Statistical Society Series B: Statistical Methodology},
  volume={65},
  number={1},
  pages={159--174},
  year={2003},
  publisher={Oxford University Press}
}

@article{omori2007stochastic,
  title={Stochastic volatility with leverage: Fast and efficient likelihood inference},
  author={Omori, Yasuhiro and Chib, Siddhartha and Shephard, Neil and Nakajima, Jouchi},
  journal={{Journal of Econometrics}},
  volume={140},
  number={2},
  pages={425--449},
  year={2007},
  publisher={Elsevier}
}

@article{kim1998stochastic,
  title={Stochastic volatility: likelihood inference and comparison with ARCH models},
  author={Kim, Sangjoon and Shephard, Neil and Chib, Siddhartha},
  journal={The review of economic studies},
  volume={65},
  number={3},
  pages={361--393},
  year={1998}
}

@article{piironen2017sparsity,
  title={Sparsity information and regularization in the horseshoe and other shrinkage priors},
  author={Piironen, Juho and Vehtari, Aki},
  journal ={Electronic Journal of Statistics},
  volume={11},
  pages={5018--5051},
  year={2017}
}

@article{baydin2018automatic,
  title={Automatic differentiation in machine learning: a survey},
  author={Baydin, Atilim Gunes and Pearlmutter, Barak A and Radul, Alexey Andreyevich and Siskind, Jeffrey Mark},
  journal={Journal of machine learning research},
  volume={18},
  number={153},
  pages={1--43},
  year={2018}
}

@article{gerlach2000efficient,
  title={Efficient {Bayesian} inference for dynamic mixture models},
  author={Gerlach, Richard and Carter, Chris and Kohn, Robert},
  journal={Journal of the American Statistical Association},
  volume={95},
  number={451},
  pages={819--828},
  year={2000},
  publisher={Taylor \& Francis}
}

@book{chopin2020introduction,
  title={An Introduction to Sequential Monte Carlo},
  author={Chopin, Nicolas and Papaspiliopoulos, Omiros},
  year={2020},
  publisher={Springer}
}

@article{makridakis1997arma,
  title={{ARMA} models and the {Box--Jenkins} methodology},
  author={Makridakis, Spyros and Hibon, Michele},
  journal={Journal of forecasting},
  volume={16},
  number={3},
  pages={147--163},
  year={1997},
  publisher={Wiley Online Library}
}

@book{box2015time,
  title={Time series analysis: forecasting and control},
  author={Box, George EP and Jenkins, Gwilym M and Reinsel, Gregory C and Ljung, Greta M},
  year={2015},
  edition = {Fifth},
  publisher={John Wiley \& Sons}
}

@article{knaus2023dynamic,
  title={The dynamic triple gamma prior as a shrinkage process prior for time-varying parameter models},
  author={Knaus, Peter and Fr{\"u}hwirth-Schnatter, Sylvia},
  journal={arXiv preprint arXiv:2312.10487},
  year={2023}
}

@article{polson2010shrink,
  title={Shrink globally, act locally: Sparse {B}ayesian regularization and prediction},
  author={Polson, Nicholas G and Scott, James G},
  journal={Bayesian statistics},
  volume={9},
  number={501-538},
  pages={105},
  year={2010},
  publisher={Oxford}
}

@article{julier2004unscented,
  title={Unscented filtering and nonlinear estimation},
  author={Julier, Simon J and Uhlmann, Jeffrey K},
  journal={Proceedings of the IEEE},
  volume={92},
  number={3},
  pages={401--422},
  year={2004},
  publisher={IEEE}
}

@inproceedings{skoglund2019iterative,
  title={On iterative unscented Kalman filter using optimization},
  author={Skoglund, Martin A and Gustafsson, Fredrik and Hendeby, Gustaf},
  booktitle={2019 22th International Conference on Information Fusion (FUSION)},
  pages={1--8},
  year={2019},
  organization={IEEE}
}

@inproceedings{skoglund2015extended,
  title={Extended {K}alman filter modifications based on an optimization view point},
  author={Skoglund, Martin A and Hendeby, Gustaf and Axehill, Daniel},
  booktitle={2015 18th International Conference on Information Fusion (Fusion)},
  pages={1856--1861},
  year={2015},
  organization={IEEE}
}

@article{garcia2015posterior,
  title={Posterior linearization filter: Principles and implementation using sigma points},
  author={Garc{\'\i}a-Fern{\'a}ndez, {\'A}ngel F and Svensson, Lennart and Morelande, Mark R and S{\"a}rkk{\"a}, Simo},
  journal={IEEE transactions on signal processing},
  volume={63},
  number={20},
  pages={5561--5573},
  year={2015},
  publisher={IEEE}
}

@article{yu2011center,
  title={To center or not to center: That is not the question—an Ancillarity--Sufficiency Interweaving Strategy (ASIS) for boosting MCMC efficiency},
  author={Yu, Yaming and Meng, Xiao-Li},
  journal={Journal of Computational and Graphical Statistics},
  volume={20},
  number={3},
  pages={531--570},
  year={2011},
  publisher={Taylor \& Francis}
}

@article{cogley2005drifts,
  title={Drifts and volatilities: monetary policies and outcomes in the post WWII US},
  author={Cogley, Timothy and Sargent, Thomas J},
  journal={Review of Economic dynamics},
  volume={8},
  number={2},
  pages={262--302},
  year={2005},
  publisher={Elsevier}
}

@article{bitto2019achieving,
  title={Achieving shrinkage in a time-varying parameter model framework},
  author={Bitto, Angela and Fr{\"u}hwirth-Schnatter, Sylvia},
  journal={Journal of Econometrics},
  volume={210},
  number={1},
  pages={75--97},
  year={2019},
  publisher={Elsevier}
}

\doublespacing 

\newpage


\title{ Supplement to \\ Time-Varying Multi-Seasonal AR Models}
\if0\blind{
\author{Ganna Fagerberg$^{a}$\thanks{Corresponding author: ganna.fagerberg@stat.su.se. $^a$Department of Statistics, Stockholm University. 
$^b$School of Business, University of New South Wales. $^c$Data Analytics Center for Resources and Environments (DARE).}, Mattias Villani$^{a}$ and Robert Kohn$^{b,c}$}
}\fi
\maketitle

\renewcommand{\theequation}{S\arabic{equation}}
\renewcommand{\thesection}{S\arabic{section}}
\renewcommand{\theproposition}{S\arabic{proposition}}
\renewcommand{\theassumption}{S\arabic{assumption}}

\renewcommand{\thethm}{S\arabic{thm}}

\renewcommand{\thelemma}{S\arabic{lemma}}
\renewcommand{\thealgocf}{S\arabic{algocf}}
\renewcommand{\thefigure}{S\arabic{figure}}
\renewcommand{\thetable}{S\arabic{table}}
\renewcommand{\thepage}{S\arabic{page}}
\renewcommand{\thetable}{S\arabic{table}}
\renewcommand{\thepage}{S\arabic{page}}
\setcounter{page}{1}
\setcounter{section}{0}
\setcounter{equation}{0}
\setcounter{algocf}{0}
\setcounter{lemma}{0}
\setcounter{assumption}{0}
\setcounter{table}{0}
\setcounter{figure}{0}
\setcounter{thm}{0}
\numberwithin{equation}{section}

\begin{abstract}
    This Supplement contains proofs and additional results for the paper '\emph{Time-Varying
Multi-Seasonal AR Models}'.
\end{abstract}

\section{Uniform distribution on the stability region}\label{app:t_transform}
\subsection{Proof of Lemma \ref{lem:uniform_t}}\label{proofStationaryLemma}

A random variable $X \sim t_{\mathrm{skew}}(a,b)$, where $a>$ and $b>0$ are parameters, follows the skew-$t$ distribution in \citet{jones2003skew} if it has density
\begin{equation}
f(x) = \frac{1}{2^ {a+b-1} \mathrm{B}(a,b)\sqrt{a+b}}
    \Bigg(1 + \frac{x}{\sqrt{a + b + x^2}}\Bigg)^{a + 1/2}
    \Bigg(1 - \frac{x}{\sqrt{a + b + x^2}}\Bigg)^{b + 1/2}.
\end{equation}
When $a = b$, $t_{\mathrm{skew}}(a,b)$ reduces to the usual symmetric t-distribution with $2a$ degrees of freedom. When $a<b$, $t_{\mathrm{skew}}(a,b)$ is negatively skewed, while for $a>b$ it is positively skewed.

One representation of this distribution is that if $Y \sim \mathrm{Beta}(a,b)$ then \citep{jones2003skew}
\begin{equation}\label{eq:skew_representation}
X := \sqrt{a+b}\frac{2Y-1}{2\sqrt{Y(1-Y)}} \sim  t_{\mathrm{skew}}(a,b).
\end{equation}
More generally, $X \sim t_{\mathrm{skew}}(a,b,\mu,\sigma)$ follows from a location-scale transformation $X \rightarrow \mu + \sigma X$.

\cite{jones1987beta} proves that if the partial autocorrelations are independently distributed as 
\begin{equation}\label{eq:beta_prior} 
r_k\sim\mathrm{Beta_{(-1,1)}}\big(\alpha=\lfloor (k+1)/2\rfloor,\beta=\lfloor k/2\rfloor+1\big), 
\end{equation}
where $\lfloor x \rfloor$ is the integer part of $x$, $k=1,\ldots,p,$ and $\mathrm{Beta}_{(-1,1)}$ denotes the Beta distribution shifted to the interval $(-1,1)$, then the implied distribution for the AR coefficients is uniform over the stability/stationary region in $\v\phi$-space. 

Let $q_k \sim \mathrm{Beta}\big(\alpha_k=\lfloor (k+1)/2\rfloor,\beta_k=\lfloor k/2\rfloor+1\big)$ follow the usual Beta distribution over $(0,1)$ so that, by definition, $r_k \overset{d}{=} 2q_k-1$, where $\overset{d}{=}$ denotes equality in distribution. Since $r_k = \theta_k/\sqrt{1+\theta^2_k}$, we have
$$ \theta_k = \frac{r_k}{\sqrt{1-r^2_k}} 
\overset{d}{=} \frac{2q_k-1}{\sqrt{1-(2q_k-1)^2}} 
            = \frac{2q_k-1}{2\sqrt{q_k(1-q_k)}}
            = \frac{1}{\sqrt{k+1}}T
$$
where
$$T=\sqrt{k+1}\frac{2q_k-1}{2\sqrt{q_k(1-q_k)}}$$ 
Since $\alpha_k + \beta_k = k + 1$ for all $k$, we have from \eqref{eq:skew_representation} that $T\sim t_{\mathrm{skew}}(\alpha_k,\beta_k)$.
This shows that $\theta_k \sim t_{\mathrm{skew}}(\alpha_k,\beta_k,0,1/\sqrt{k+1})$ for all $k$. When $k$ is even, we have $\alpha_k=k/2$ and $\beta_k = (k+2)/2$. When $k$ is odd, we have $\alpha_k = \beta_k$ and the skew-t distribution reduces to the symmetric student-t distribution with $2\alpha_k = k+1$ degrees of freedom \citep{jones2003skew}, and hence $\theta_k \sim t(k+1,0,1/\sqrt{k+1})$, which proves the result in Lemma \ref{lem:uniform_t}.

\subsection{Normal approximation}\label{normal_approx_uniform}

\begin{figure}
 \centering
 \includegraphics[width=1\textwidth]{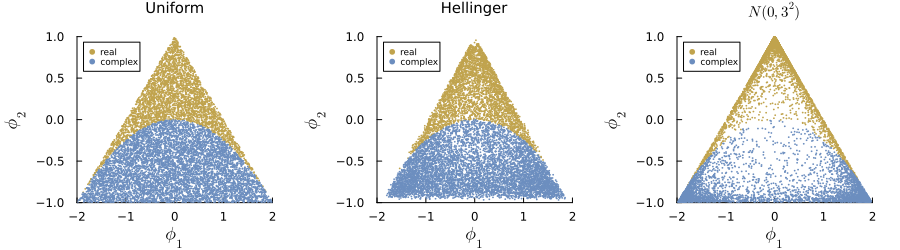}
\caption{Illustrating \num{10000} draws from the uniform distribution in Lemma \ref{lem:uniform_t} (left) and the optimal normal approximation in Table \ref{table:NormalApproxUniformPrior} (middle). The implied distribution from a $\mathrm{N}(0,3^2)$ prior for both $\theta_1$ and $\theta_2$ is shown as reference (right). The beige points are draws with real roots in the AR polynomial while the blue points are draws with   complex roots.} \label{fig:normalapprox2uniform}
\end{figure}

The FFBSx algorithm requires a multivariate normal distribution for the initial state $\v\theta_0$, or a mixture of multivariate normals if a more precise approximation is needed. We approximate the student-$t$ and skew-$t$ distributions in Lemma \ref{lem:uniform_t} by independent univariate normal distributions by finding the closest normal approximation to a $t$ or skew $t$ as measured by the Hellinger distance
\begin{equation}
    H^2(p,q) = \frac{1}{2}\int_{\mathcal{X}} \Big( \sqrt{p(x)} - \sqrt{q(x)}\Big) dx. 
\end{equation}
For the symmetric student-$t$ case, we minimize with respect to the standard deviation, while for the  skew-$t$, we minimize with respect to both the mean and standard deviation of the approximating normal. The optimal normal approximation is listed in Table \ref{table:NormalApproxUniformPrior} for up to $10$ AR lags. 
Figure~\ref{fig:normalapprox2uniform} shows that the implied distribution on the stability region from the normal approximation (middle) is close to the ideal uniform distribution (left). The right panel shows the implied distribution on $\phi_1$ and $\phi_2$ from a $\mathrm{N}(0,3^2)$ distribution for each $\theta_k$ as a reference, showing that this prior puts most of its mass close to the boundaries of the stability region. 

\begin{table}
\begin{center}
\begin{tabular}{l rrrrrrrrrr}
\hline
  & $\theta_1$ & $\theta_2$ & $\theta_3$ & $\theta_4$ & $\theta_5$ & $\theta_6$ & $\theta_7$ & $\theta_8$ & $\theta_9$ & $\theta_{10}$ \\
\cline{2-11}
\textbf{mean} & 0 & -0.53 & 0 & -0.264 & 0 & -0.175 & 0 & -0.13 & 0 & -0.103 \\
\textbf{stdev} & 1.042 & 0.858 & 0.622 & 0.558 & 0.475 & 0.441 & 0.397 & 0.375 & 0.348 & 0.332 \\
\hline
\end{tabular}
\caption{Normal approximation of the student-t and skew-t distributions in Lemma \ref{lem:uniform_t} that minimizes the Hellinger distance.}
\label{table:NormalApproxUniformPrior}
\end{center}
\end{table}

\section{Prior evolution of $\theta$, $r$ and $\phi$}
\label{app:prior_evol}
\begin{figure}
 \centering
\includegraphics[width=0.3\textwidth]{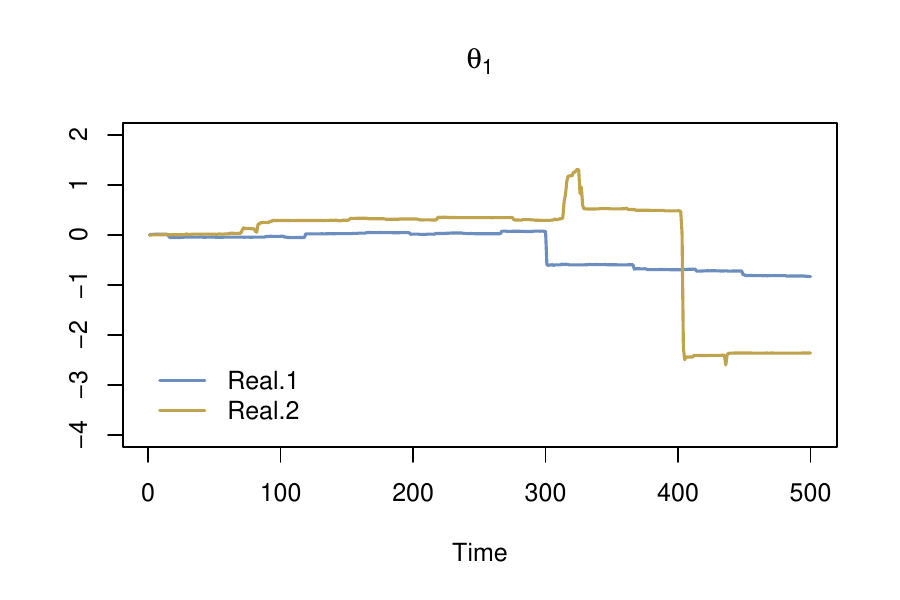}
\includegraphics[width=0.3\textwidth]{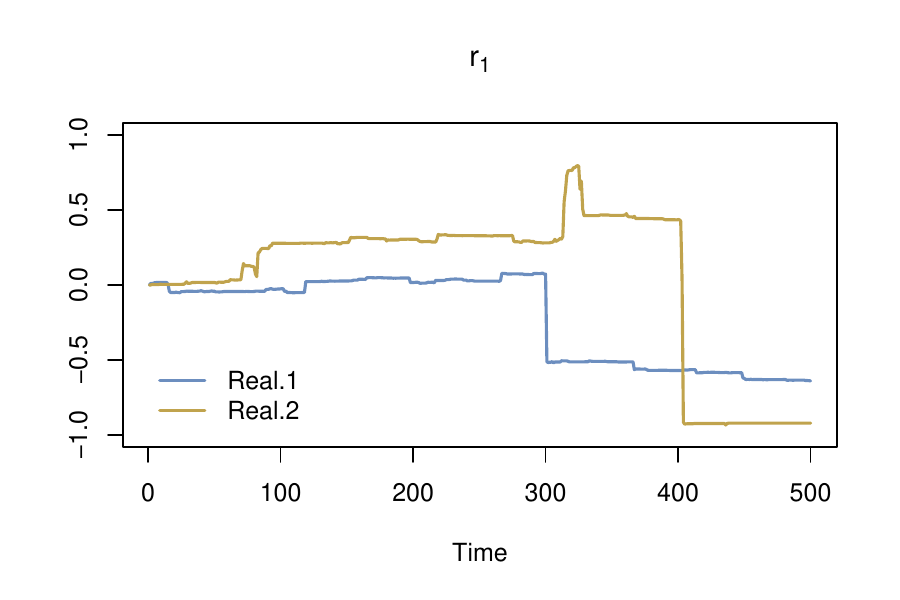}
\includegraphics[width=0.3\textwidth]{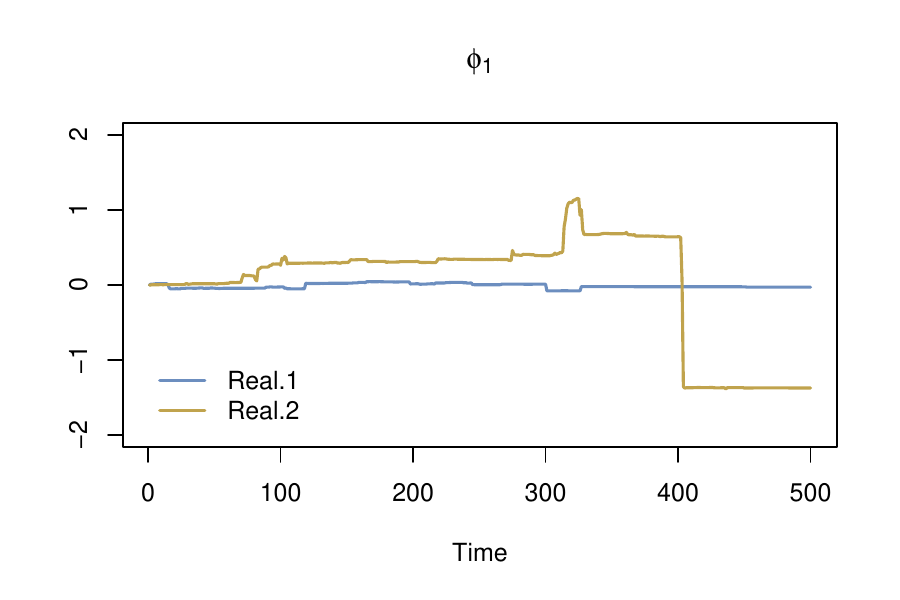}

\includegraphics[width=0.3\textwidth]{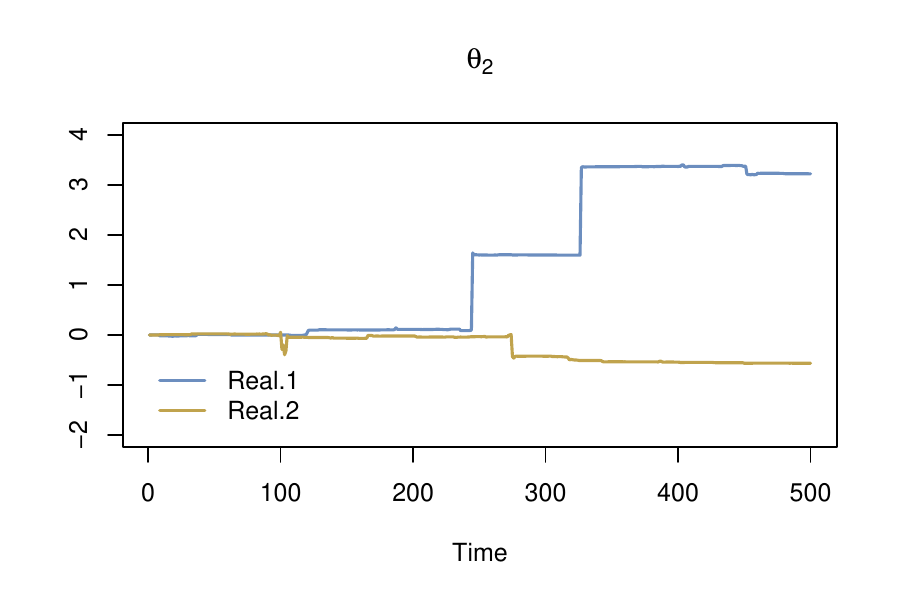}
\includegraphics[width=0.3\textwidth]{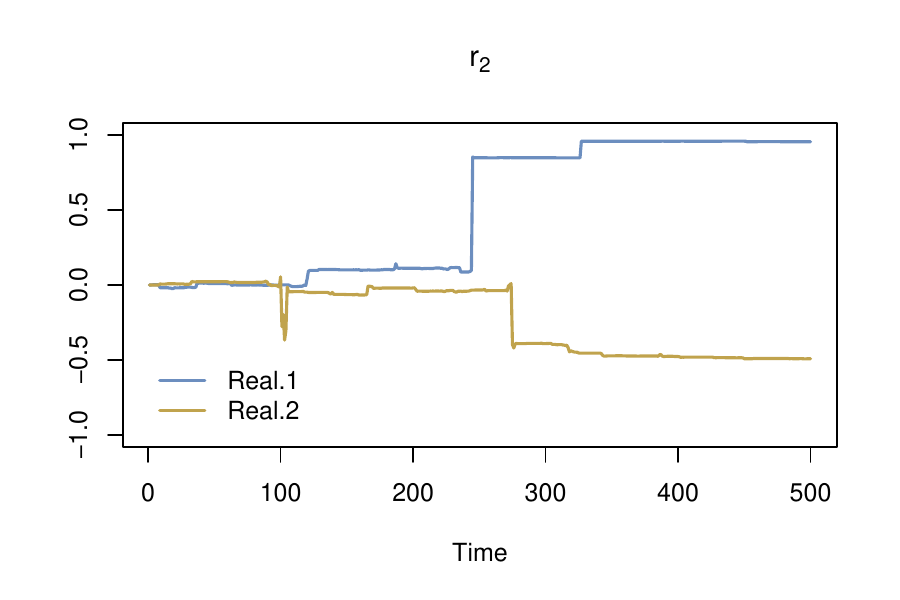}
\includegraphics[width=0.3\textwidth]{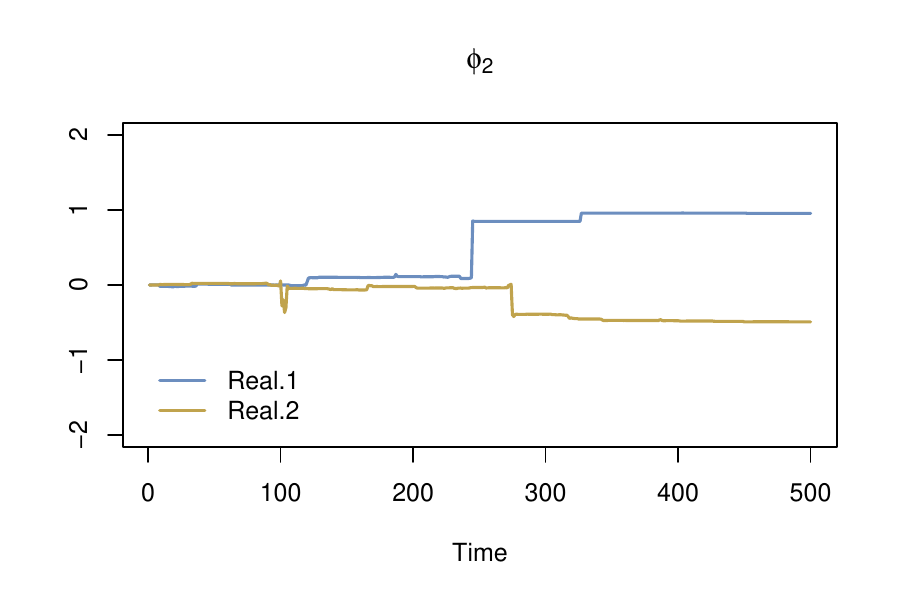}

\caption{AR(2): Two realizations of time-varying trajectories for the unrestricted parameters $\theta_1$ and $\theta_2$ (left) following the DSP prior with global mean $\mu=-15$, the corresponding partial autocorrelations $r_1$ and $r_2$ (center), and the AR coefficients $\phi_1$ and $\phi_2$ (right). The persistence parameter is fixed at the prior mean $\kappa=0.5$ in the first realization and at $\kappa=0.8$ in the second realization.} \label{fig:AR1_ev1}
\end{figure}
\begin{figure}
 \centering
\includegraphics[width=0.4\textwidth]{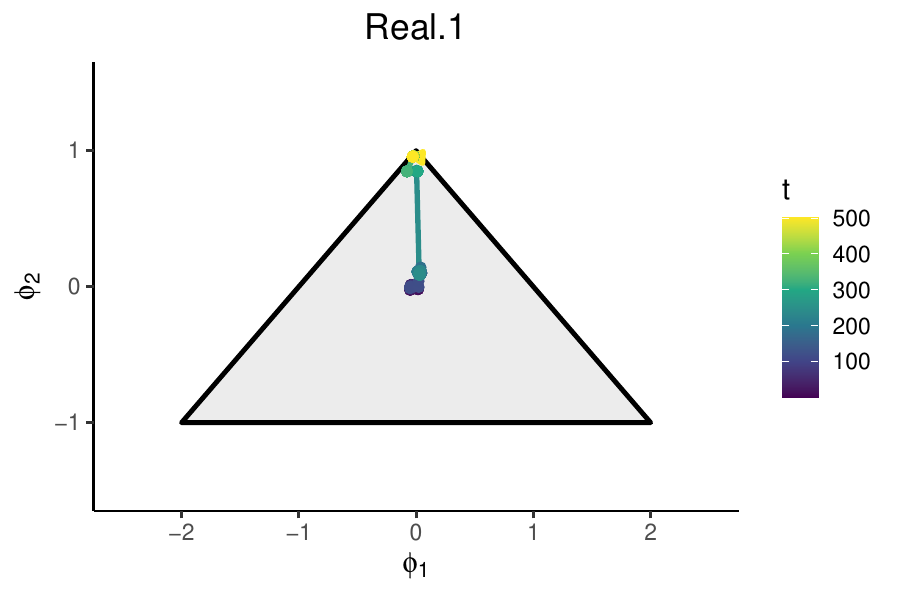}
\includegraphics[width=0.4\textwidth]{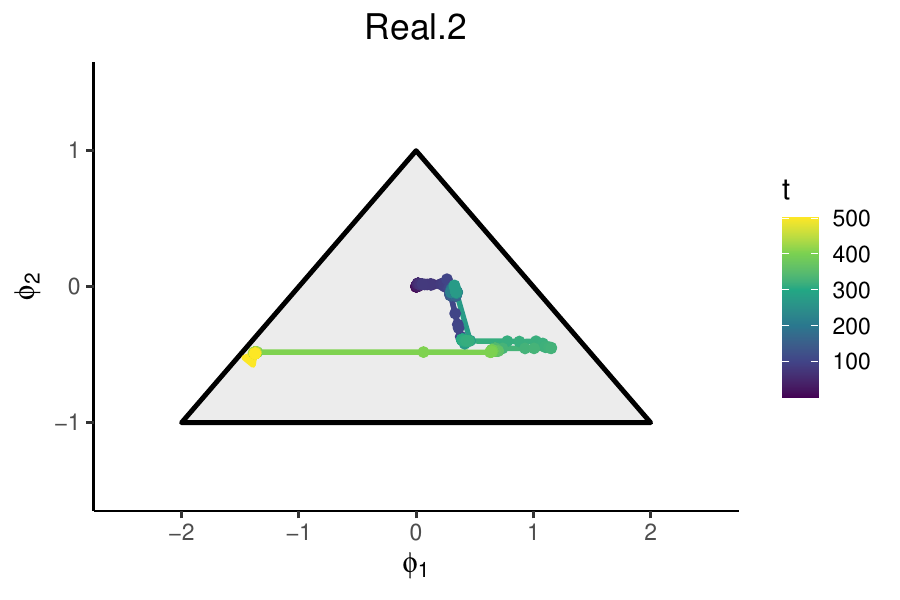}
\caption{AR(2):  The plots show the joint evolution of $(\phi_1, \phi_2)$ within the stability region, indicated by the triangle, across Real.1 and Real.2.} \label{fig:AR1_ev1}
\end{figure}

\newpage

\section{Gibbs sampling updating steps}\label{app:GibbsUpdating}

The Gibbs sampling algorithm in summarized in Algorithm \ref{alg:gibbs_algorithm}
This section describes the updating steps for the parameters not covered in the main text. The updates are independent across parameters and will therefore be described for one parameter without the subscript $k$.

\begin{algorithm}\label{alg:Gibbs_sampler}
\KwInput{data $y_{1:T}$}
\hspace{1.4cm}initial $\mu^{(0)}$, defaulting to $(\mu_0,\ldots,\mu_0)^\top$, where $\mu_0$ is the prior mean\\
\hspace{1.4cm}initial $\kappa^{(0)}$, defaulting to $(\kappa_0,\ldots,\kappa_0)^\top$, where $\kappa_0$ is the prior mean\\
\hspace{1.4cm}initial $\v h^{(0)}_{1:T}$, defaulting to $h_{kt}^{(0)}=\mu_0\text{ for all }k,t $\\
\hspace{1.4cm}initial $\sigma_{1:T}^{(0)}$, defaulting to an estimate from the static seasonal AR \\
\hspace{1.4cm}initial P\'olya-Gamma variables $\v\xi_{0:T}^{(0)}$, defaulting to $1 \text{ for all }k,t$\\
\hspace{1.4cm}tuning parameters for PGAS \\
\hspace{1.4cm}the number of posterior draws $J$

\BlankLine

Run the particle filter to compute initial reference trajectory $\textcolor{orange}{\v \theta^{(0)}_{0:T}}$ for PGAS 
\BlankLine
\For{$j = 1$ \KwTo $J$}{

    \BlankLine
    
    \textcolor{blue}{\textbackslash\textbackslash\hspace{0.1cm}draw the parameter evolutions using PGAS or FFBSx} \\
    $\v \theta^{(j)}_{0:T}\leftarrow \texttt{PGAS/FFBSx}(\v \theta_{0:T} \vert \v h^{(j-1)}_{0:T}, \sigma_{1:T}^{(j-1)}, y_{1:T},  \textcolor{orange}{\v \theta^{(j-1)}_{0:T}} )$ \\
    \BlankLine

     \textcolor{blue}{\textbackslash\textbackslash\hspace{0.1cm}draw the error standard deviations } \\
    $\v\sigma_{1:T}^{(j)} \leftarrow \ p(\v\sigma_{1:T} \vert \v\theta^{(j)}_{0:T}, y_{1:T})$\\
    \BlankLine

\ParFor{$k = 1$ \KwTo $r$}{
\BlankLine

    \textcolor{blue}{\textbackslash\textbackslash\hspace{0.1cm}draw mixture of normals allocations} \\
   $\v a^{(j)}_{k,1:T} \leftarrow \ p(\v a_{k,1:T}\vert \v \theta^{(j)}_{k,0:T}, h^{(j-1)}_{k,0:T})$\\
   \BlankLine
   
    \textcolor{blue}{\textbackslash\textbackslash\hspace{0.1cm}draw log-volatilities} \\
   $\v h^{(j)}_{k,0:T} \leftarrow \ p(\v h_{k,0:T}\vert \v \theta^{(j)}_{k,0:T}, \v a^{(j)}_{k,1:T}, \kappa_k^{(j-1)}, \mu_k^{(j-1)}, \v\xi_{k, 1:T}^{(j-1)})$\\
   \BlankLine
   
    \textcolor{blue}{\textbackslash\textbackslash\hspace{0.1cm}draw P\'olya-Gamma variables} \\
    $\v\xi_{k, 0:T}^{(j)} \leftarrow \ p(\v\xi_{k, 0:T}\vert \v h_{k, 0:T}^{(j)},\kappa_k^{(j-1)},\mu_k^{(j-1)})$\\
    \BlankLine
    
    \textcolor{blue}{\textbackslash\textbackslash\hspace{0.1cm}draw global mean log-volatility} \\
    $\mu_k^{(j)} \leftarrow \ p(\mu_k \vert \v h_{k,1:T}^{(j)}, \kappa_k^{(j-1)}, \v\xi_{k, 1:T}^{(j)})$\\
    \BlankLine
    
    \textcolor{blue}{\textbackslash\textbackslash\hspace{0.1cm}draw global log-volatility persistence} \\
    $\kappa_k^{(j)} \leftarrow \ p(\kappa_k \vert \v h_{k, 1:T}^{(j)}, \mu_k^{(j)}, \v\xi_{k, 1:T}^{(j)})$\\
    \BlankLine

    }
} 
\KwOutput {
draws from the joint posterior of $\v\theta_{0:T},\v h_{0:T},\sigma_{1:T}, \v\mu$ and $\v\kappa$.
}

\BlankLine
\caption{Gibbs sampling from joint posterior $p(\v\theta_{0:T},\v h_{0:T},\sigma_{1:T}, \v\mu,\v\kappa \vert y_{1:T})$ \label{alg:gibbs_algorithm}}
\end{algorithm}

\subsection{Updating the log volatility paths}\label{subsec:update_h}

Conditional on the static parameters $\v\mu$ and $\v\kappa$ and the parameter paths $\v\theta_{0:T}$, the sampling step for stochastic volatility models in  \citet{rue2001fast} and \citet{kastner2014ancillarity} can be used to efficiently sample the $r$ columns of $\v h_{0:T}$ independently from multivariate normal distributions with a tri-diagonal covariance matrix. Since the log-volatilities for different parameters are drawn independently, we focus on one parameter and drop the subscript $k$ and write $h_t$ for the rest of this subsection. The updating step for $\v h_{0:T}$ is the same as in \citet{kowal2019dynamic} and is given here to explain the role of the  P\'olya-Gamma variables in Algorithm \ref{alg:gibbs_algorithm}. The updating of $\v\mu$ and $\v\kappa$ are slightly different from \citet{kowal2019dynamic} due to different priors.

The sampling of $h_{0:T}$ in \citet{kowal2019dynamic} follows the log-volatility estimation in \citet{kim1998stochastic} where the model for the parameter innovations $\nu_t = \theta_t - \theta_{t-1}$
 \begin{equation}
 \nu_{t} = \exp(h_{t}/2)\epsilon_{t}, \quad \epsilon_t \overset{\mathrm{iid}}{\sim} \mathrm{N}(0,1)
 \end{equation}
is rewritten by squaring and taking logs on both sides

\begin{equation}\label{eq:likelihood_h}
 \log \nu_{t}^2 = h_{t} + \log \epsilon_{t}^2,
 \end{equation}
so that $h_{t}$ becomes an additive term. The error term $\log \epsilon_{t}^2$ is however $\log \chi_1^2$-distributed, which \citet{kim1998stochastic} solves by approximating the $\log\chi_1^2$ distribution with a mixture of normals, thereby turning the model into a Gaussian model conditional on the latent mixture allocation variables $a_{1:T}$; we use the 10-component mixture in \citet{omori2007stochastic}. 

The prior for the log-volatility $h_{0:T}$ series follows the dynamic shrinkage process 
\begin{equation}
h_{t}=\mu+\kappa(h_{t-1}-\mu)+\eta_{t},\qquad\eta_{t}\overset{\mathrm{iid}}{\sim}\mathrm{Z(1/2,1/2,0,1),\text{ for }}t=1,\ldots,T,
\end{equation}
with initial condition $h_0=\mu+\eta_0$ and $\eta_0\sim Z(1/2,1/2,0,1)$.
\citet{kowal2019dynamic} show that the prior can be augmented with P\'olya-Gamma variables $\xi_{0:T}$ to become conditionally Gaussian
\begin{align}\label{eq:polyagamma_augment}
h_{t} &=\mu+\kappa(h_{t-1}-\mu)+\eta_{t} \nonumber \\
\eta_{t}\vert\xi_{t} &\overset{\mathrm{iid}}{\sim}\mathrm{N}\big(0,\xi_{t}^{-1}\big) \nonumber \\
\xi_{t}	&\overset{\mathrm{iid}}{\sim}\mathrm{PG}(1,0), 
\end{align}
where $\mathrm{PG}$ is the P\'olya-Gamma distribution \citep{polson2013bayesian}. 
Now, conditional on the mixture allocation $a_{1:T}$ and the P\'olya-Gamma variables $\xi_{0:T}$, \citet{kowal2019dynamic} combine the conditionally Gaussian likelihood from \eqref{eq:likelihood_h} with the conditionally Gaussian prior process in \eqref{eq:polyagamma_augment} to derive the posterior of $\v h_{0:T}$ as multivariate normal. The inverse covariance matrix is tri-diagonal, which can be elegantly exploited for efficient sampling, see \citet{rue2001fast} and \citet{kowal2019dynamic} for details. 

To avoid numerical issues it is common to add some small positive offset before taking logs in \eqref{eq:likelihood_h}
\begin{equation}\label{logvolatility_offset}
    \log (\nu_{t}^2 +\mathrm{offset}) = h_{t} + \log \epsilon_{t}^2,
\end{equation}
With $\nu_{t}$ being the change in the  parameter corresponding to $h_t$, this offset turns out to be important here since the dynamic shrinkage prior encourages parameters to remain constant for extended periods, so $\nu_{t}$ tends to be close to zero for some $t$. The \texttt{dsp} package in R that accompanies \citet{kowal2019dynamic} uses a different offset for each parameter and in each Gibbs sampling iteration following the rule 
\begin{equation}\label{dsp_offset} 
    \mathrm{offset} = 
    \begin{cases}
    \max\big(10^{-8}, 10^{-6}\cdot \mathrm{mad}(\nu_{1:T})\big) & \text{if } \nu_{t}^2 < 10^{-16} \text{ for any } t\\
    0 & \text{otherwise},
    \end{cases} 
\end{equation}
where $\mathrm{mad}(\v x)$ is the median absolute deviation for the elements in the vector $\v x$. Section \ref{sec:simulations} further discusses the effect of the offset. 

\subsection{Updating $\v\kappa$}\label{subsec:update_kappa} 

The full conditional posteriors of the $\kappa_k$ are independent and we now give the distribution for a typical element $\kappa$ of $\v\kappa$ conditional on its log-volatility process, which we denote by the generic $h_{0:T}$. The relevant part of the model for updating one of the $\kappa$ parameters is the P\'olya-Gamma augmented model in \eqref{eq:polyagamma_augment}.

Conditioning on the P\'olya-Gamma variables $\xi_{0:T}$, the parameter $\kappa$ is the regression coefficient in a heteroscedastic Gaussian regression model. Multiplying both sides of $h_{t} =\mu+\kappa(h_{t-1}-\mu)+\eta_{t}$ in \eqref{eq:polyagamma_augment} by $\xi_t^{1/2}$ makes the model homoscedastic and a standard derivation for Bayesian linear regression shows that the full conditional posterior is
\begin{equation}
 \kappa \vert \mu,\xi_{0:T},h_{0:T} \sim \mathrm{TN}(\kappa_{T},\psi_{T}^{2},-1,1)
\end{equation}
where $\psi_{T}^{-2}=\tilde{\boldsymbol{h}}_{-1}^{\top}\tilde{\boldsymbol{h}}_{-1}+\psi_{0}^{-2}$, $\kappa_{T}=w\hat{\kappa}+(1-w)\kappa_{0}$, $\hat{\kappa}=(\tilde{\boldsymbol{h}}_{-1}^{\top}\tilde{\boldsymbol{h}}_{-1})^{-1}\tilde{\boldsymbol{h}}_{-1}^{\top}\tilde{\boldsymbol{h}}$, $w=\frac{\tilde{\boldsymbol{h}}_{-1}^{\top}\tilde{\boldsymbol{h}}_{-1}}{\tilde{\boldsymbol{h}}_{-1}^{\top}\tilde{\boldsymbol{h}}_{-1}+\psi_{0}^{-2}}$, using the notation  $\tilde{\boldsymbol{h}}=(\tilde{h}_{1},\ldots,\tilde{h}_{T})^{\top}$ and $\tilde{\boldsymbol{h}}_{-1}=(\tilde{h}_{0},\ldots,\tilde{h}_{T-1})^{\top}$.

\subsection{Updating $\v\mu$}\label{subsec:update_mu}

The full conditional posteriors for the elements in $\v\mu$ are independent and we will describe the updating step for one such generic element $\mu$. We can rewrite the model in \eqref{eq:polyagamma_augment} as
\begin{equation}
    z_{t}=\mu+\tilde{\eta}_{t}
\end{equation}
where 
\begin{equation}
    z_{t}=\frac{h_{t}-\kappa h_{t-1}}{1-\kappa},
\end{equation}
\begin{equation}
    \tilde{\eta}_{t}=\eta_{t}/(1-\kappa)\sim \mathrm{N}\left(0,\tilde{\xi}_{t}^{-1}\right)
\end{equation}
with $\tilde{\xi}_{t} = (1-\kappa)^2\xi_{t}$ and $z_{0}=h_{0}=\mu+\eta_{0}$ with $\eta_{0}\sim \mathrm{N}(0,\xi_{t}^{-1})$.
This shows that $\mu$ is the mean of the transformed data $z_{t}$. A standard derivation of the posterior of a mean in a Gaussian model with known heteroscedastic variances shows that 
\begin{equation}
    \mu \vert \kappa,\xi_{0:T},h_{0:T}\sim \mathrm{N}(\mu_{T},\sigma_{T}^{2}),
\end{equation}
where $\sigma_{T}^{-2}=\sum_{t=0}^{T}\tilde{\xi}_{t}+\sigma_{0}^{-2}$ and $\mu_{T}=v\hat{\mu}+(1-v)\mu_{0}$ with 
\begin{equation}
    v=\frac{\sum_{t=0}^{T}\tilde{\xi}_{t}}{\sum_{t=0}^{T}\tilde{\xi}_{t}+\sigma_{0}^{-2}}.
\end{equation}

\subsection{Updating the $\log \chi_1^2$ mixture allocation $\v a_{1:T}$}\label{subsec:update_mixture_allocation} 
This is a standard step based on the 10-component mixture in \citet{omori2007stochastic}; see \citet{kowal2019dynamic} for details.

\subsection{Updating $\v\xi_{1:T}$}\label{subsec:update_xi} 
The updating of the P\'olya-Gamma variables is identical to the update in \citet{kowal2019dynamic}.

\subsection{Updating the observational noise variance}\label{subsec:update_sigma}
For the static variance case we use the conjugate prior $\sigma^2 \sim \mathrm{Inv-}\chi^2(v_0, s_0^2)$, the scaled inverse chi-square distribution. The posterior distribution is $\mathrm{Inv-}\chi^2(v_T, s_T^2)$, with $v_T=v_0 + T$ and $s_T^2 = (v_0 s_0^2 + \sum_{i=1}^T \hat\epsilon_t^2)/(v_0 + T)$.
   
In the time-varying error variance case, we follow \cite{kowal2019dynamic} and use a stochastic volatilty (SV) model with the updating step in \cite{kastner2014ancillarity}, similarly to the sampling of $h_{kt}$ in 
Section~\ref{subsec:update_h}.

\section{Particle Gibbs with Ancestor Sampling}\label{app:pgas}

A general SSM  with state vector $\v\theta_t$ and measurements $y_t$ is of the form
\begin{equation}\label{eq: Markov_SSM}
\v \theta_0 \sim f_0(\v  \theta_0), \qquad  \v  \theta_{t} \sim f(\v \theta_{t}| \v \theta_{t-1}),  \qquad y_{t} \sim g(y_{t} \vert \v \theta_{t}),
\end{equation}
where $f_0$ is the initial prior, $f(\v \theta_{t}| \v \theta_{t-1})$ is the state transition model and $g(y_{t} \vert \v \theta_{t})$ is the measurement model.

Particle filters approximate the target filtering distribution, $p(\v \theta_t\vert y_{1:t})$ by a set of $N$ weighted particles, $ \{\v \theta_t^i, w_t^i\}^N_{i=1}$
\begin{equation}
   \widehat p(\v \theta_t|y_{1:t})= \sum_{i=1}^N{w_t^i \delta_{\v\theta_t^i}(\v \theta_t)},
\end{equation}
where $\sum_i w_t^i = 1$ and $\delta(\cdot)$ is the point mass Dirac delta function. The particles evolve in time based on a Markov proposal density $q_t(\v \theta_t|\v \theta_{t-1},y_t)$ and are assigned importance weights
\begin{equation}\label{eq:imp_weights}
    w_t^{i} \propto \frac{g(y_t|\v \theta_{t}^i)f(\v \theta_t^i|\v \theta_{t-1}^i)}{q_t(\v \theta_{t}^i|\v \theta_{t-1}^i,y_t)}w_{t-1}^i,
\end{equation}
where $g(y_t|\v \theta_{t})$ is the measurement density and $f(\v \theta_t|\v \theta_{t-1})$ is the transition density. The particles at step $t$ are resampled with replacement with probabilities proportional to $w_t^{i}$ to form an unweighted sample from the filtering distribution $p(\v \theta_t|y_{1:t})$. The bootstrap particle filter uses the model's transition density $f(\v \theta_{t}| \v \theta_{t-1})$ as a proposal and the importance weights in (\ref{eq:imp_weights}) simplify to 
\begin{equation}
    w_t^{i} \propto g(y_t|\v  \theta_{t}^i)w_{t-1}^i.  
\end{equation}
The resampling step can be expressed as the sampling of ancestor indexes, $a_t^i$, where the outcome $a_t^i = j$ means that the $i$th particle at time $t$ is propagated from the $j$th particle at time $t-1$ \citep{Andriue@ParticleGibbs2010}. 

Particle filters often face degeneracy problems with only a few particles carrying significant weight. The resampling step partially addresses this problem, but frequent resampling can lead to a loss of particle diversity. To address this, it is common to resample only when the estimated effective sample size
\begin{equation}\label{ESS}
    \mathrm{ESS}_{t-1} := \frac{1}{\sum_{i=1}^N{(w_{t-1}^i)^2}}
\end{equation}
is below a certain threshold $\mathrm{ESS}_{\min}$. We use $\mathrm{ESS}_{\min}=N/2$ and systematic resampling \citep[Ch.~9]{chopin2020introduction} as default settings.

\cite{Andriue@ParticleGibbs2010} introduce particle MCMC (PMCMC) methods that combine MCMC and SMC. One such PMCMC sampler is the particle Gibbs sampler that uses a particular conditional SMC update. This update is similar to the usual SMC approximation, but is conditioned on a prespecified reference trajectory $\v \theta^\star_{0:T}$ that is ensured to survive all resampling steps. \cite{Andriue@ParticleGibbs2010} show that conditioning on the reference trajectory makes the update a valid Markov kernel that leaves the conditional posterior $p(\v \theta_{0:T}|y_{1:T}, \cdot)$ invariant, and can therefore be used in a Gibbs sampling step.

\begin{algorithm}
\SetAlgoLined
\KwInput{data $y_{1:T}$, reference trajectory $\v \theta^\star_{0:T}$, number of particles $N$, initial proposal distribution $q_0(\v \theta_0)$ and resampling threshold $\mathrm{ESS}_{\min}$.}
\BlankLine
\textcolor{blue}{\textbackslash\textbackslash\hspace{0.1cm}$t=0$}\\
\BlankLine
$\v \theta_0^i \sim q_0(\v \theta_0)$ for $i=1,\ldots,N-1$\\
$\v \theta_0^N = \v\theta^\star_0$ \\
$w_0^i \propto \frac{f_0(\v\theta_0^i)}{q_0(\v \theta_0^i)}$ for $i=1,\ldots,N$ \\
\BlankLine
\For{$t = 1$ \KwTo $T$}{
  \BlankLine
    \textcolor{blue}{\textbackslash\textbackslash\hspace{0.1cm}resample}  \\
  \uIf{$\mathrm{ESS}_{t-1}<\mathrm{ESS}_{\min}$}{
    \BlankLine
    draw ${a_t^i}$ with $P(a_t^i=j)\propto w_{t-1}^j$ for $i=1,\ldots,N-1$
    \BlankLine
    \textcolor{blue}{\textbackslash\textbackslash\hspace{0.1cm}sample the ancestor of reference particle} \\
    draw $a_t^N$ with $P(a_t^N=j) \propto {w_{t-1}^j}f(\v \theta^\star_t\vert \v \theta_{t-1}^j$)\\
    \BlankLine
    set $w_{t-1}^i=\frac{1}{N}$ for $i=1,\ldots,N$
  }
  \Else{
    set ${a_t^i} = i$ for $i=1,\ldots,N$
  }
    \BlankLine\BlankLine
    \textcolor{blue}{\textbackslash\textbackslash\hspace{0.1cm}propagate the particles forward one step} \\
    draw $\v \theta_t^i \sim f(\v \theta_t^i\vert \v \theta_{t-1}^{a_{t}^i})$  for $i=1,\ldots,N-1$\\
    set $\v \theta_t^N=\v \theta^\star_t$ \\
    set $\v \theta_{1:t}^i=\{\v {\theta}_{1:t-1}^{^{a_{t}^i}}, \v \theta_t^i\}$ for $i=1,\ldots,N$ \\ 
     \BlankLine\BlankLine
    \textcolor{blue}{\textbackslash\textbackslash\hspace{0.1cm}compute the importance weights } \\
     $w_t^i  \propto w_{t-1}^ig(y_t|\v \theta_{t}^i)$ for $i=1,\ldots,N$
    \BlankLine
} 
\BlankLine
Draw $J$ with $P(J=i)\propto w_T^i$ \\
\Return $\v \theta_{0:T}^J$ \\
\BlankLine
\BlankLine
\caption{PGAS Markov kernel for $p(\v \theta_{0:T} | y_{1:T} $)   \label{alg:pgas_algorithm}}
\end{algorithm}

The particle Gibbs sampler can mix poorly due to path degeneracy in the underlying SMC sampler. The Particle Gibbs with Ancestor Sampling (PGAS) algorithm in \citet{lindsten2014particle} improves the mixing by sampling the ancestor of the reference trajectory at each time $t$ based on the ancestor weights
\begin{equation}\label{eq: AS_weight}
\tilde{w}_{t-1|T}^i\propto {w_{t-1}^i}f(\v \theta^\star_t\vert \v \theta_{t-1}^i) \text{ for }  i=1,\ldots,N,    
\end{equation}
in the special case of Markovian SSMs using the bootstrap filter. \cite{lindsten2014particle} show that the ancestral sampling step does not compromise the validity of the Markov kernel, and can greatly improve mixing, even with few particles. The PGAS algorithm approximates the joint smoothing density with a single forward sweep, unlike other methods that often require explicit forward and backward sweeps \citep{lindsten2013backward}. Algorithm \ref{alg:pgas_algorithm} summarizes the PGAS sampler with the bootstrap filter, and we refer to \cite{lindsten2014particle} for further details.

The initial proposal distribution $r_0(\v\theta_0)$ is often taken as the prior $f_0(\v\theta_0)$, but we find it useful for faster convergence to instead use the smoothing posterior at $t=0$ from a short ($500$ draws) preliminary run of the approximate FFBSx described in the main article. 

The PGAS algorithm can suffer from particle degeneracy in near-degenerate models \citep{lindsten2015particle}, for example when the innovations in the transition model have close to zero variance for some $t$.
A potential solution is to rejuvenate both the ancestors and the reference trajectory itself; see \citet{lindsten2015particle} in the context of PGAS. However, we find that these modifications require careful adjustments to suit our specific model, indicating that further research is needed to adapt them effectively. We instead develop an approximate sampling step that replaces the Kalman filter in the FFBS sampler with the extended Kalman filter (EKF), as explained in our article.

\section{Additional details and results from the simulation experiments}

\subsection{Experiment 1}\label{app: exp1}

The first experiment simulates data from the time-varying TVSAR$(2,2)_{12}$
\begin{equation}\label{eq:exp1}
   (1-\phi_{1t} L -\phi_{2t} L^2)(1-\Phi_{1t} L^{12} -\Phi_{2t} L^{24})y_t = \varepsilon_t,
\end{equation}
where $\varepsilon_t \sim \mathrm{N}(0,1)$. 

The time evolution of the unrestricted non-seasonal parameters is
\begin{align}\label{eq: Exp1_nonseas}
\theta_{1t} = \begin{cases}
       \phantom{-}0.8 \sin(\frac{\pi t}{T}) & \text{ for } t=1,\ldots,500 \\
       -0.8\sin(\frac{\pi t}{T}) & \text{ for }  t=501,\ldots,1000
\end{cases} & & 
\theta_{2t} = -0.8 \text{ for all }t .
\end{align}
and the time evolution of the unrestricted seasonal parameters is
\begin{align}\label{eq: Exp1_seas}
\Theta_{1t} = \begin{cases}
       -0.70 & \text{ for } t=1,\ldots,300 \\
       \hspace{0.8cm}0 & \text{ for } t=301,\ldots,700 \\
       \phantom{-}0.95 & \text{ for } t=701,\ldots,1000
\end{cases} & &
\Theta_{2t} = -0.9 \text{ for all }t .
\end{align}

\begin{figure}
 \centering
 \includegraphics[width=0.8\textwidth]{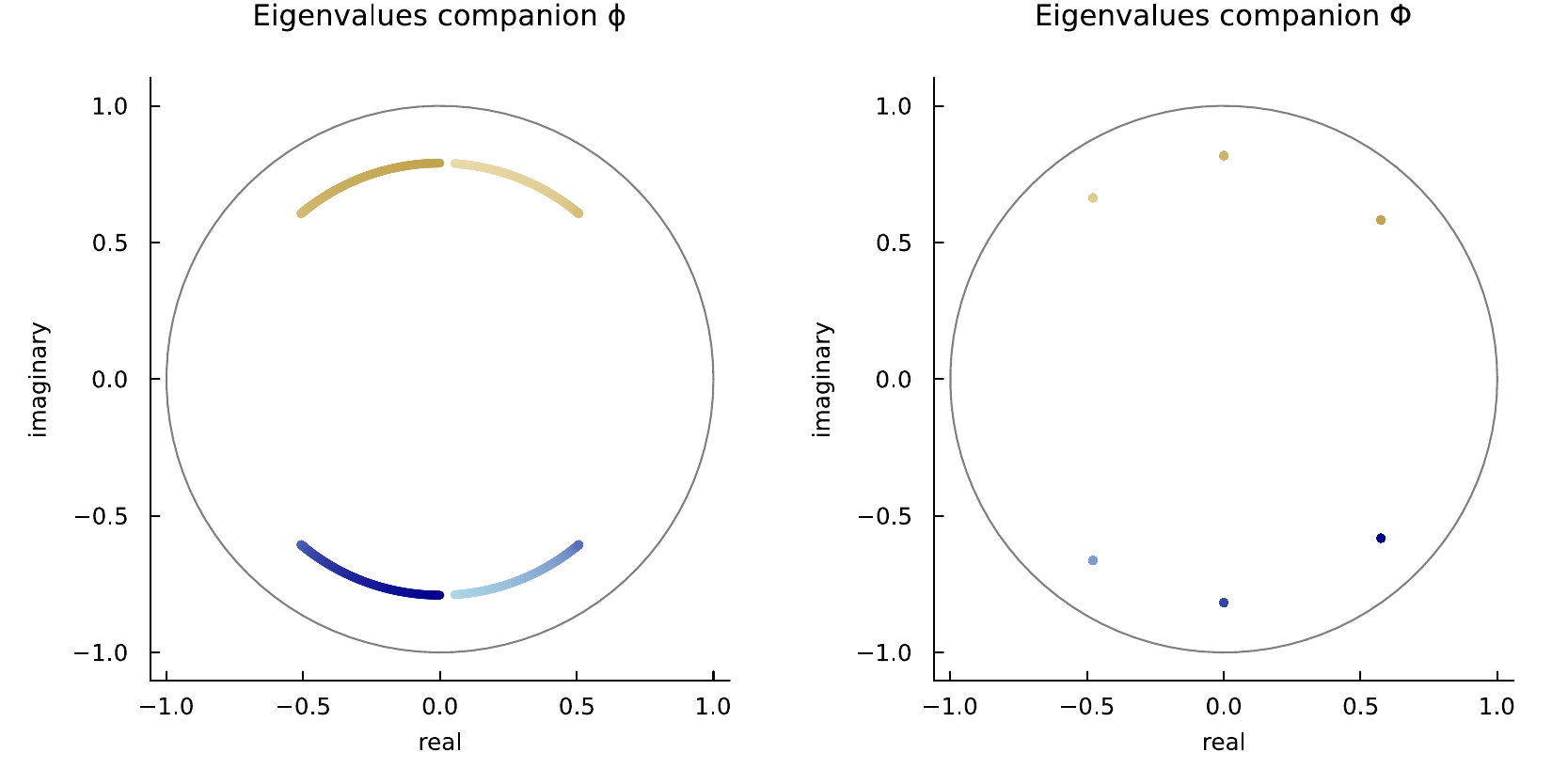}
\caption{Experiment 1. Eigenvalues for the companion matrix of the regular and seasonal polynomials over time for the data generating process. The color scale for the eigenvalues represents time, with darker colors indicating later points in the time sequence.}\label{fig:Exp1Eigen}
\end{figure}

The left graph of Figure \ref{fig:Exp1Eigen} shows the time evolution of the eigenvalues of the companion matrix for the regular AR polynomial
\begin{equation}
    \begin{pmatrix}
    \phi_{1t} \, \phi_{2t}  \, \cdots  \, \phi_{pt}  \\
    \boldsymbol{I}_{p-1} \quad \boldsymbol{0}_{p-1} 
    \end{pmatrix}.
\end{equation}
The right graph of Figure \ref{fig:Exp1Eigen} shows the time evolution of the eigenvalues of the companion matrix for the seasonal AR polynomial.


\begin{table}
\begin{center}
\begin{tabular}{c c c c c c c c c c c c }
\hline
\hspace{1.5cm} &  \multicolumn{3}{c}{\textbf{FFBSx-m}}  &  \hspace{0.5cm}  & \multicolumn{3}{c}{\textbf{FFBSx-l}} & \hspace{0.5cm} &  \multicolumn{3}{c}{\textbf{PGAS(100)}}  \\ [0.5ex] 
& $\pi/4$ & $\pi/2$ & $3\pi/4$ &  & $\pi/4$ & $\pi/2$ & $3\pi/4$ &  & $\pi/4$ & $\pi/2$ & $3\pi/4$\\ [0.5ex] 
\cline{2-4} \cline{6-8} \cline{10-12}
$t=100$ & 377 & 330 & 460 & & 506 & 475 & 533 & & 38 & 35 & 41 \\ 
 
$t=400$ & 418 & 454 & 436 & & 548 & 620 & 595 & & 167 & 216 & 179 \\
 
$t=800$ & 375 & 412 & 361 & & 418 & 534 & 474 & & 324 & 412 & 348 \\
 \hline
\end{tabular}
\caption{Experiment 1. Effective sample size (ESS) for estimating the spectral density for three algorithms. ESS for the spectral density is computed at three selected time points (rows) and at three selected frequencies (columns). The ESS is from an initial posterior sample of \num{10000} draws thinned down to \num{1000} draws. An offset of $10^{-16}$ is used.}
\label{table:Experiment1_ESS}
\end{center}
\end{table}

\begin{figure}
 \centering
 \includegraphics[width=0.95\textwidth]{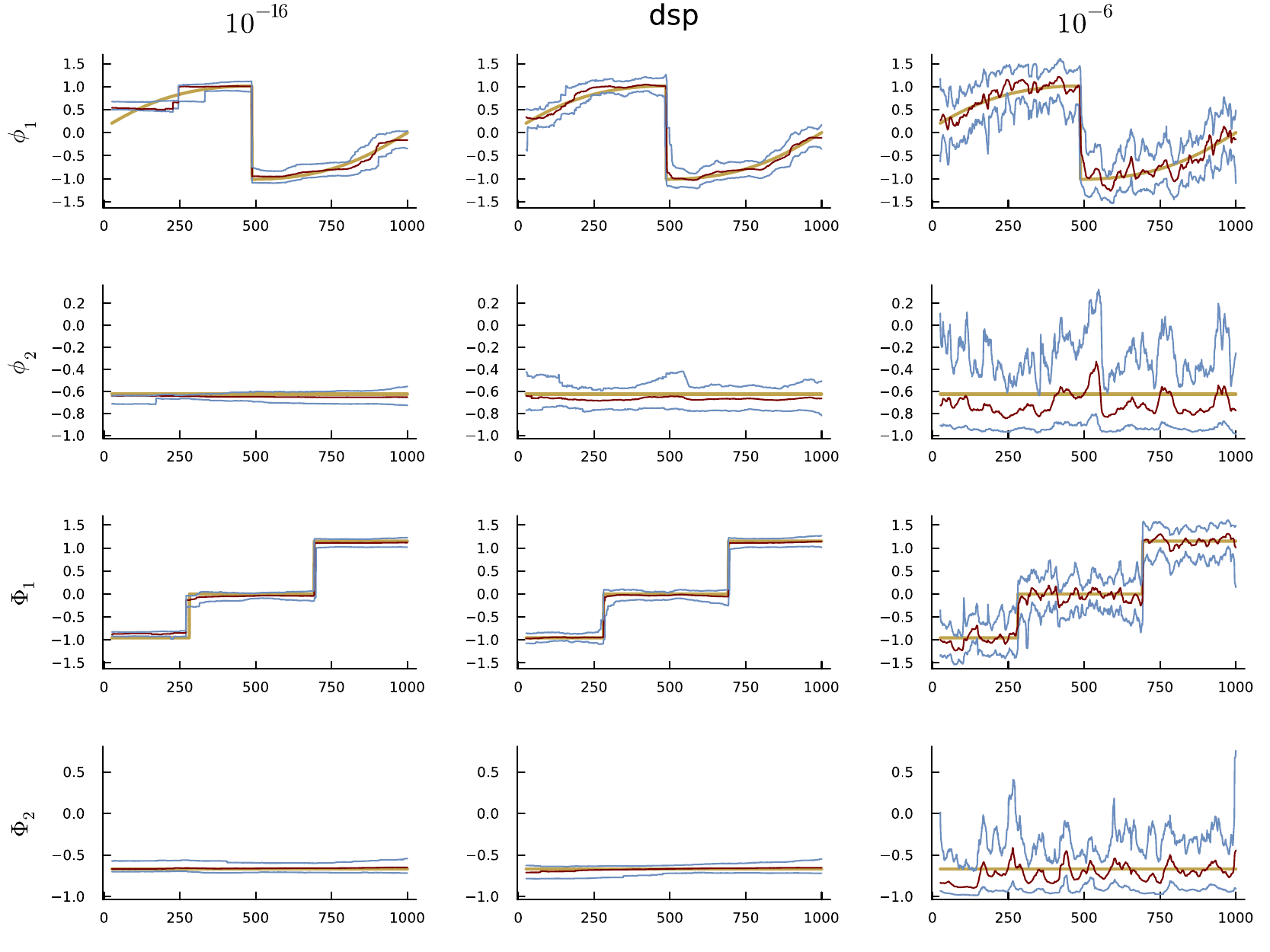}
\caption{Experiment 1. Posterior median and 95\% credible intervals from PGAS(100) from one run for the three different offsets in the log-volatility model for $\v h_t$ }\label{fig:exp1_postOffsets}
\end{figure}

\begin{figure}
 \centering
 \includegraphics[width=1\textwidth]{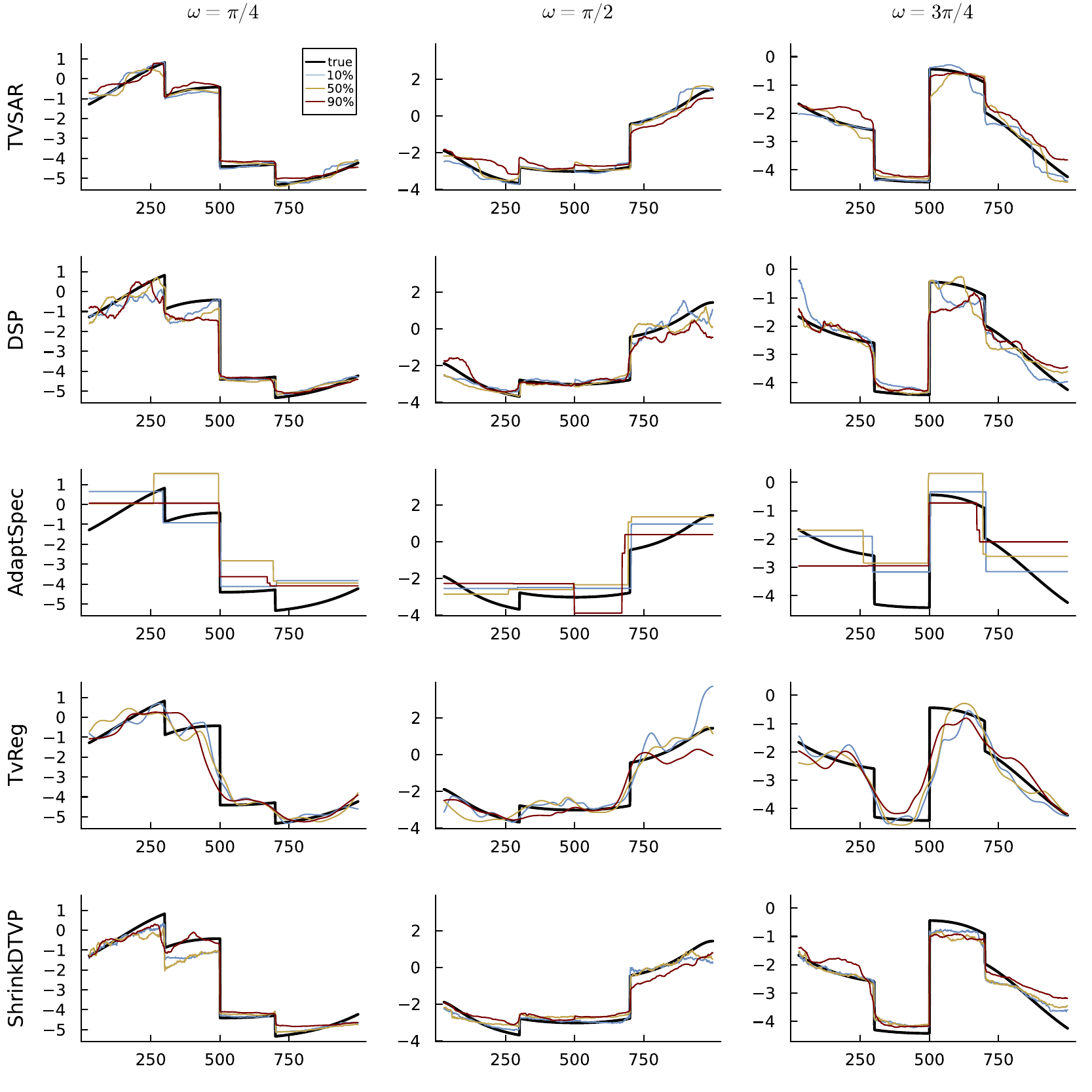}
    \caption{Experiment 1. Time evolution of the log spectral density at three different frequencies (columns) for several different models (rows). The black line is the true spectral density and the three colored lines are the posterior median estimates from three different datasets chosen from the MSE percentiles.}\label{fig:exp1spectralovertime}
\end{figure}

\begin{figure}
 \centering
 \includegraphics[width=0.9\textwidth]{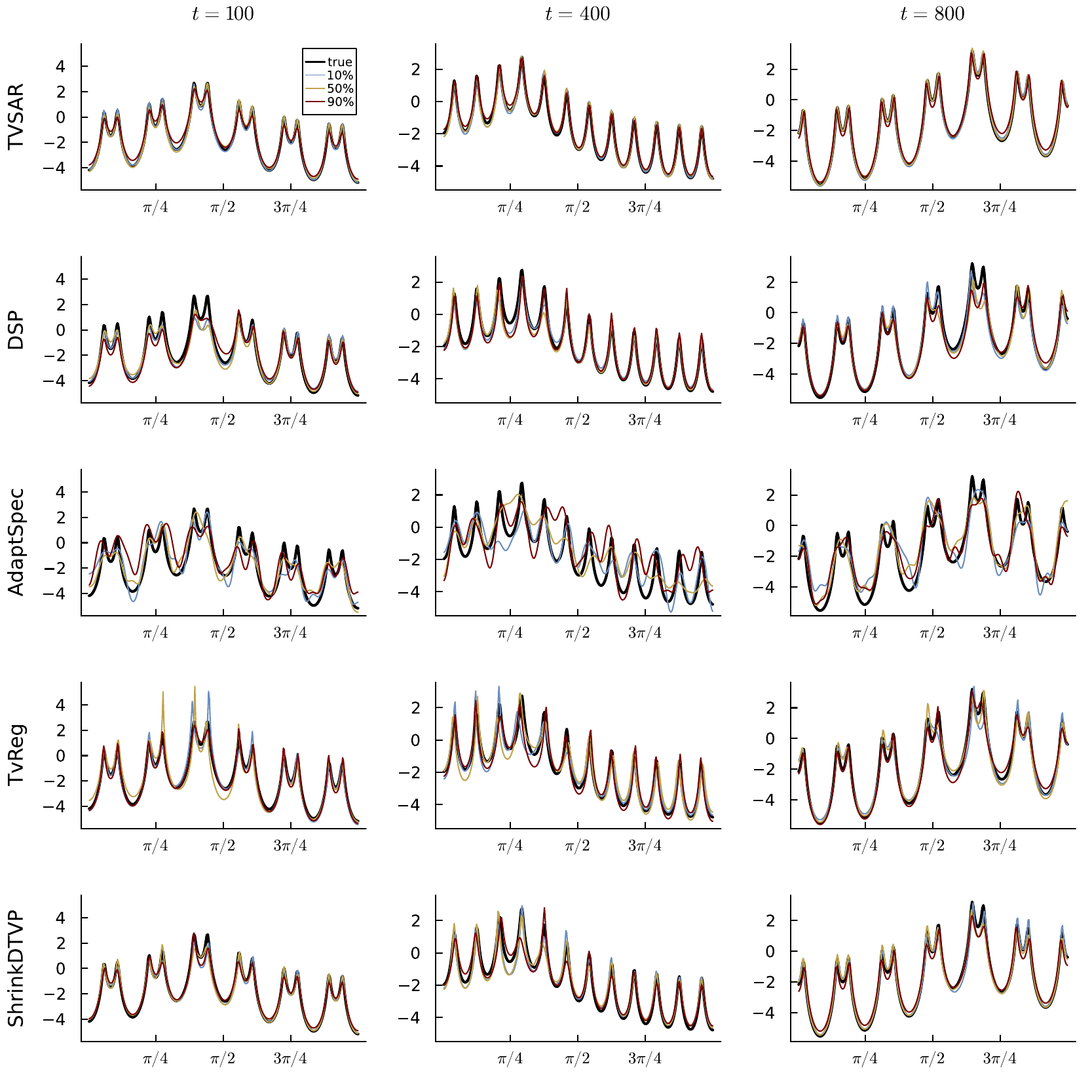}
    \caption{Experiment 1. Log spectral density at three different time points. The black line is the true spectral density and the three colored lines are the posterior medians from three different datasets chosen from the MSE percentiles.}\label{fig:exp1spectraloverfreq}
\end{figure}

Figure~\ref{fig:exp1spectralovertime} gives an alternative view to the heatmaps in Figure~\ref{fig:exp1heatmaps} by plotting the estimated time-evolution at some selected frequencies. 
Figure~\ref{fig:exp1spectraloverfreq} plots the fitted log spectral densities at three different time points, again showing the fit for three different datasets selected from the MSE percentiles for each method. Clearly, our TVSAR model gives less wiggly fits than DSP, while TvReg oversmooths.

For completeness, we also compare the performance of the TVSAR model with a SARMA model with time invariant parameters in this experiment. Figure~\ref{fig:exp1MSE} does not show the MSE for this model, but it does, as expected, perform very poorly with a median MSE of $5.22$. The static SARMA model is fitted with \texttt{auto.arima} in the \texttt{forecast} package in R, which automatically determines the optimal AR and MA lag orders; we allow for a maximal order of $5$ for all regular and seasonal lags. 

\subsection{Experiment 1 - one lag version}\label{app: sar11}

\begin{figure}
 \centering
 \includegraphics[width=0.8\textwidth]{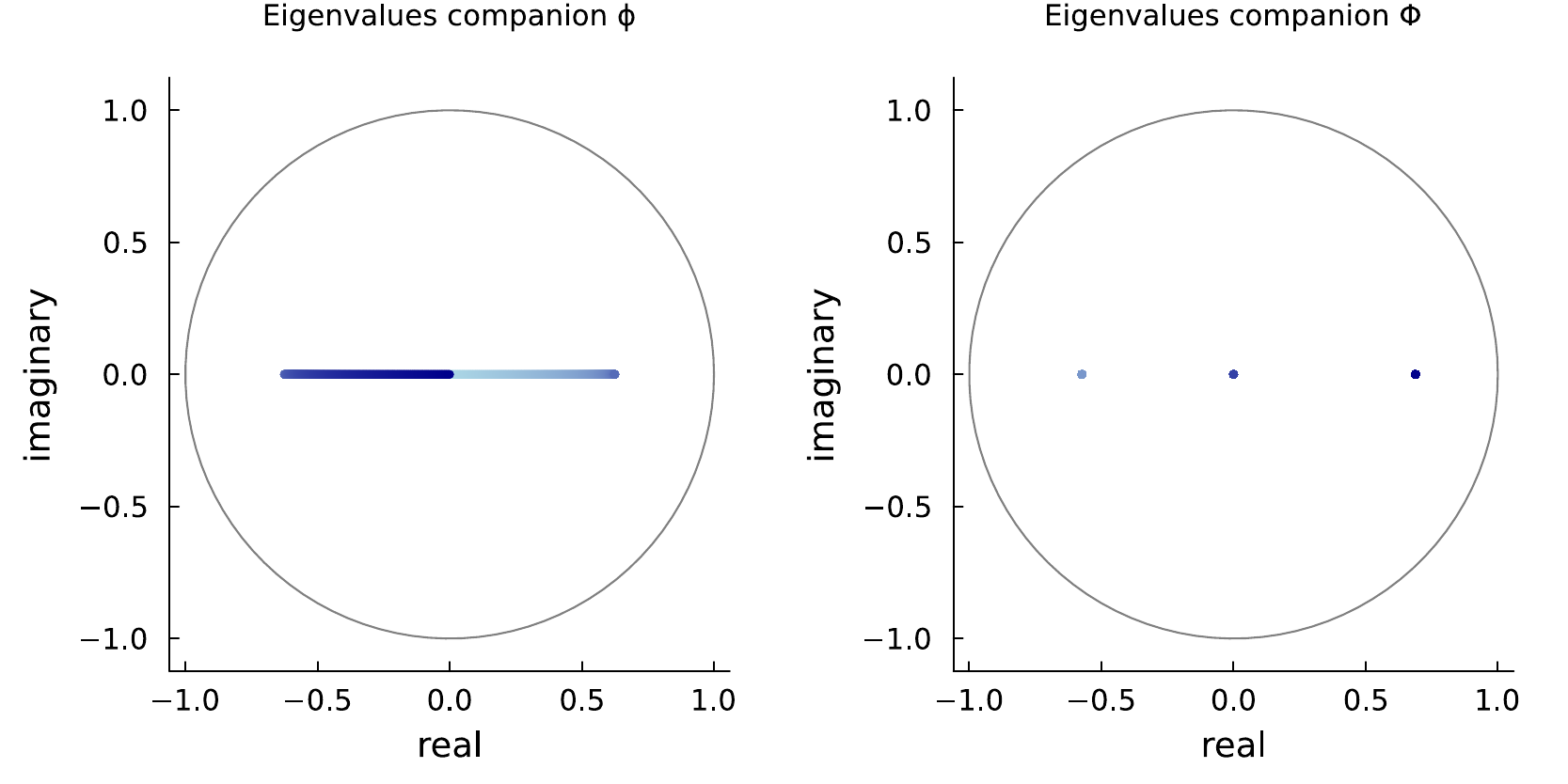}
\caption{SAR(1,1). Eigenvalues for the companion matrix of the regular and seasonal polynomials over time for the data generating process. The color scale for the eigenvalues represents time, with darker colors indicating later time points.}\label{fig:sar11Eigen}
\end{figure}

In this experiment, data is simulated from the following one-lag restricted version of the DGP in 
Experiment~1, a time-varying TVSAR$(\mathbf{s}, \mathbf{p})$ with $\mathbf{s}=(1,12)$ and $\mathbf{p}=(1,1)$,
\begin{align}\label{eq:sar11}
   (1-\phi_{1t} L)(1-\Phi_{1t} L^{12})y_t &= \varepsilon_t,
\end{align}
where $\varepsilon_t \sim \mathrm{N}(0,1)$. The time evolution of the parameters are the same as for the first regular and seasonal lag in Experiment 1. Figure \ref{fig:sar11Eigen} shows the time evolution of the eigenvalues of the companion matrix of the regular and seasonal AR coefficients.

\begin{figure}
 \centering
 \includegraphics[width=0.7\textwidth]{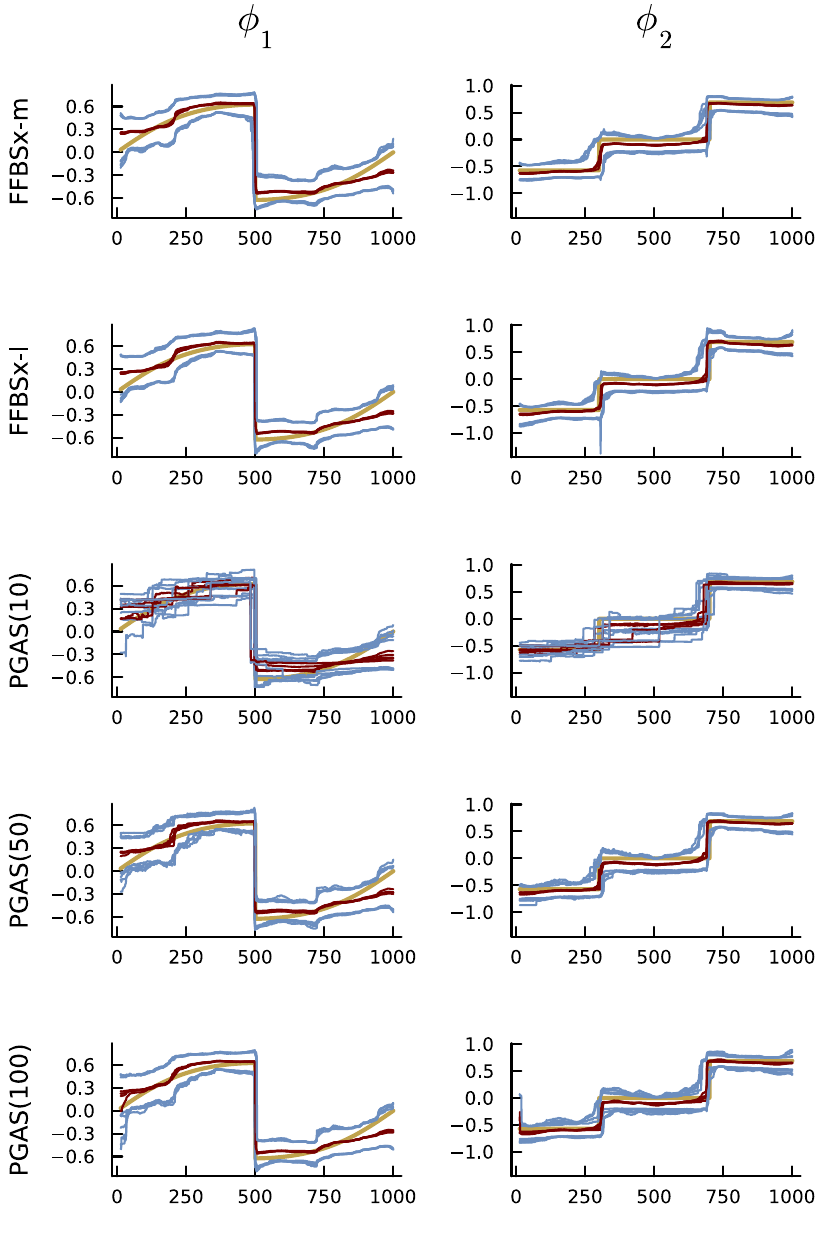}
    \caption{SAR(1,1). Experiment 1. Assessing MCMC convergence by re-estimating the model on the same dataset with five different initial values. The red and blue lines are posterior medians and $95\%$ equal tail credible bands over time for each of the five runs. The beige line is the true parameter evolution in the $\phi$-parameterization.}
    \label{fig:sar11convergence} 
\end{figure}

\begin{figure}
 \centering
 \includegraphics[width=0.32\textwidth]{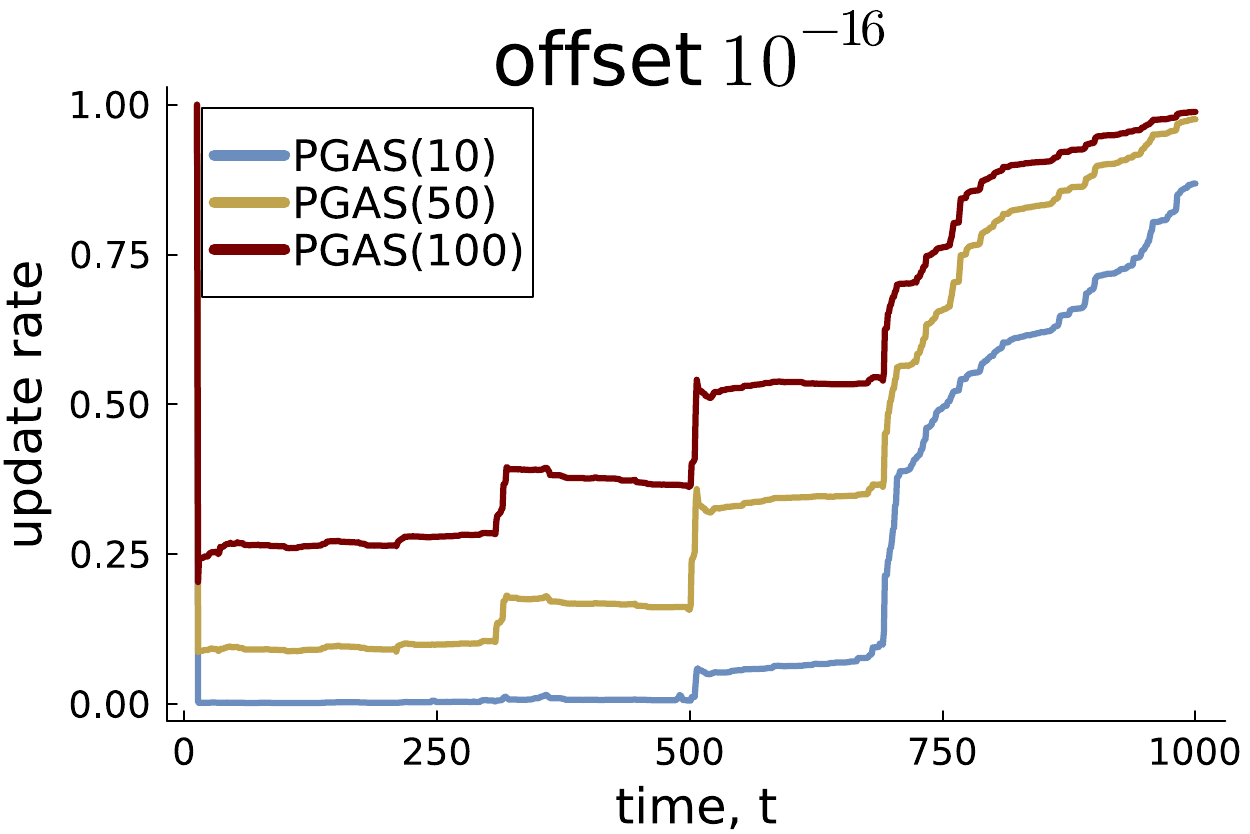}
 \includegraphics[width=0.32\textwidth]{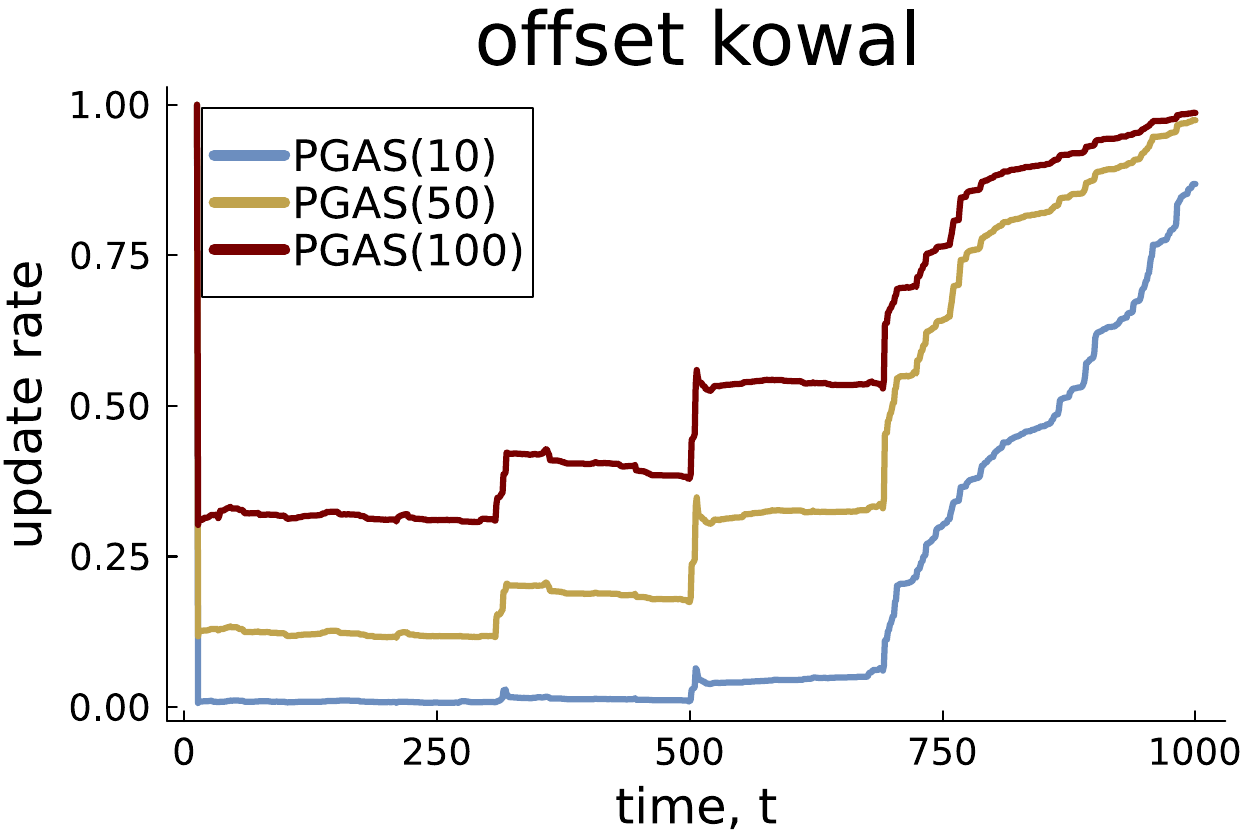}
 \includegraphics[width=0.32\textwidth]{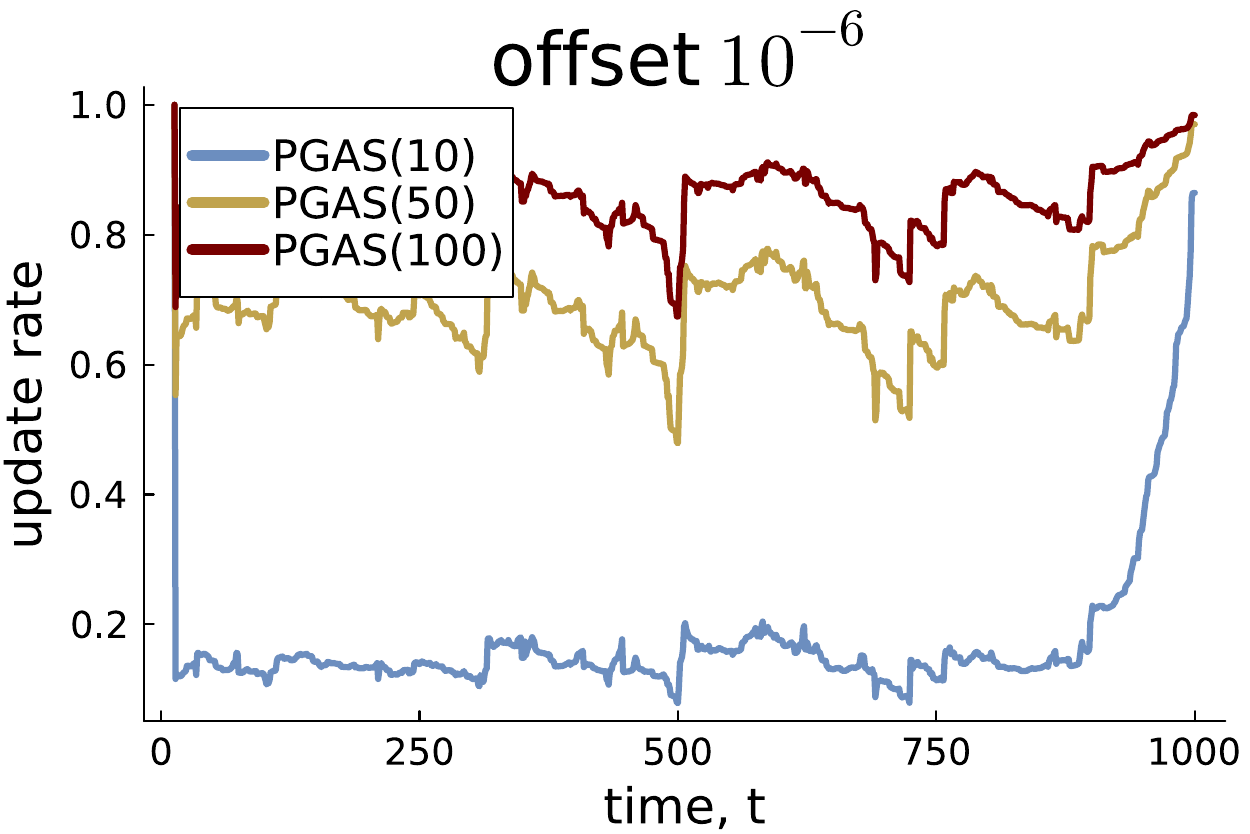}
\caption{SAR(1,1). PGAS update rates for three offsets in the log-volatility model for $\v h_t$ for different numbers of particles.}
\label{fig:sar11updaterates}
\end{figure}

Figures~\ref{fig:sar11convergence} and \ref{fig:sar11updaterates} show that PGAS converges much faster than for the SAR(2,2) model in Experiment 1, even for an offset of $10^{-16}$. 

\newpage

\subsection{Experiment 2}\label{app: exp2}

The second experiment simulates data from a multi-seasonal TVSAR with two seasonal periods $\v s=(4,12)$, non-seasonal order $p=1$ and seasonal orders $\v P=(1, 1)$ (i.e., one lag for each seasonal period). This corresponds to the TVSAR$(p=1, \v P = [1,1])_{[4,12]}$ model
\begin{equation}\label{eq: exp2}
   (1-\phi_{11t} L) (1-\Phi_{11t} L^4)(1-\Phi_{21t} L^{12})y_t = \varepsilon_t, 
\end{equation}
with $\varepsilon_t \sim \mathrm{N}(0,1)$. The time evolution of the unrestricted non-seasonal parameter is given by
\begin{equation}\label{eq: Exp2_nonseas}
    \theta_{11t} = 0.8 \sin(2\pi t/T) \text{ for } t=1,\ldots,1000, 
\end{equation} 
while the unrestricted parameters in the two seasonal periods follow
\begin{align}\label{eq: Exp2_seas2}
    \Theta_{21t} = 0.5 \text{ for all }t    & &
    \Theta_{31t} = 
    \begin{cases}
       -0.50 & \text{ for } t=1,\ldots,250 \\
       \hspace{0.8cm}0 & \text{ for } t=251,\ldots,750 \\
       \phantom{-}0.95 & \text{ for } t=751,\ldots,1000.
    \end{cases}
\end{align}

\begin{figure}
 \centering
 \includegraphics[width=0.8\textwidth]{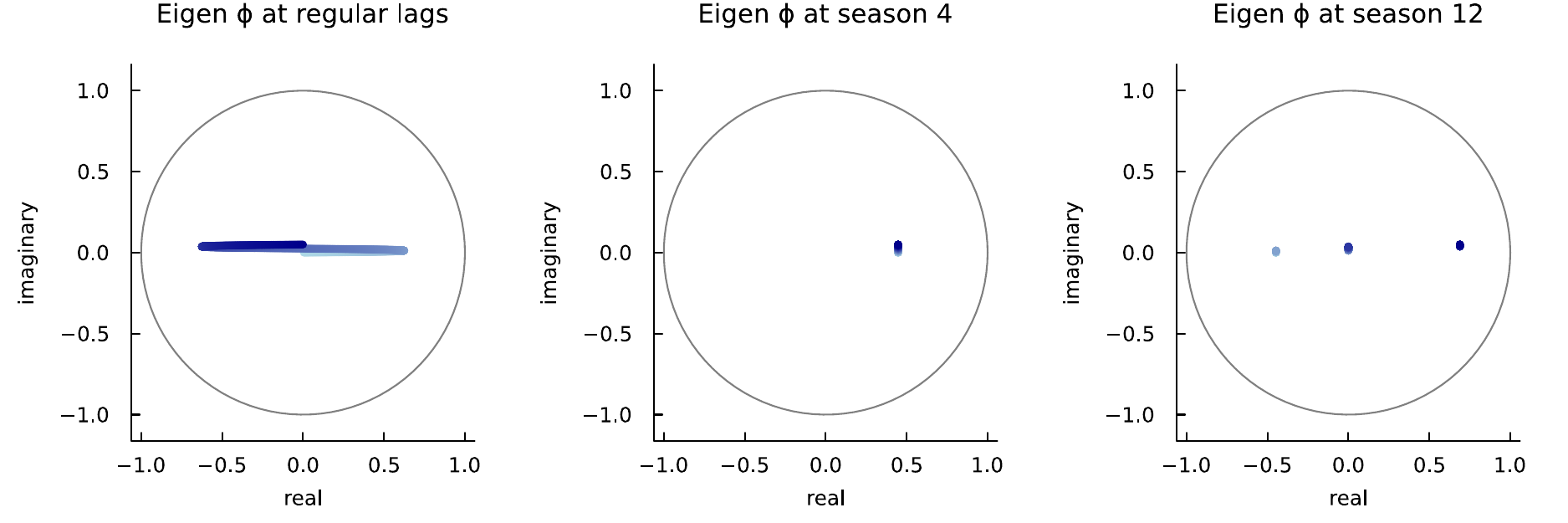}
\caption{Experiment 2. Eigenvalues for the companion matrix of the regular and seasonal polynomials over time for the data generating process. The color scale for the eigenvalues represents time, with darker colors indicating later time points.}\label{fig:Exp2Eigen}
\end{figure}

\begin{figure}
 \centering
 \includegraphics[width=0.9\textwidth]{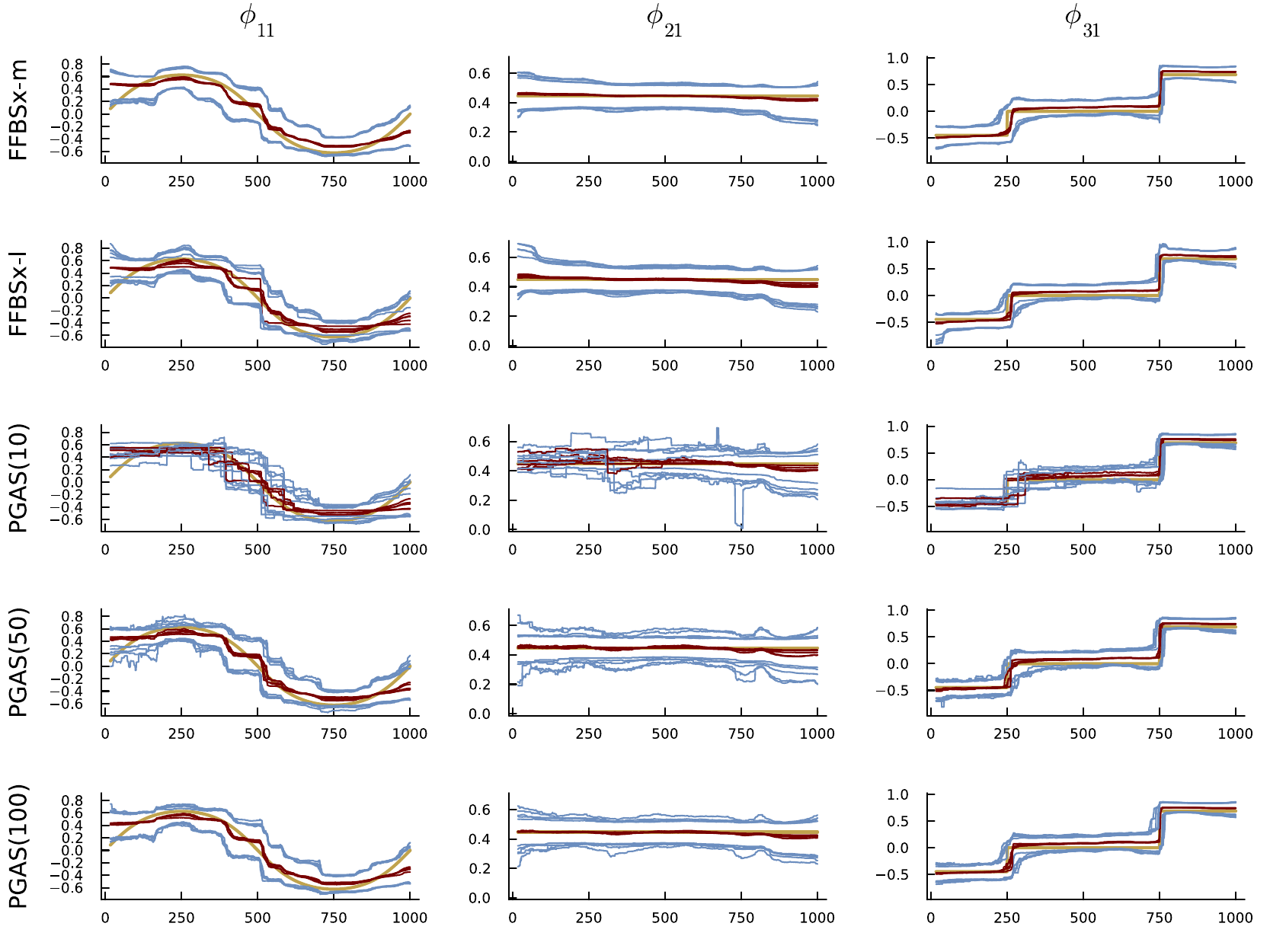}
    \caption{Experiment 2. Assessing MCMC convergence by re-estimating the model on the same dataset with five different initial values. The red and blue lines are posterior medians and $95\%$ equal tail credible bands over time for each of the five repeated runs. The beige line is the true parameter evolution in the $\phi$-parameterization.}\label{fig:exp2convergence}
\end{figure}

\begin{figure}
 \centering
 \includegraphics[width=1\textwidth]{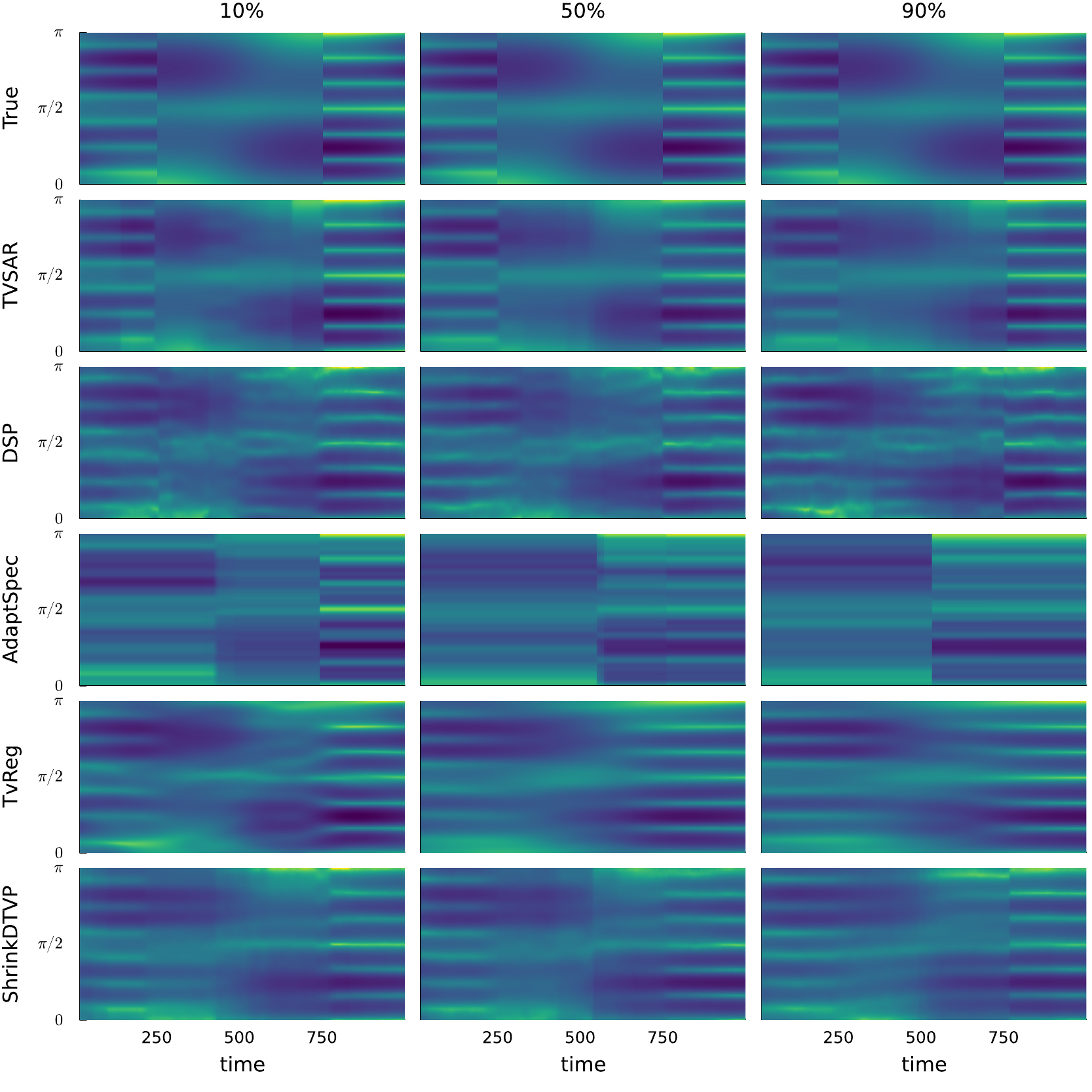}
    \caption{Experiment 2. Heat maps of the estimated spectrogram for the different methods. The columns correspond to different datasets chosen from the percentiles of each model's MSE distribution, to show the performance of each model when it performs well ($10\%$), average ($50\%$) and poorly ($90\%$).}\label{fig:exp2heatmaps}
\end{figure}

\begin{figure}
 \centering
 \includegraphics[width=0.32\textwidth]{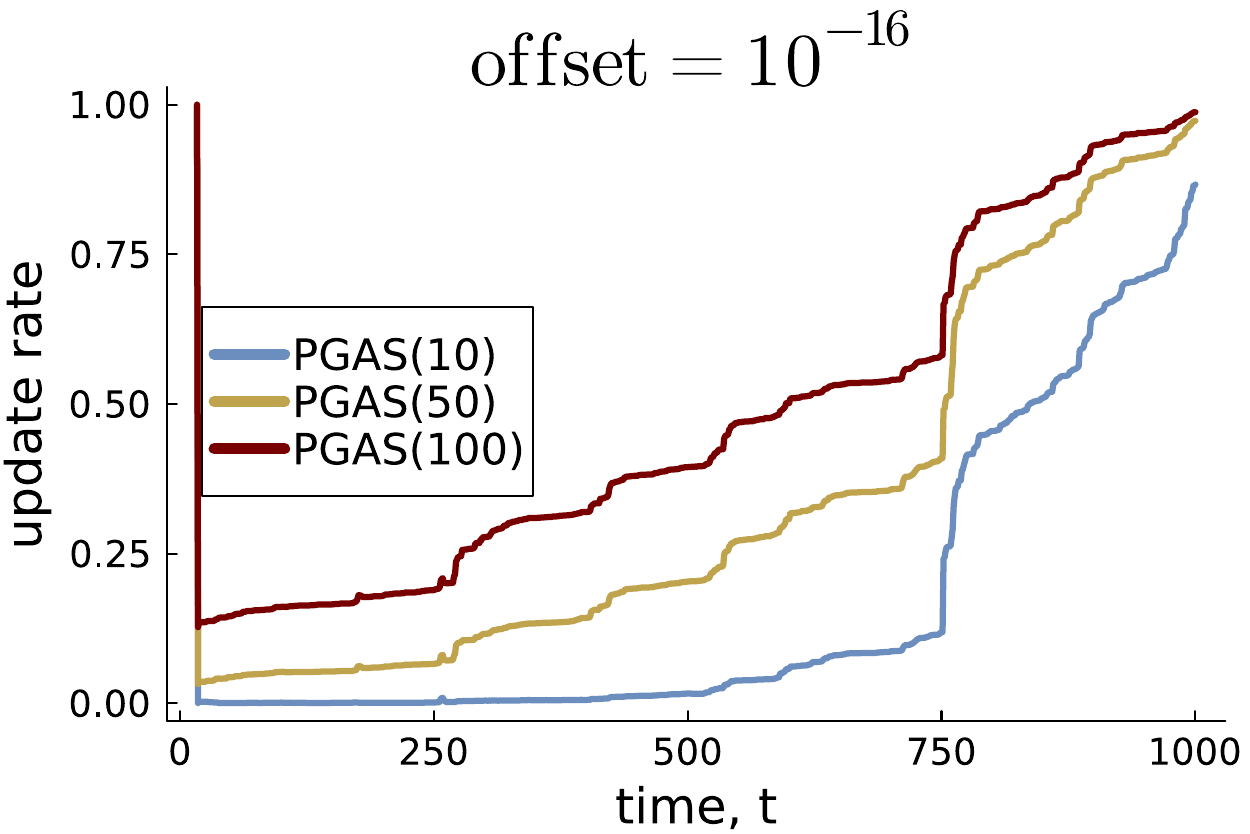}
\caption{Experiment 2. PGAS update rates for draws using the smallest offset $10^{-16}$ in the log-volatility model for $\v h_t$ for different numbers of particles.}\label{fig:exp2updaterates}
\end{figure}

\begin{figure}
 \centering
 \includegraphics[width=0.9\textwidth]{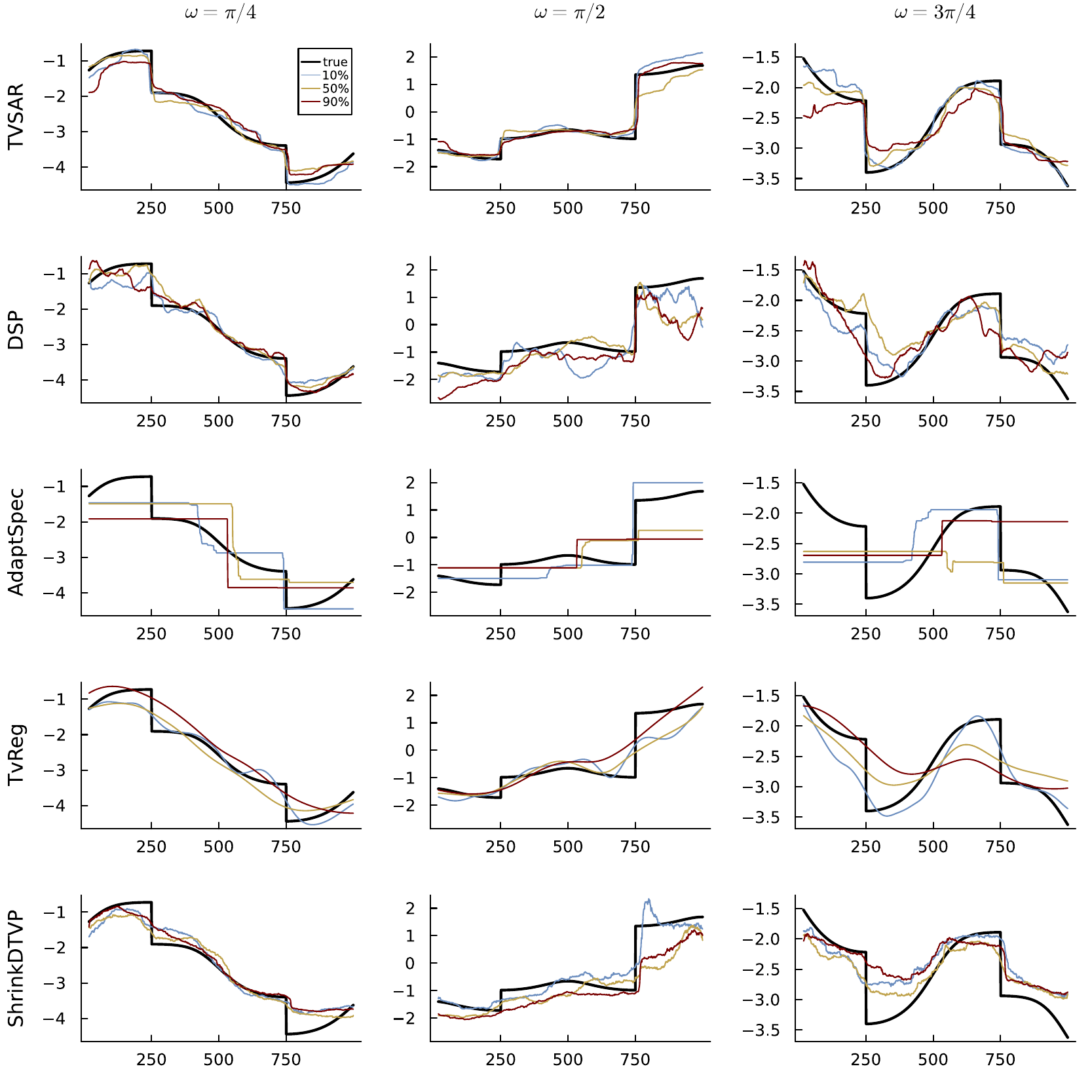}
    \caption{Experiment 2. Time evolution of the log spectral density at three frequencies. The black line is the true spectral density and the three colored lines are the posterior median from three different datasets chosen from the MSE percentiles.}\label{fig:exp2spectralovertime}
\end{figure}

\begin{figure}
 \centering
 \includegraphics[width=0.9\textwidth]{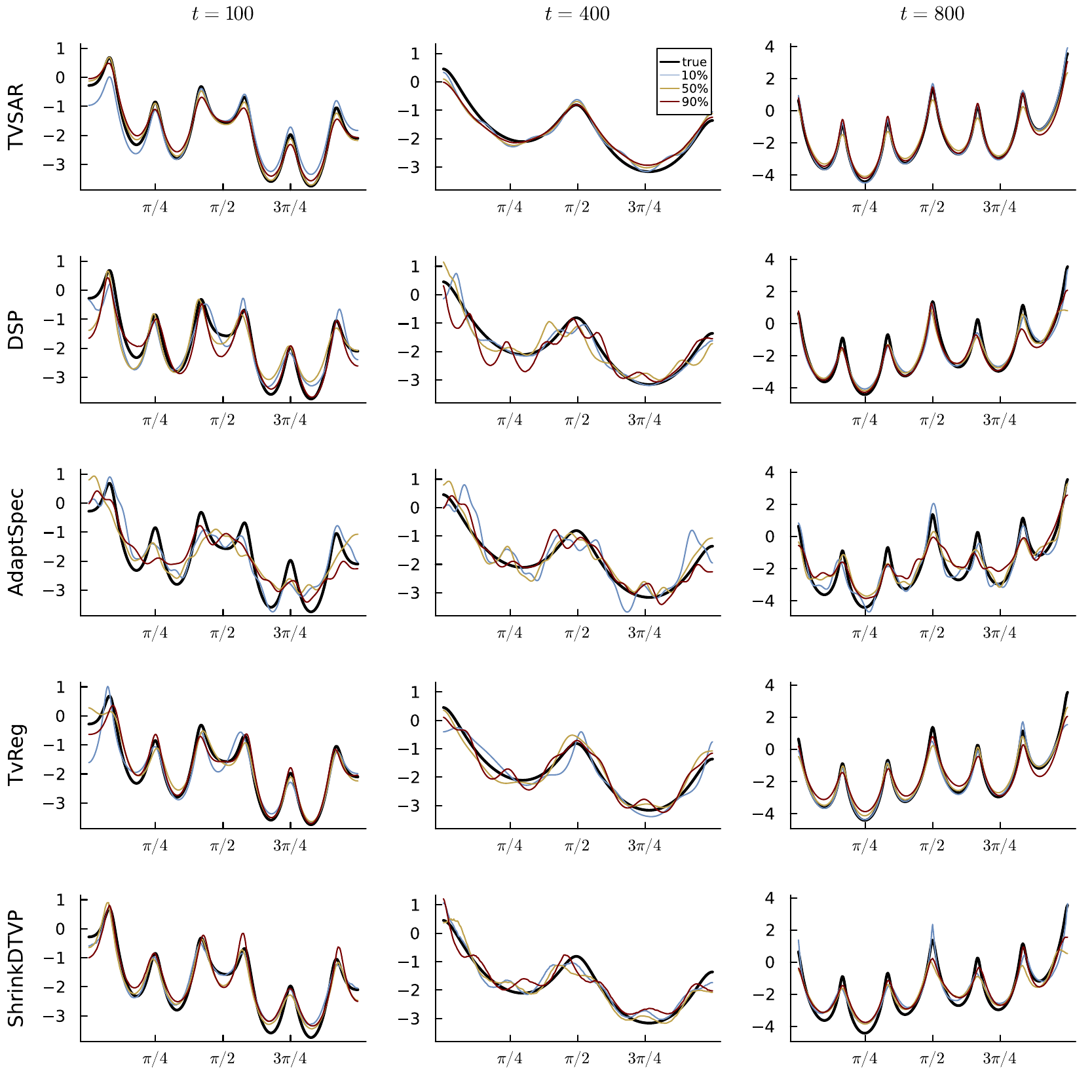}
    \caption{Experiment 2. Log spectral density at three time points. The black line is the true spectral density and the three colored lines are the posterior median from three different datasets chosen from the MSE percentiles.}\label{fig:exp2spectraloverfreq}
\end{figure}

Figure~\ref{fig:Exp2Eigen} shows the time evolution of the eigenvalues of the companion matrix of the regular and seasonal AR coefficients; the eigenvalues are slightly jittered to avoid complete overlap. Table~\ref{table:Experiment2_ESS} displays effective sample sizes for spectral density estimates for the different posterior algorithms for the TVSAR model. 
Figure~\ref{fig:exp2convergence} explores the convergence of estimates from different algorithm over five different initial values; similarly to Experiment 1, the FFBSx posteriors are similar from different initial values, and PGAS needs up to $100$ particles to give stable results. Figures \ref{fig:exp2heatmaps}-\ref{fig:exp2spectraloverfreq} give additional results for Experiment 2.


\begin{table}
\begin{center}
\begin{tabular}{c c c c c c c c c c c c }
\hline
\hspace{1.5cm} &  \multicolumn{3}{c}{\textbf{FFBSx-m}}  &  \hspace{0.5cm}  & \multicolumn{3}{c}{\textbf{FFBSx-l}} & \hspace{0.5cm} &  \multicolumn{3}{c}{\textbf{PGAS(100)}}  \\ [0.5ex] 
& $\pi/4$ & $\pi/2$ & $3\pi/4$ &  & $\pi/4$ & $\pi/2$ & $3\pi/4$ &  & $\pi/4$ & $\pi/2$ & $3\pi/4$\\ [0.5ex] 
\cline{2-4} \cline{6-8} \cline{10-12}
$t=100$ & 498 & 514 & 482 & & 593 & 606 & 587 & & 226 & 237 & 215\\ 
 
$t=400$ & 368 & 542 & 337 & & 411 & 563 & 375 & & 257 & 333 & 238 \\
 
$t=800$ & 483 & 597 & 505 & & 496 & 767 & 536 & & 412 & 620 & 457 \\
 \hline
\end{tabular}
\caption{Experiment 2. Effective sample size (ESS) for estimating the spectral density for three algorithms. ESS for the spectral density is computed at three selected time points (rows) and at three selected frequencies (columns). The ESS is from an initial posterior sample of \num{10000} draws thinned down to \num{1000} draws. An offset of $10^{-16}$ is used.}
\label{table:Experiment2_ESS}
\end{center}
\end{table}

\subsection{Experiment 3 - Redundant lags}\label{app:exp3}

\begin{figure}
 \centering
 \includegraphics[width=0.6\textwidth]{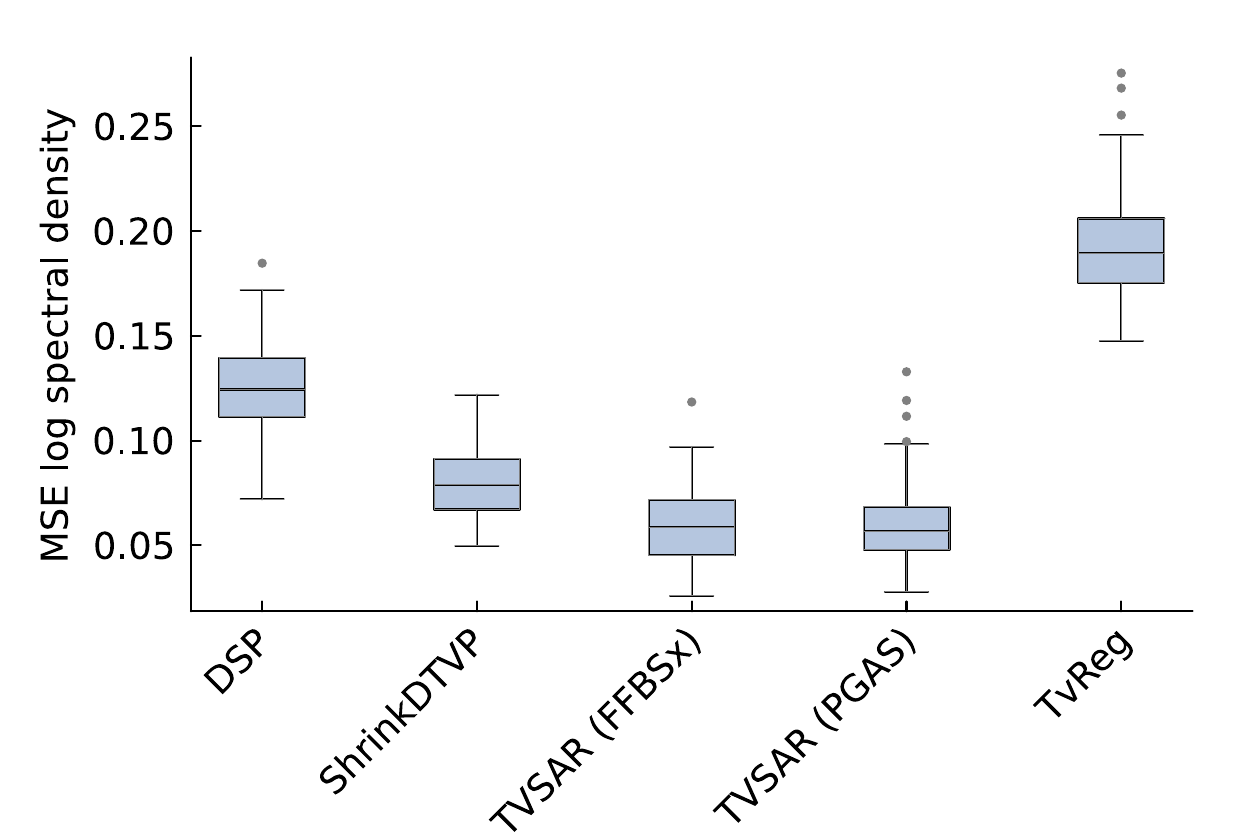}
    \caption{Experiment 3. Box plots of the MSE for the log spectral density over time and frequency from 100 simulated datasets.}\label{fig:exp3MSE}
\end{figure}

Experiments 1 and 2 use the correct number of regular and seasonal lags. The empirical application in Section \ref{sec:applications} uses the log predictive scores to determine the optimal number of lags. This third and final experiment shows that the dynamic shrinkage prior is able to effectively shrink redundant lags to zero in the TSVAR model when it has redundant lags. We simulate data from the model in Experiment~1, but with the coefficients for the second regular and second seasonal lags set to zero for all $t$. We then fit misspecified models with two regular and two seasonal lags. 

\begin{figure}
 \centering
 \includegraphics[width=0.9\textwidth]{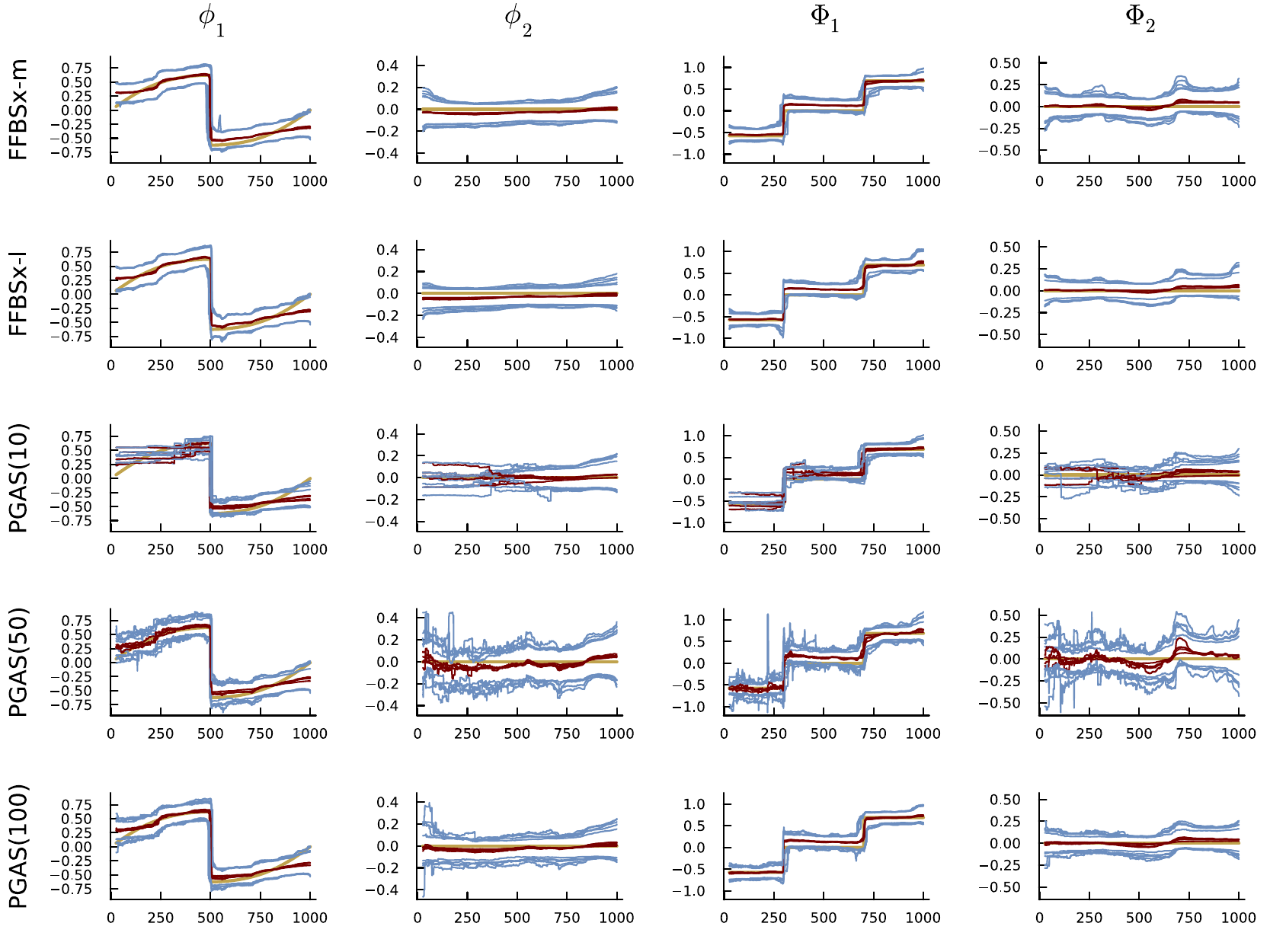}
    \caption{Experiment 3. Assessing MCMC convergence by re-estimating the model using five different initial values drawn from the prior of the global hyperparameters $\v\mu$ and $\v\kappa$. All runs are for the same dataset. The red and blue lines are posterior medians and $95\%$ equal tail credible bands over time for each of the five repeated runs. The beige line is the true parameter evolution.}\label{fig:exp3convergence}
\end{figure}

\begin{figure}
 \centering
 \includegraphics[width=1\textwidth]{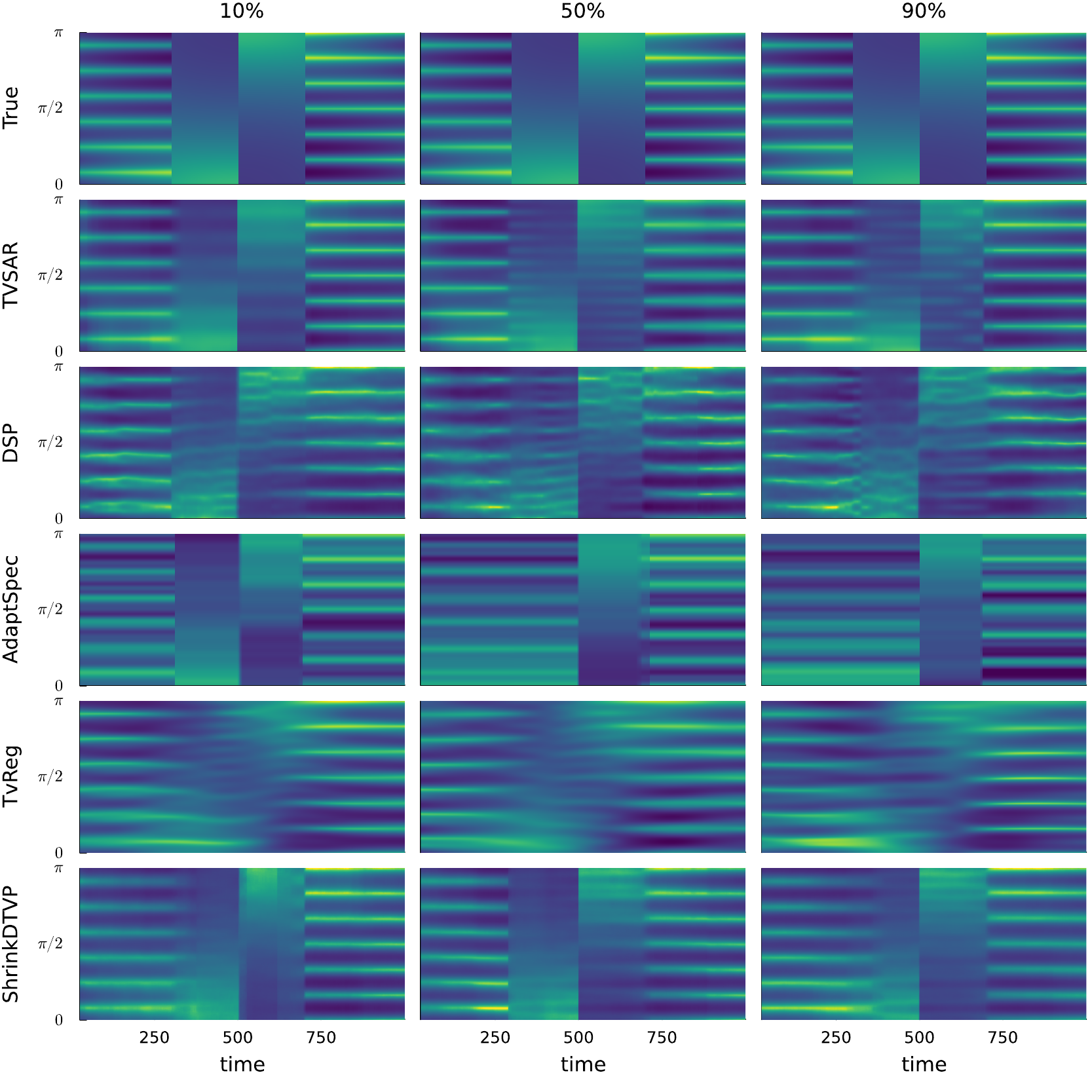}
    \caption{Experiment 3. Heatmaps of the estimated spectrogram for the different methods. The columns correspond to different datasets chosen from the percentiles of each model's MSE distribution, to show the performance of each model when it performs well ($10\%$), average ($50\%$) and poorly ($90\%$).}\label{fig:exp3heatmaps}
\end{figure}

\begin{figure}
 \centering
 \includegraphics[width=0.9\textwidth]{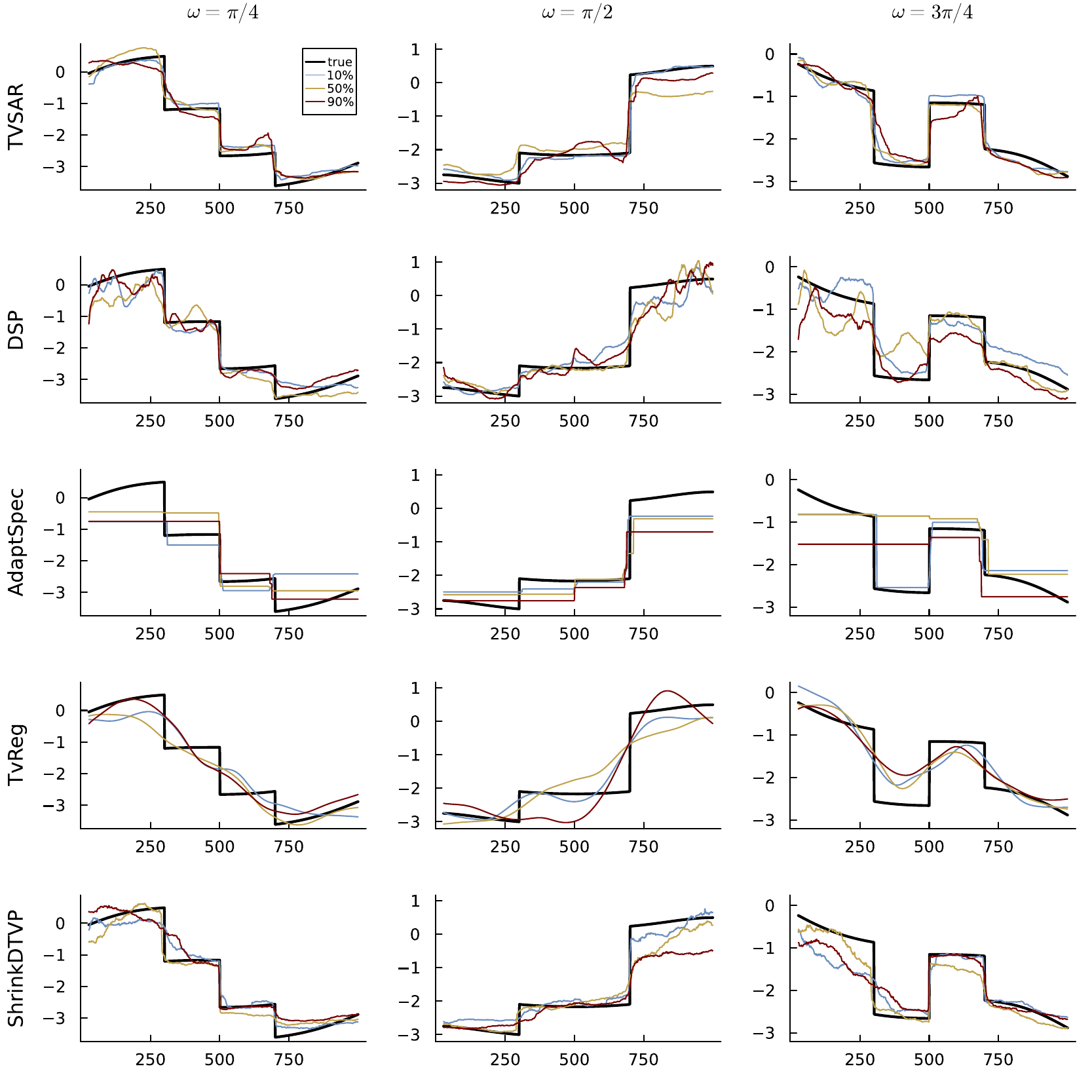}
    \caption{Experiment 3. Time evolution of the log spectral density at three different frequencies. The black line is the true spectral density and the three colored lines are the posterior median from three datasets chosen from the MSE percentiles.}\label{fig:exp3spectralovertime}
\end{figure}

\begin{figure}
 \centering
 \includegraphics[width=0.9\textwidth]{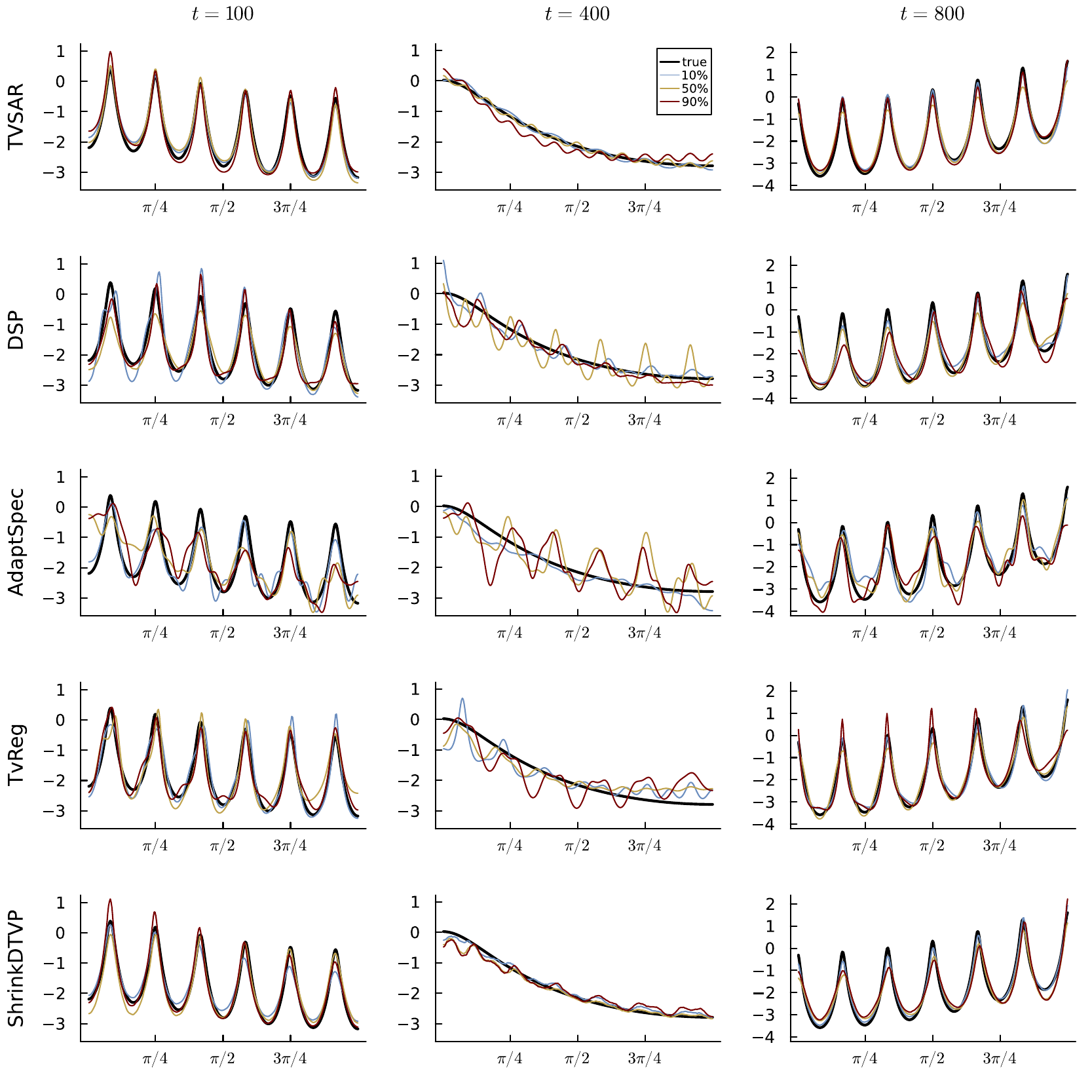}
    \caption{Experiment 3. Log spectral density at three different time points. The black line is the true spectral density and the three colored lines are the posterior median from three datasets chosen from the MSE percentiles.}\label{fig:exp3spectraloverfreq}
\end{figure}

Figure~\ref{fig:exp3MSE} shows that fitting a TVSAR$(2,2)_{12}$ to data from a TVSAR$(1,1)_{12}$ still performs well, i.e.~the TVSAR model is able to shrink the time trajectory of redundant lags to zero for all $t$; see Figure \ref{fig:exp3convergence}. TvReg, which is the only time-domain model without shrinkage, clearly suffers from the redundant parameters and is substantially worse than ShrinkDTVP here. Figures \ref{fig:exp3heatmaps}-\ref{fig:exp3spectraloverfreq} give additional results for Experiment 3.

To further demonstrate that the model can effectively shrink redundant lags to zero, we conduct additional simulation experiments.  Figures S27 - S30 summarize the results of fitting several over-specified models - TVSAR$(3,3)_{12}$,  TVSAR$(4,4)_{12}$,  TVSAR$(p=1,\v P=[1, 1])_{[4,12]}$, TVSAR$(p=1,\v P=[2, 2])_{[4,12]}$ and TVSAR$(1,1)_{4 }$ - to data generated from the true TVSAR$(1,1)_{12}$ in Experiment 3.  Each experiment was replicated five times to assess the stability of the estimation procedure.

Figures S24 - S26 show that when fitting TVSAR$(3,3)_{12}$ to data from TVSAR$(1,1)_{12}$, the model consistently identifies the extra lags as irrelevant across the three different datasets considered. 

Figure S27 provides a closer look at fitting TVSAR(4,4) for a single dataset. Here the model correctly shrinks three redundant lags to zero; however, when the number of unnecessary lags increases, the MCMC sampler exhibits slower convergence, indicating some loss of stability, particularly at the fourth lag. 

Figures S29 - S30 illustrate fits of models with extra seasonal lags with period four, TVSAR$(p=1,\v P=[1, 1])_{[4,12]}$ and TVSAR$(p=1,\v P=[2, 2])_{[4,12]}$, respectively. In both cases, the posteriors successfully drive the redundant coefficients to zero. 

Finally, Figure S30 demonstrates a seasonally misspecifed fit, where TVSAR$(1,1)_4$ is applied to data from TVSAR$(1,1)_12$. 

Overall, the model performs well in detecting and shrinking redundant lags but shows some limitations when the degree of over-specification becomes large. Poor  mixing of the dynamic shrinkage priors may contribute to this behavior.

\begin{figure}
\label{p:redundant_1}
 \centering
 \includegraphics[width=0.3\textwidth]{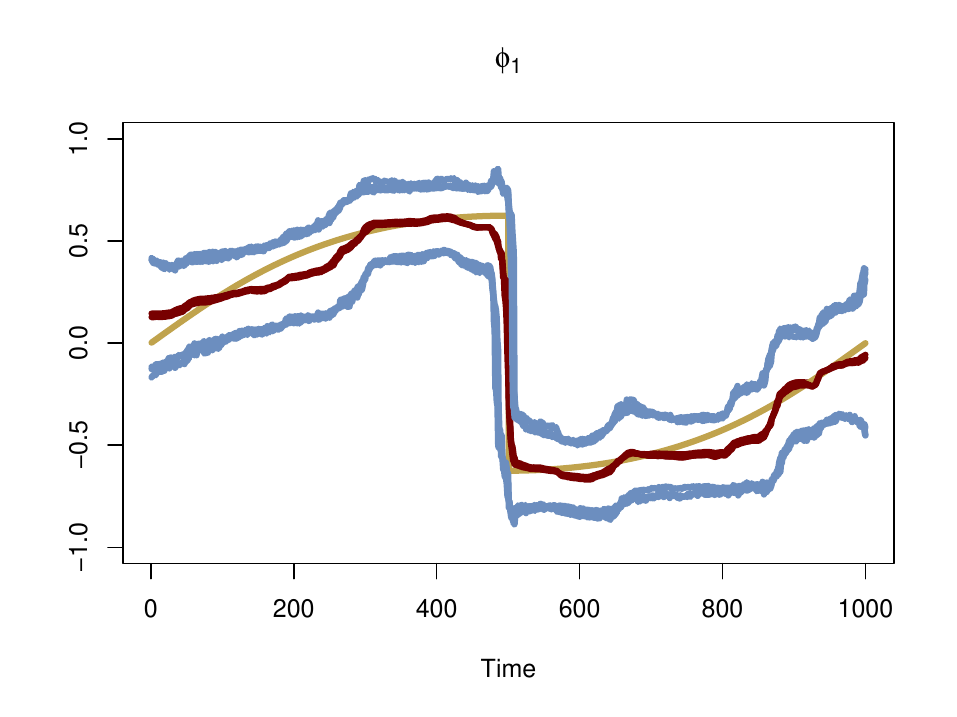}
  \includegraphics[width=0.3\textwidth]{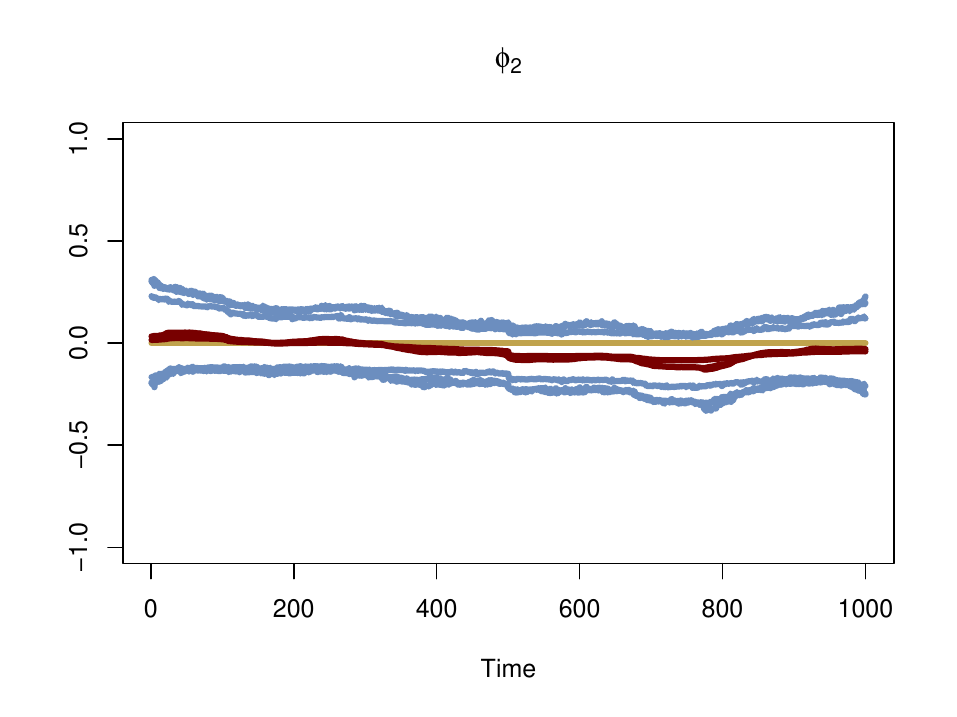}
 \includegraphics[width=0.3\textwidth]{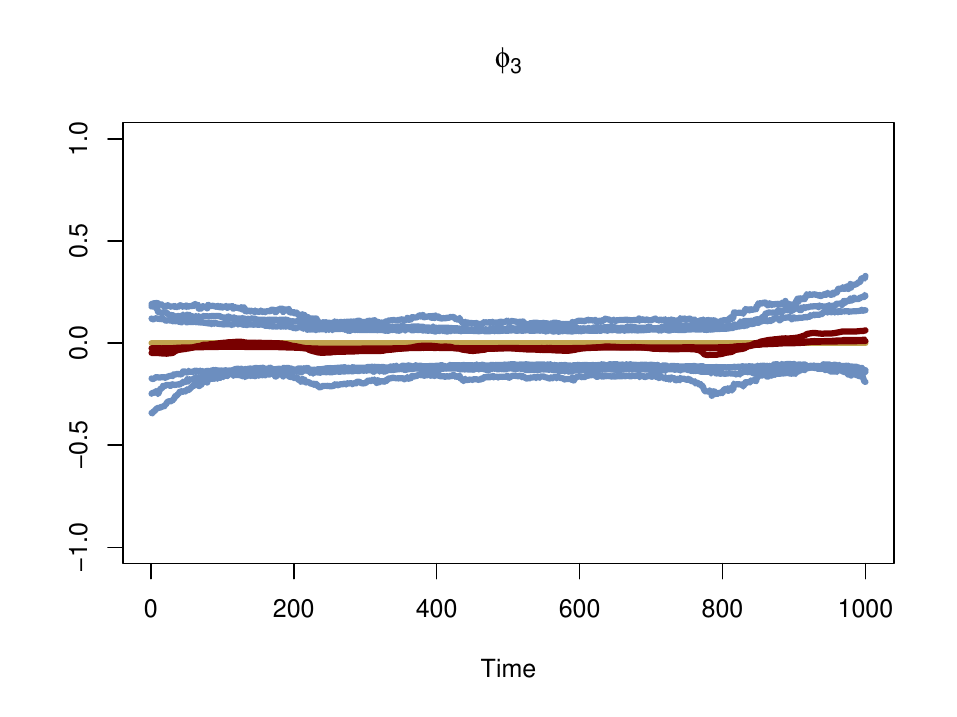}
 \includegraphics[width=0.3\textwidth]{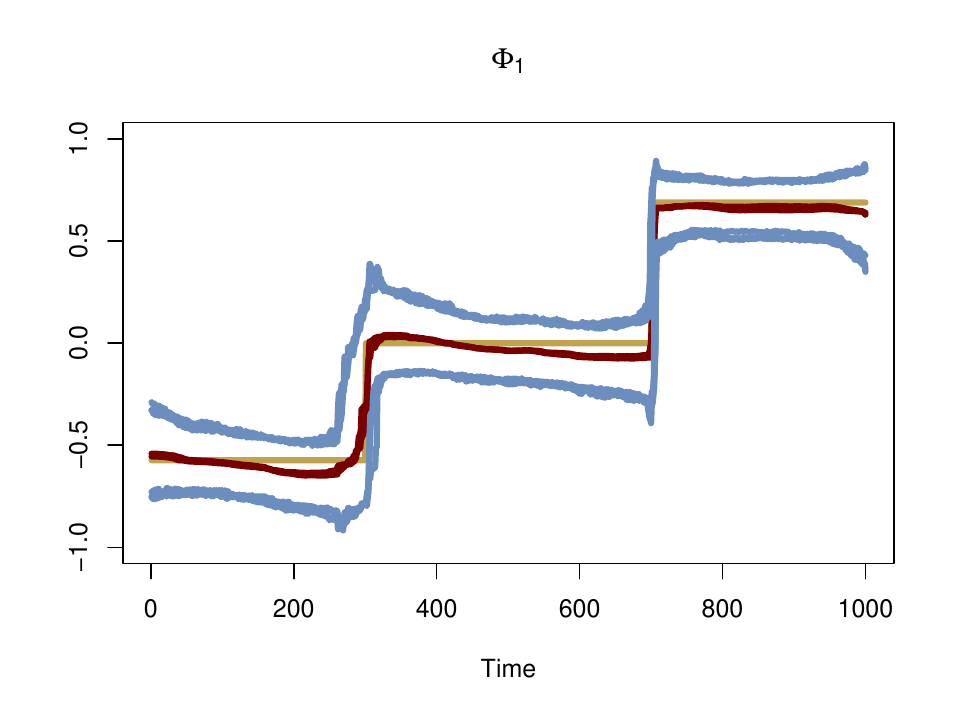}
\includegraphics[width=0.3\textwidth]{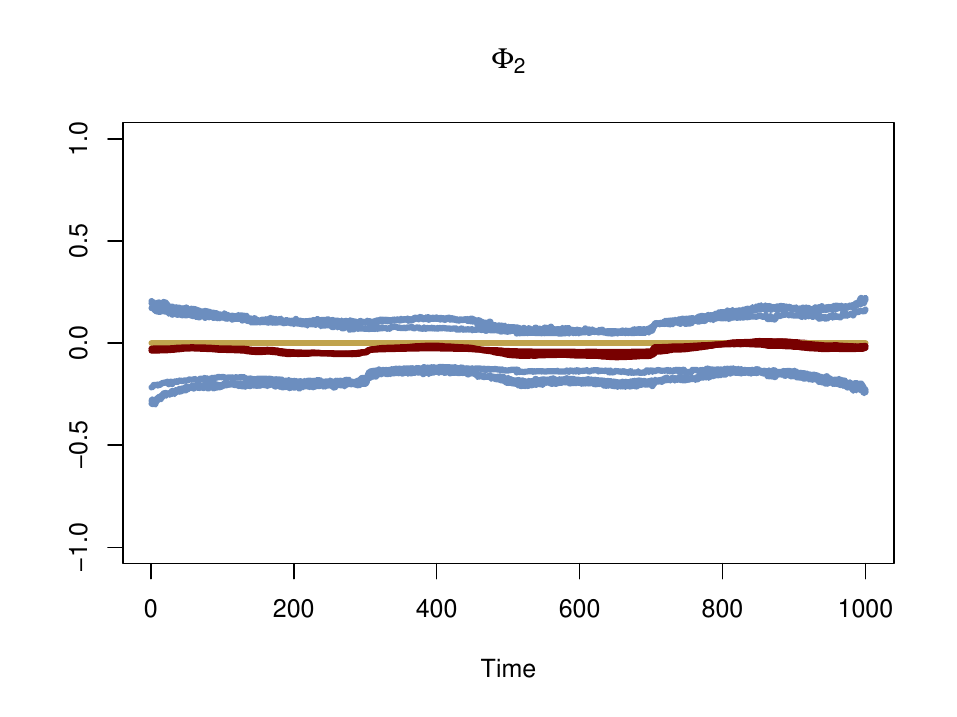}
 \includegraphics[width=0.3\textwidth]{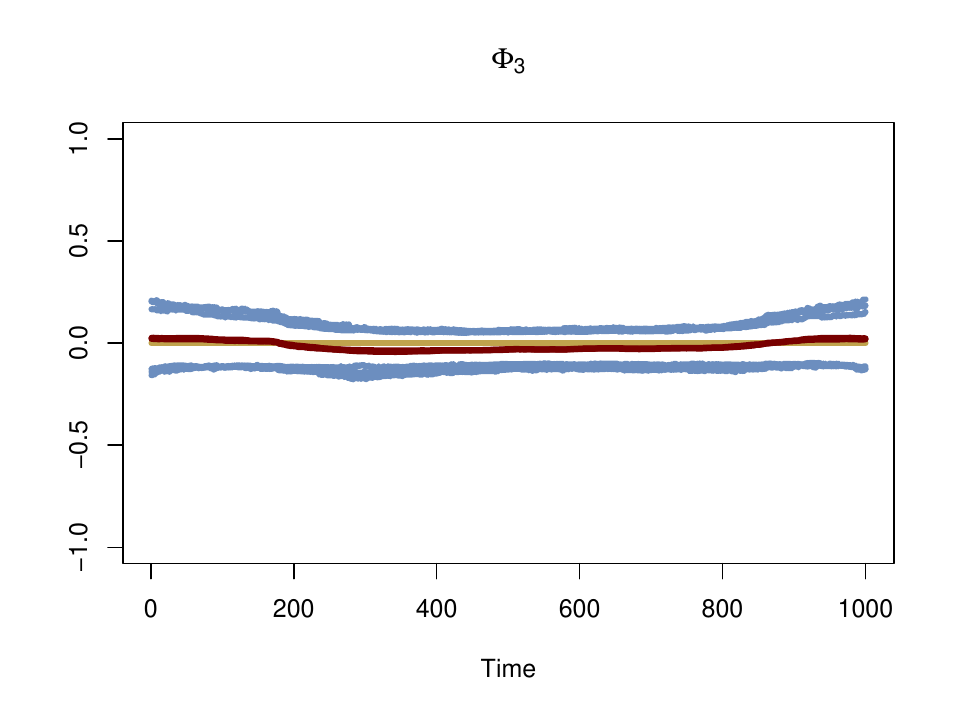}

\caption{Fitting TVSAR$(3,3)_{12}$ model to data generated from a TVSAR$(1,1)_{12}$.  MCMC convergence is assessed by re-estimating the model using three different seeds for the same dataset. The red and blue lines are posterior medians and $95\%$ HDIs over time for each of the three repeated
runs. The beige line is the true parameter evolution.} \label{fig:AR1_ev1}
\end{figure}

\begin{figure}
\label{p:redundant_2}
 \centering
 \includegraphics[width=0.3\textwidth]{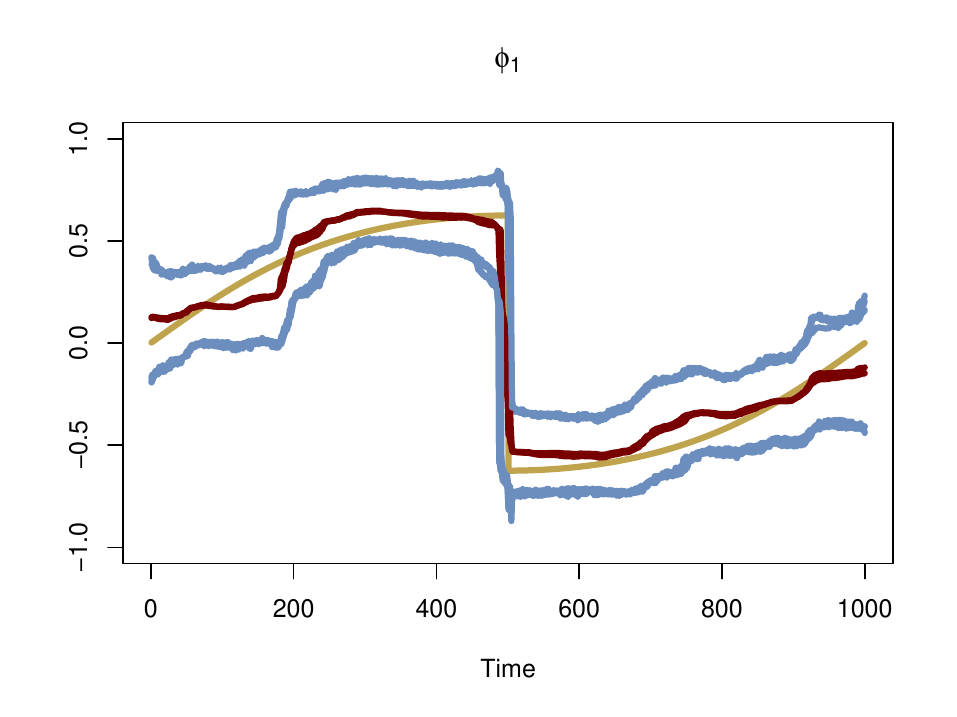}
  \includegraphics[width=0.3\textwidth]{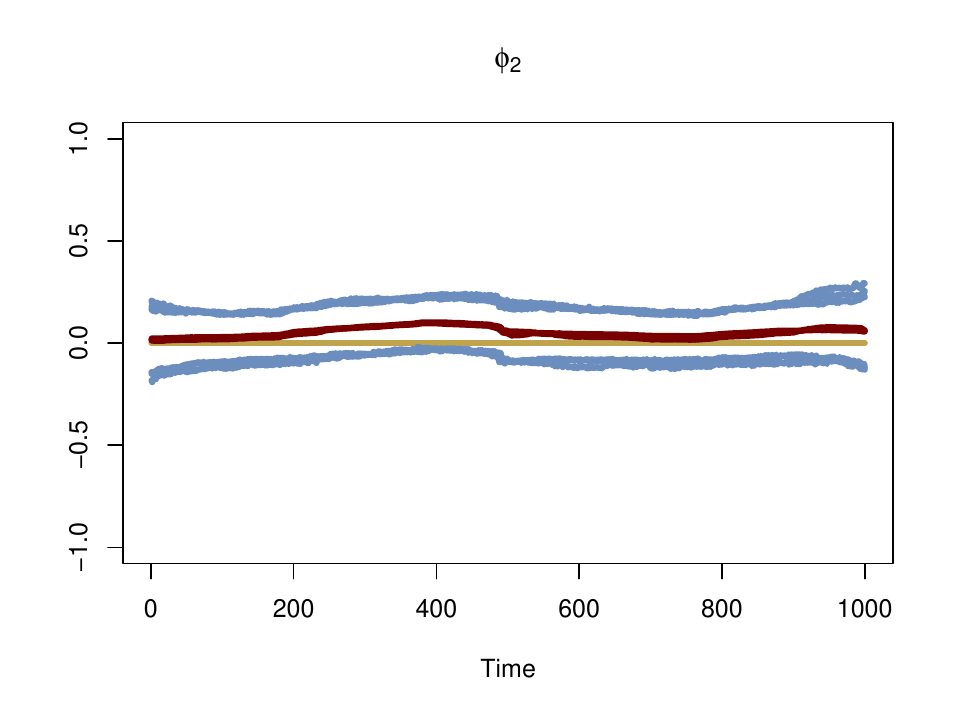}
 \includegraphics[width=0.3\textwidth]{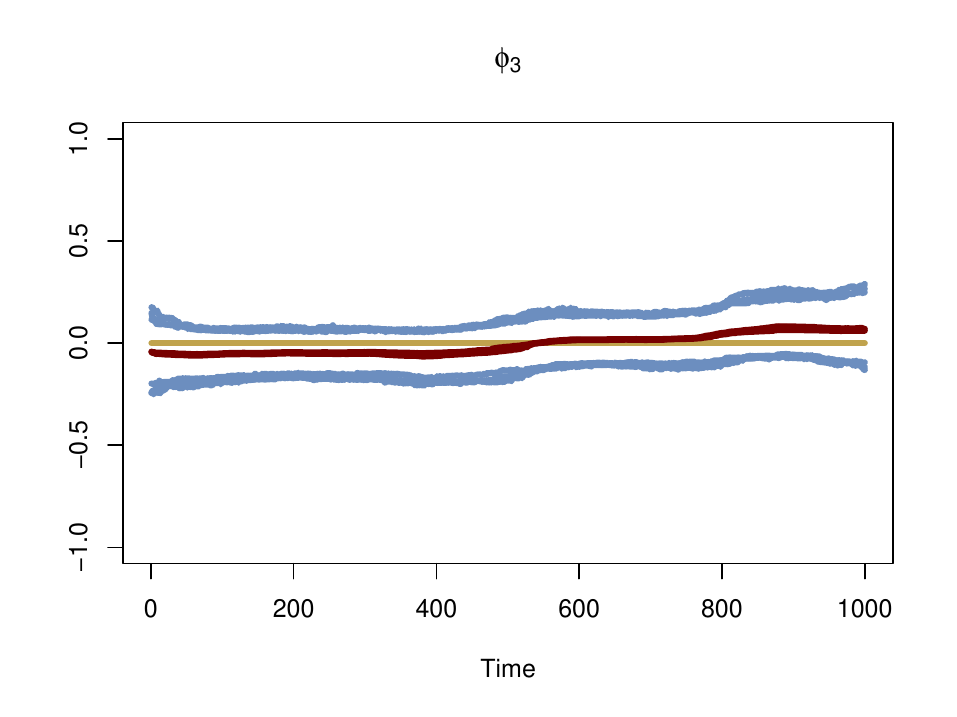}
 \includegraphics[width=0.3\textwidth]{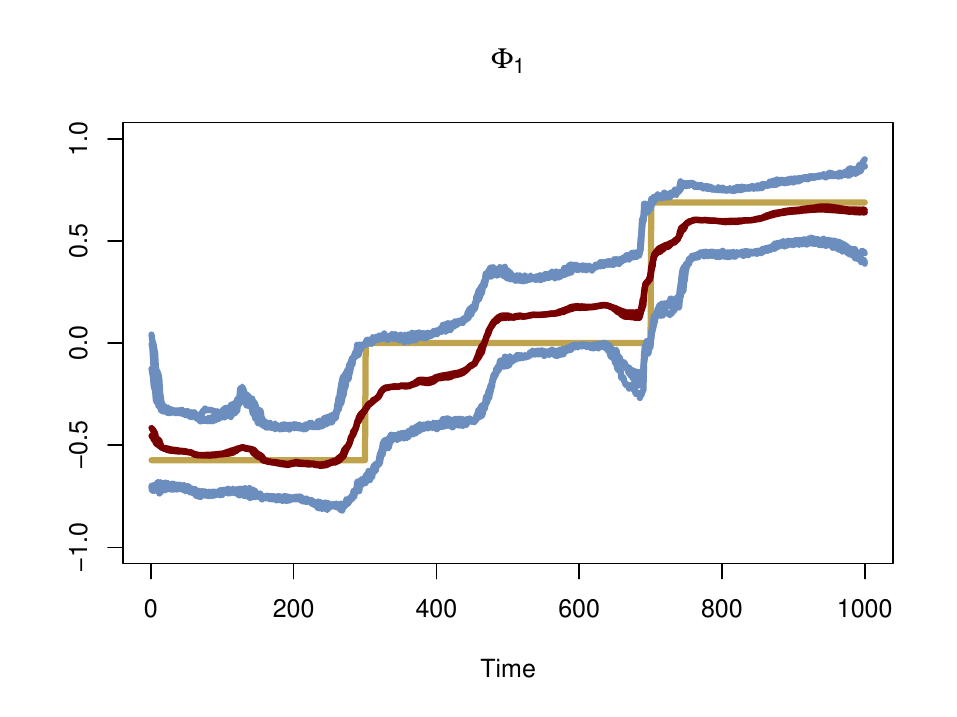}
\includegraphics[width=0.3\textwidth]{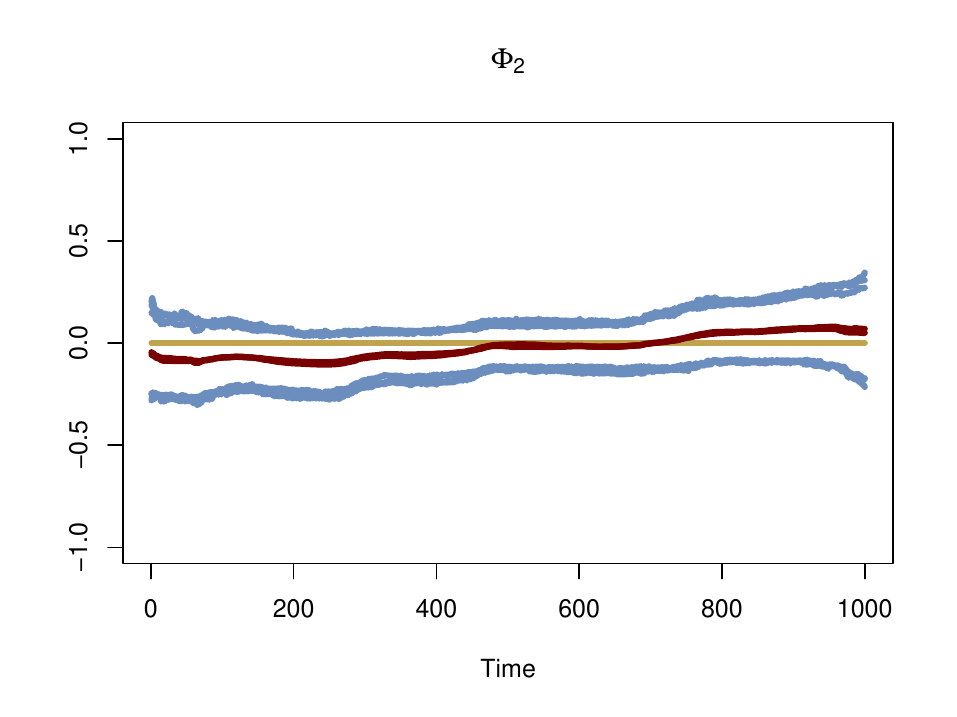}
 \includegraphics[width=0.3\textwidth]{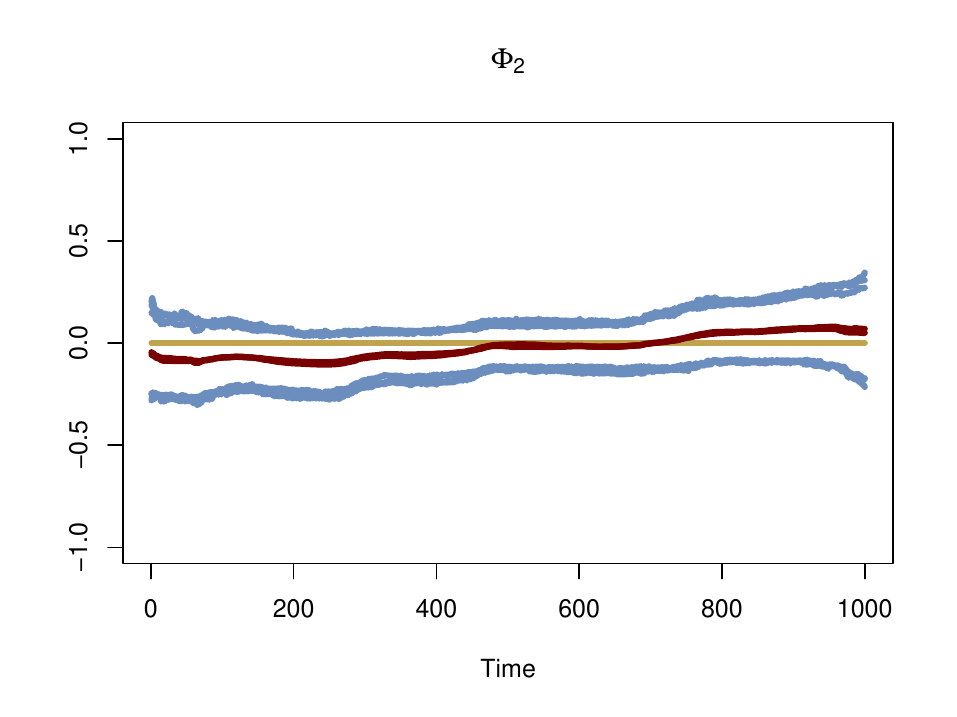}

\caption{Fitting TVSAR$(3,3)_{12}$ model to data generated from a TVSAR$(1,1)_{12}$.   MCMC convergence is assessed by re-estimating the model using three different seeds for the same dataset. The red and blue lines are posterior medians and $95\%$ HDIs over time for each of the three repeated
runs. The beige line is the true parameter evolution.} \label{fig:AR1_const}
\end{figure}

\begin{figure}
\label{p:redundant_1}
 \centering
  \includegraphics[width=0.3\textwidth]{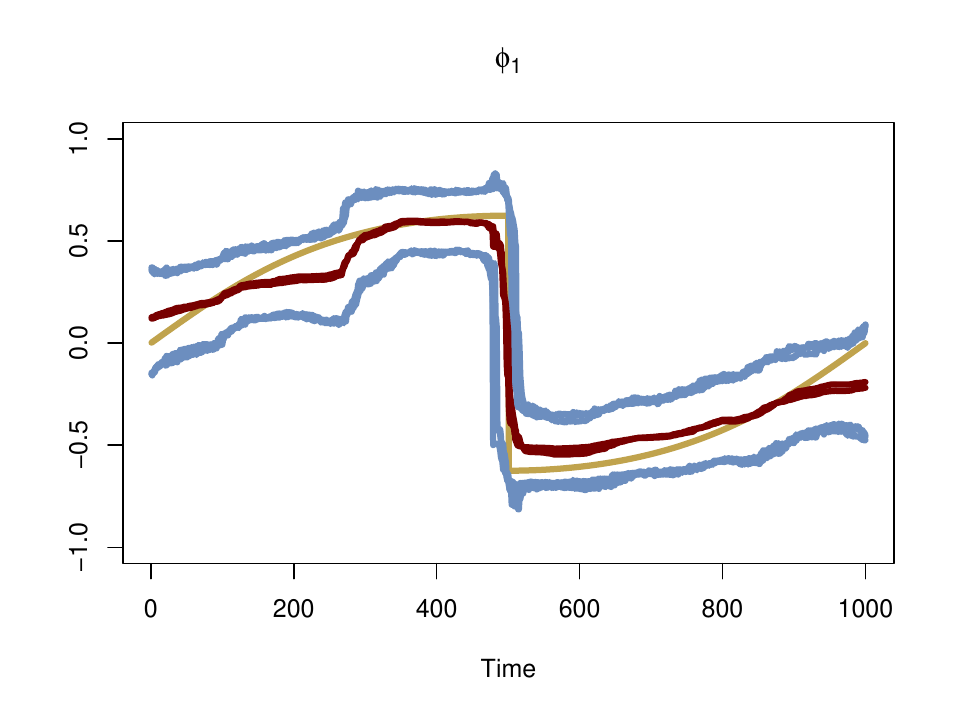}
  \includegraphics[width=0.3\textwidth]{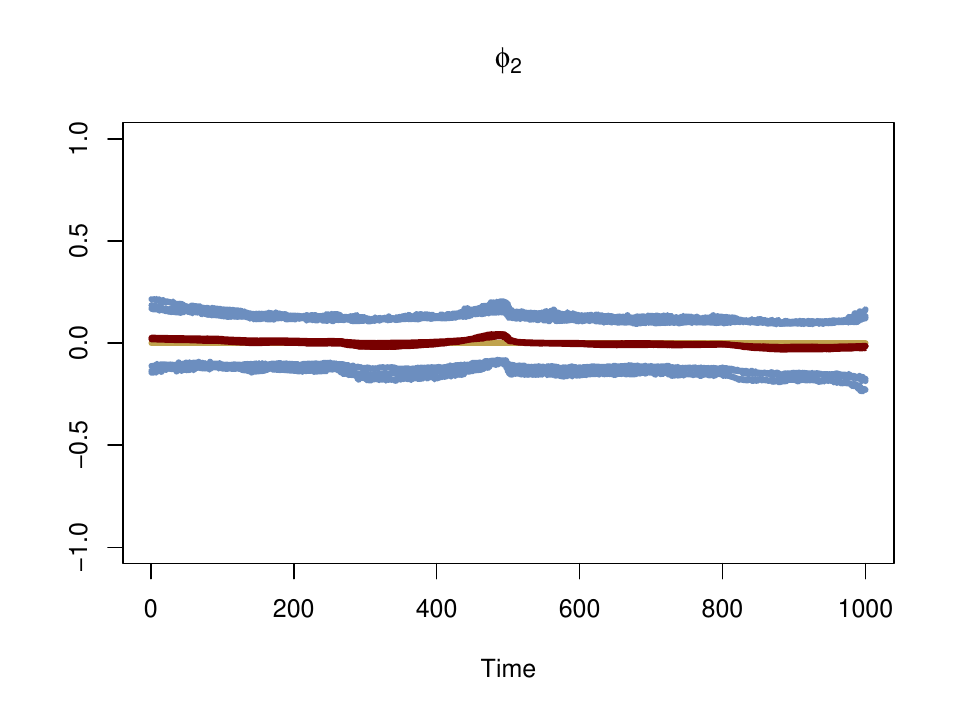}
 \includegraphics[width=0.3\textwidth]{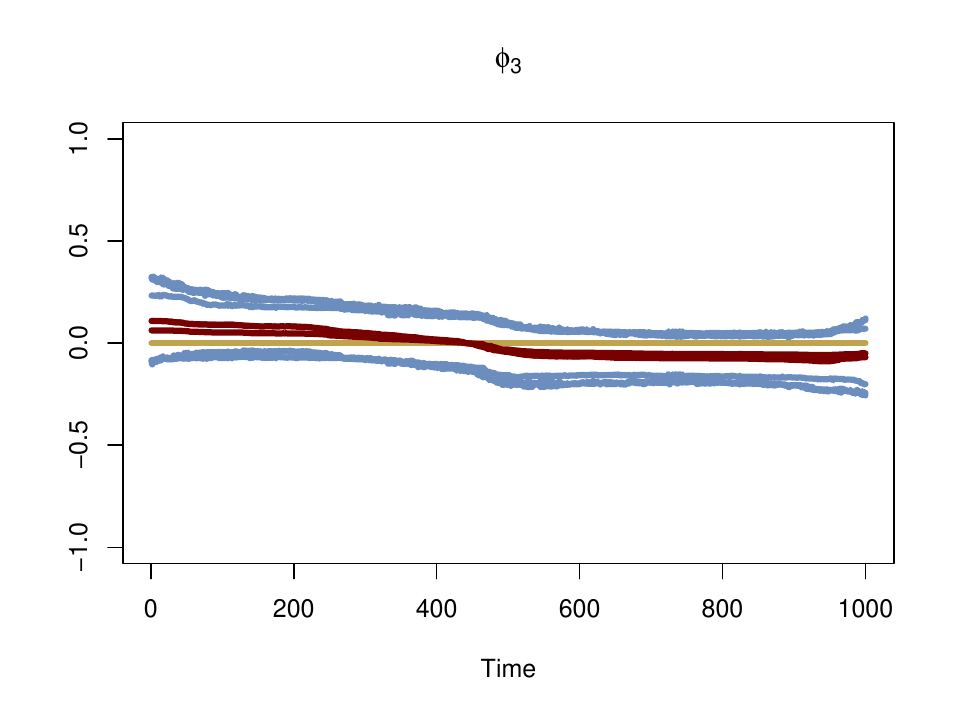}
 \includegraphics[width=0.3\textwidth]{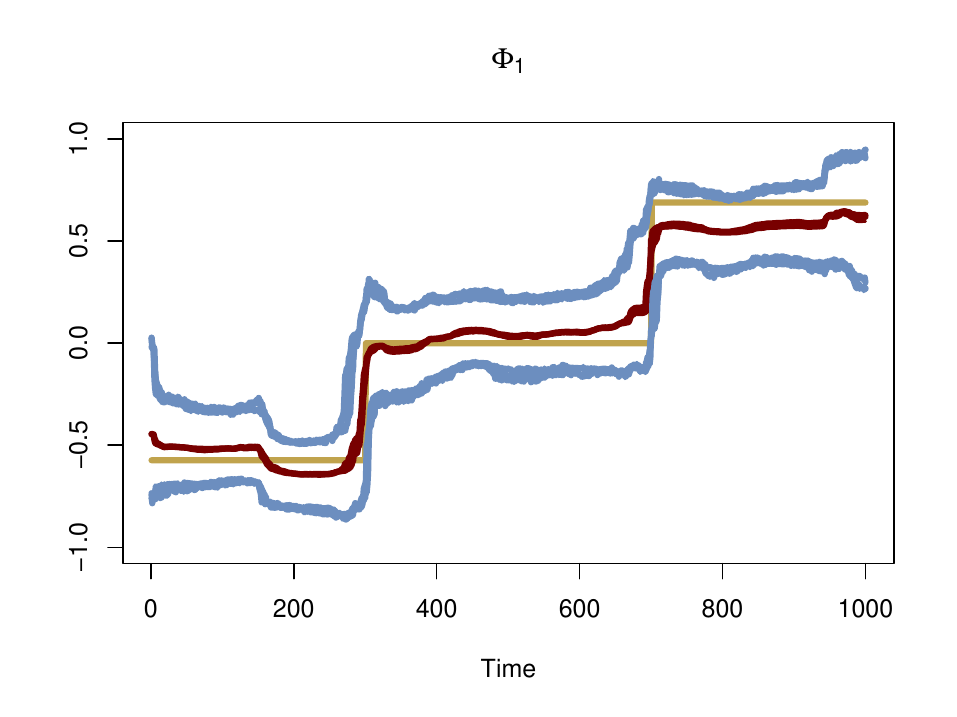}
\includegraphics[width=0.3\textwidth]{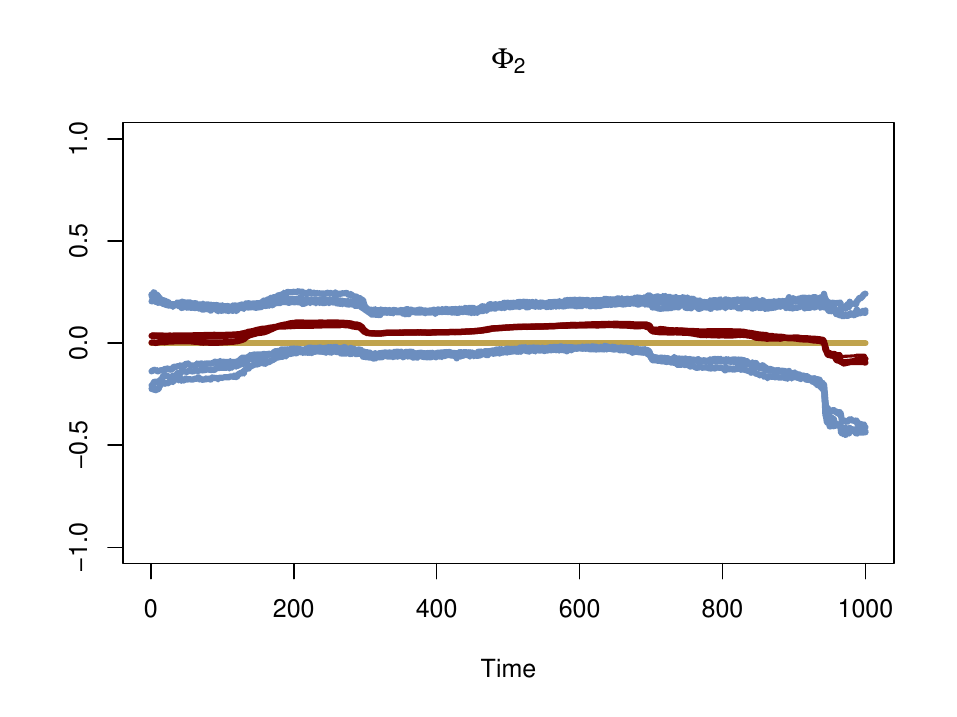}
 \includegraphics[width=0.3\textwidth]{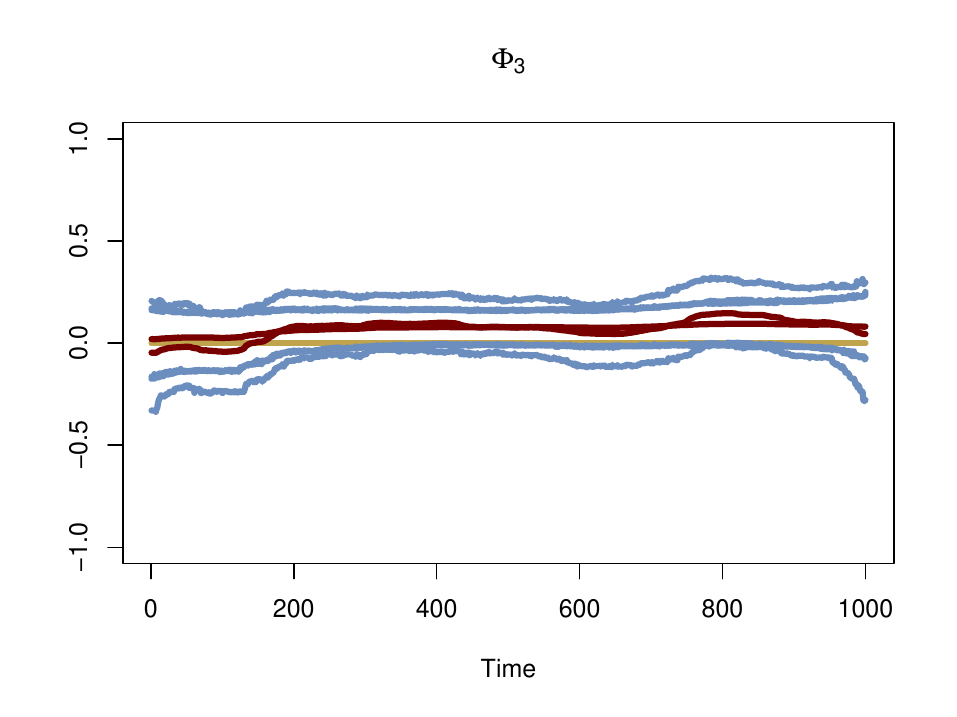}

\caption{Fitting TVSAR$(3,3)_{12}$ model to data generated from a TVSAR$(1,1)_{12}$.   MCMC convergence is assessed by re-estimating the model using three different seeds for the same dataset. The red and blue lines are posterior medians and $95\%$ HDIs over time for each of the three repeated
runs. The beige line is the true parameter evolution.} \label{fig:AR2_ev1}
\end{figure}

\begin{figure}
\label{p:redundant_2}
 \centering
  \includegraphics[width=0.22\textwidth]{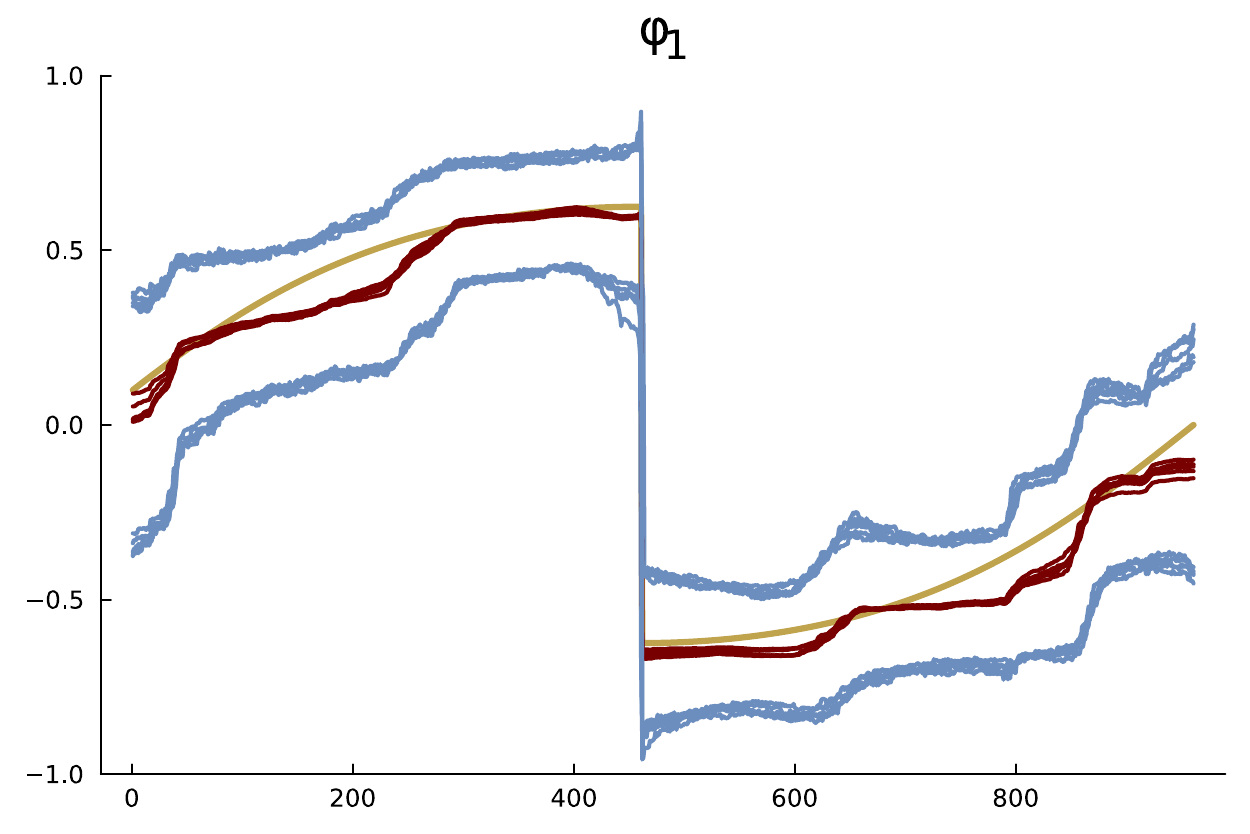}
  \includegraphics[width=0.22\textwidth]{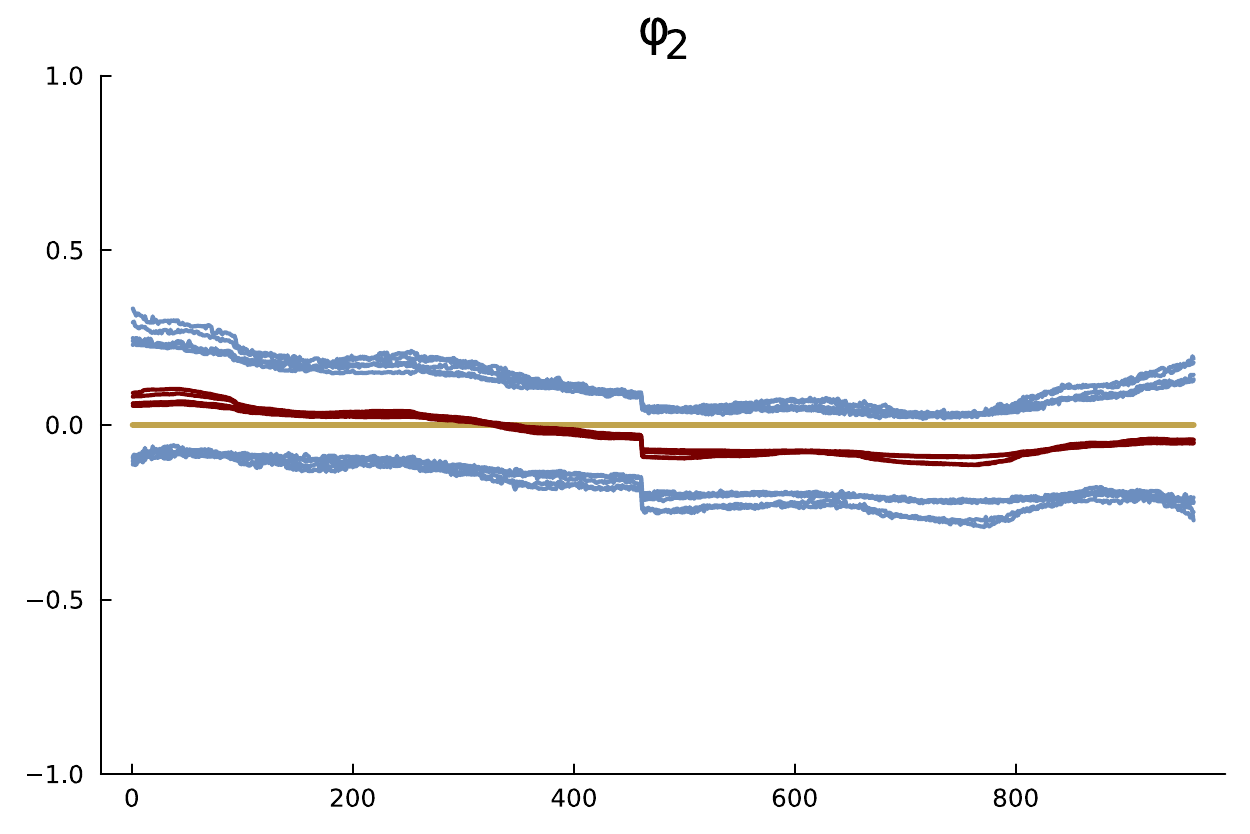}
 \includegraphics[width=0.22\textwidth]{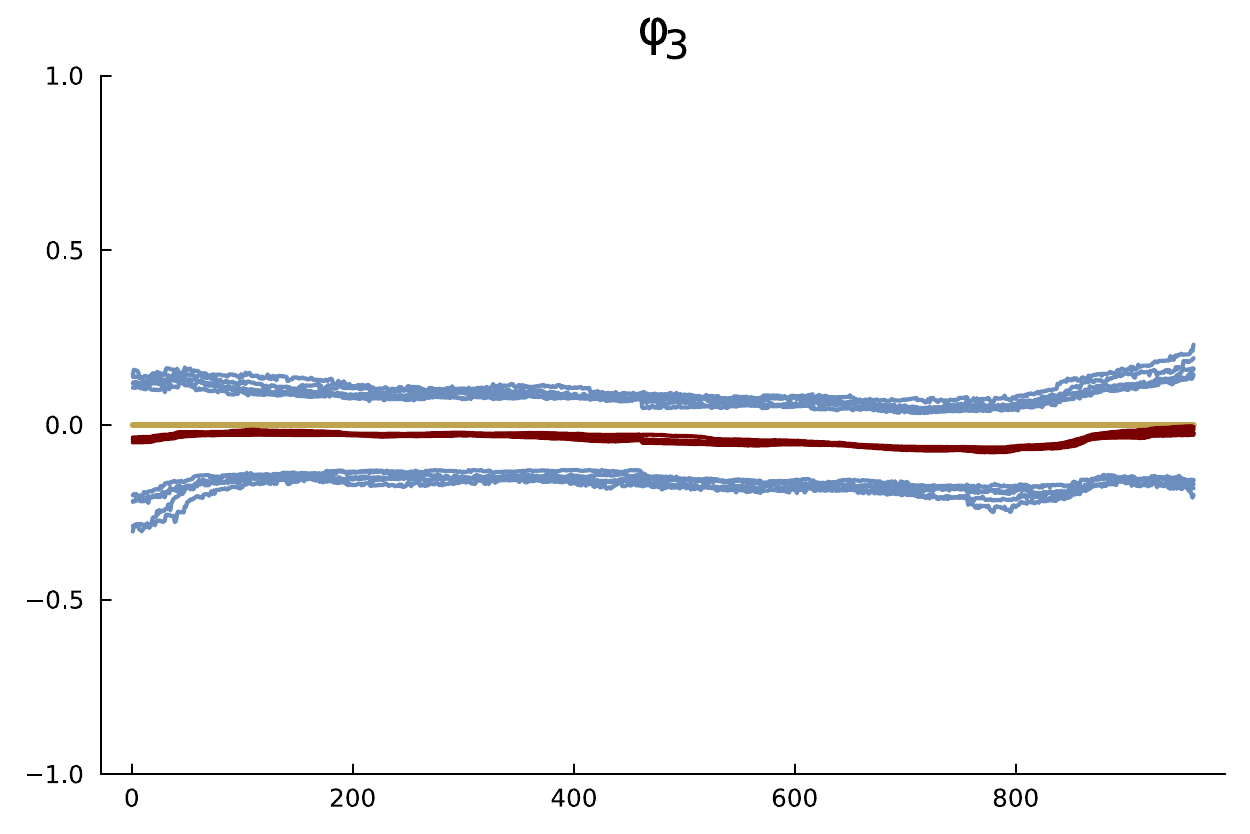}
 \includegraphics[width=0.22\textwidth]
{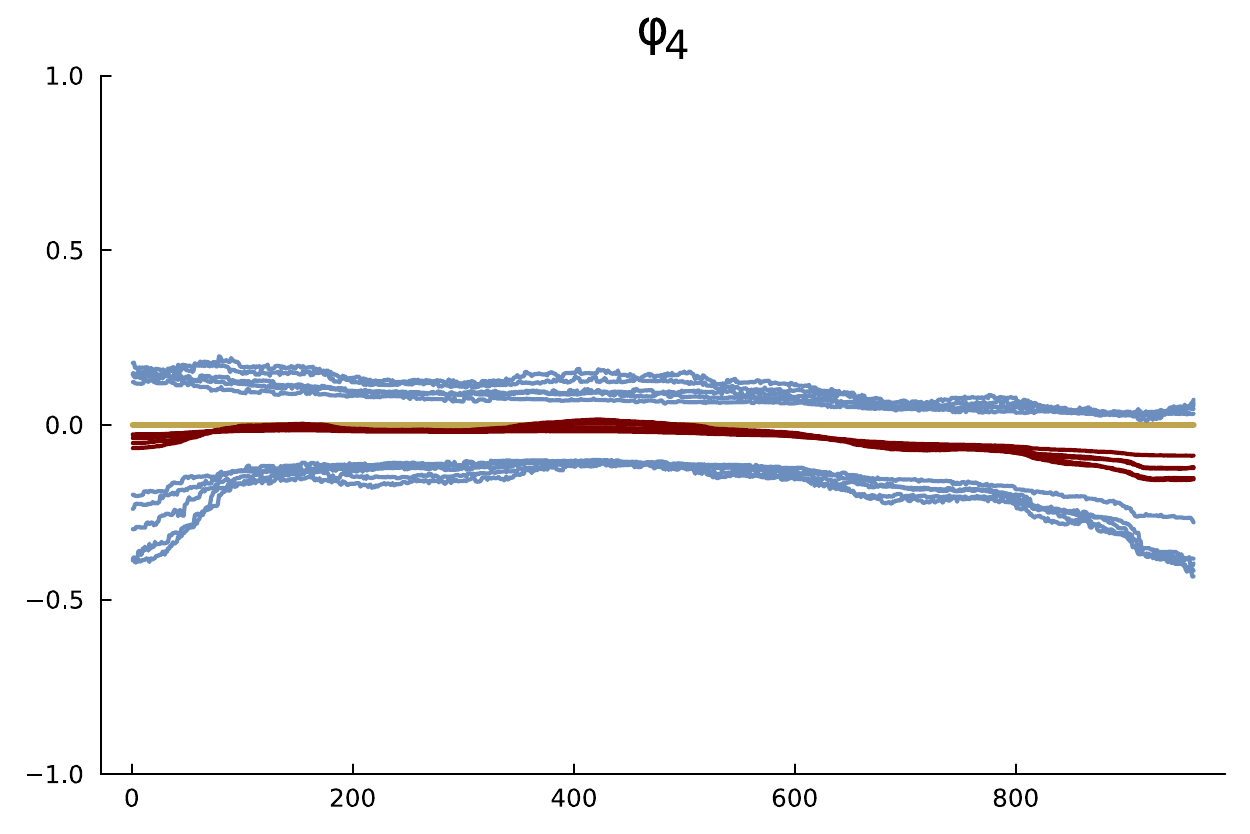}
\includegraphics[width=0.22\textwidth]{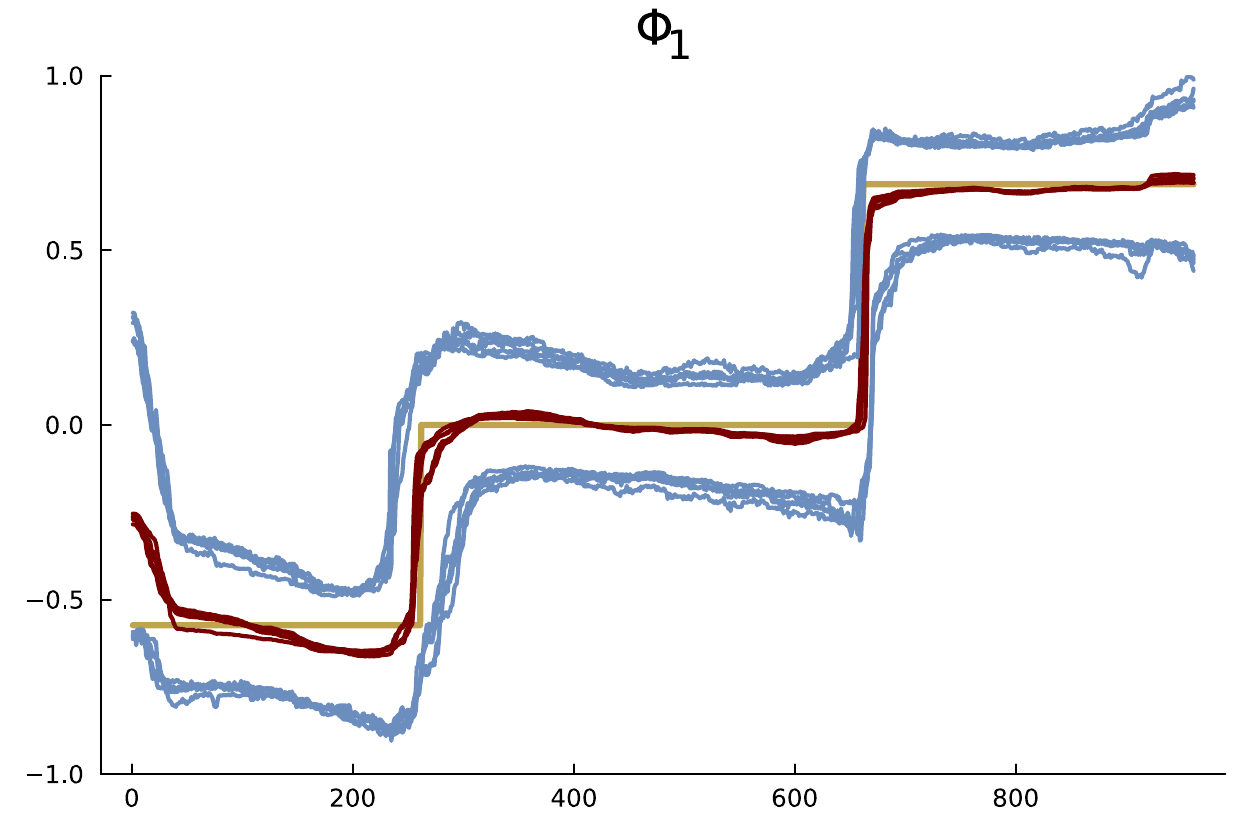}
 \includegraphics[width=0.22\textwidth]{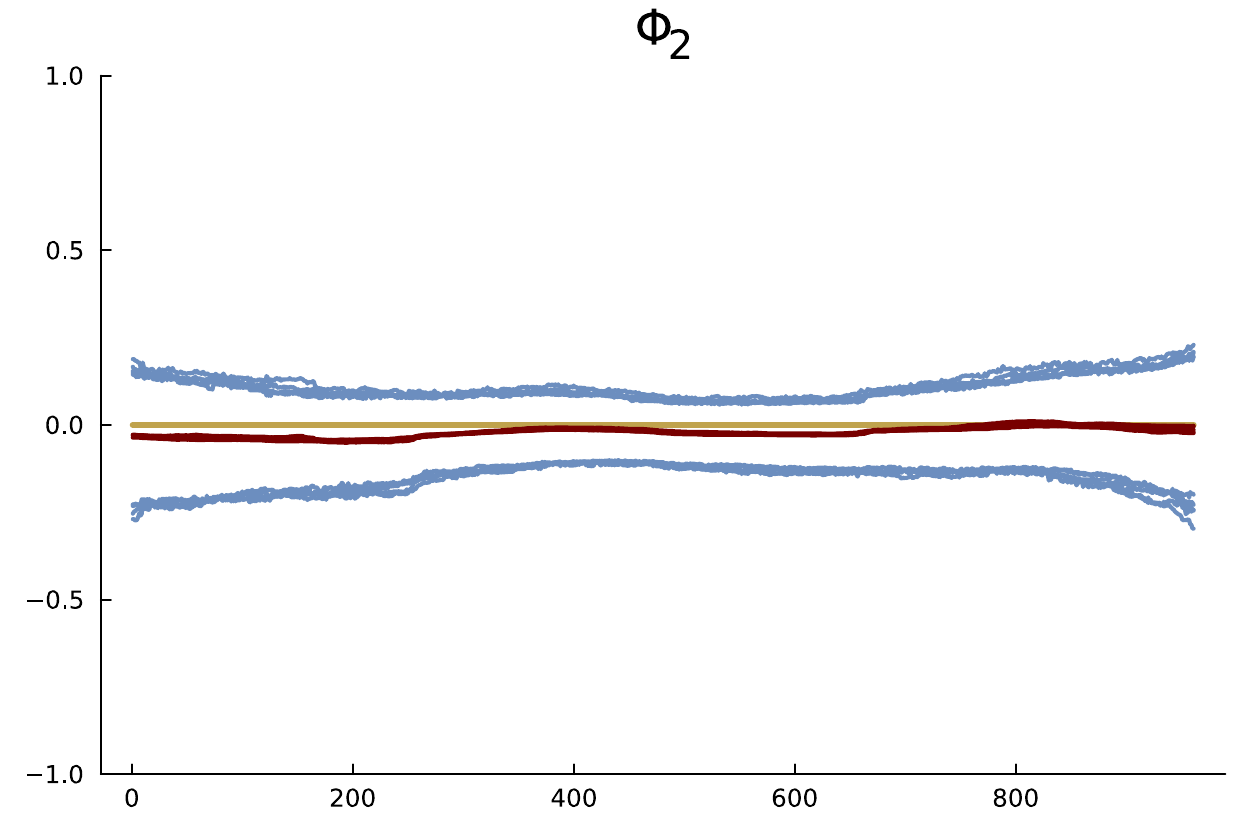}
 \includegraphics[width=0.22\textwidth]{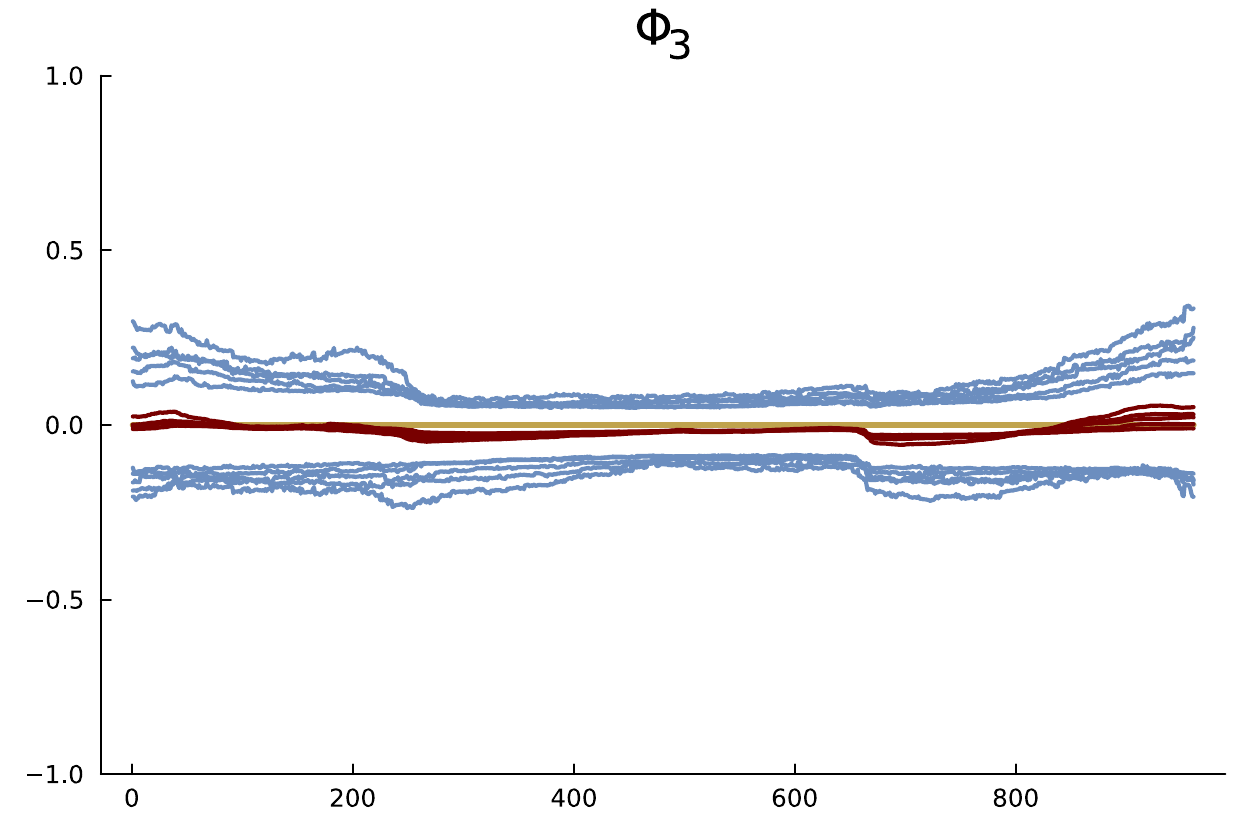}
 \includegraphics[width=0.22\textwidth]{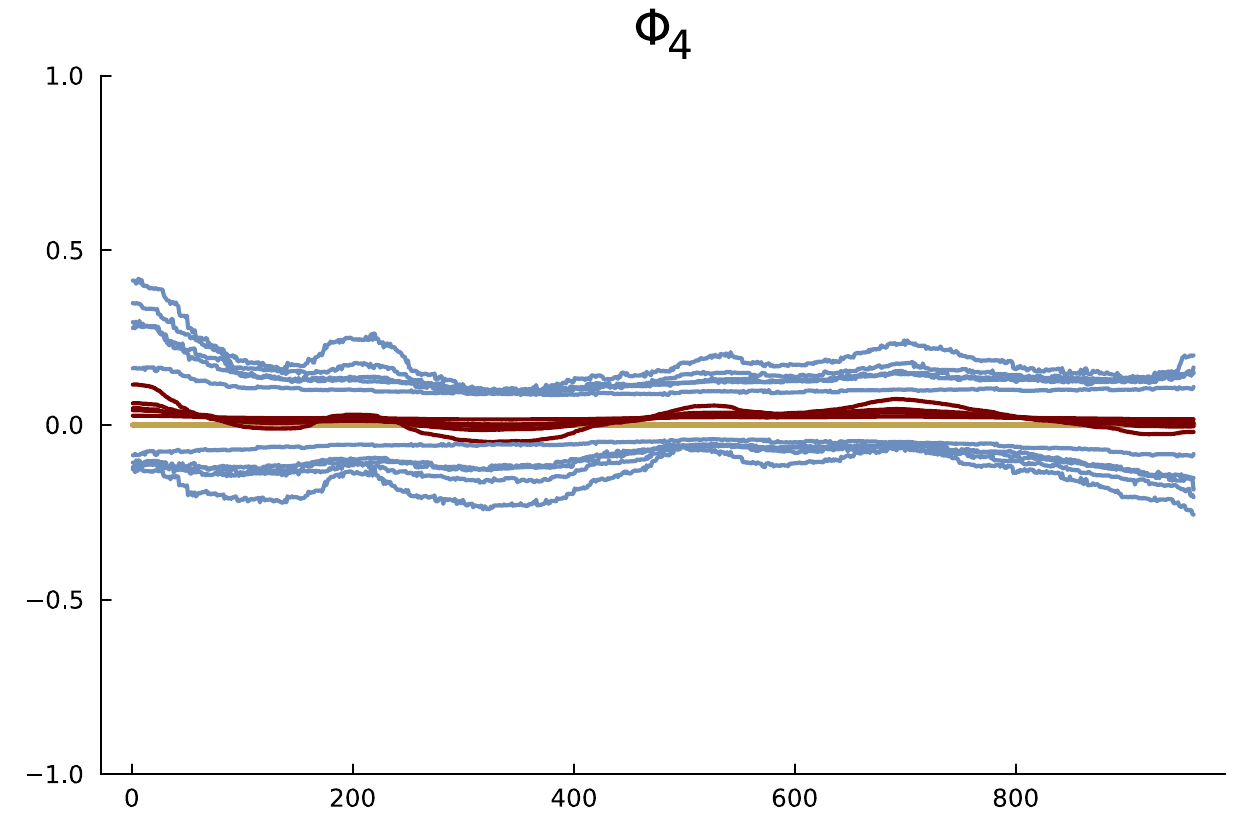}

\caption{Fitting TVSAR$(4,4)_{12}$ model to data generated from a TVSAR$(1,1)_{12}$.   MCMC convergence is assessed by re-estimating the model using three different seeds for the same dataset. The red and blue lines are posterior medians and $95\%$ HDIs over time for each of the three repeated
runs. The beige line is the true parameter evolution.} \label{fig:AR2_ev1}
\end{figure}

\begin{figure}
\label{p:redundant_3}
 \centering
  \includegraphics[width=0.3\textwidth]{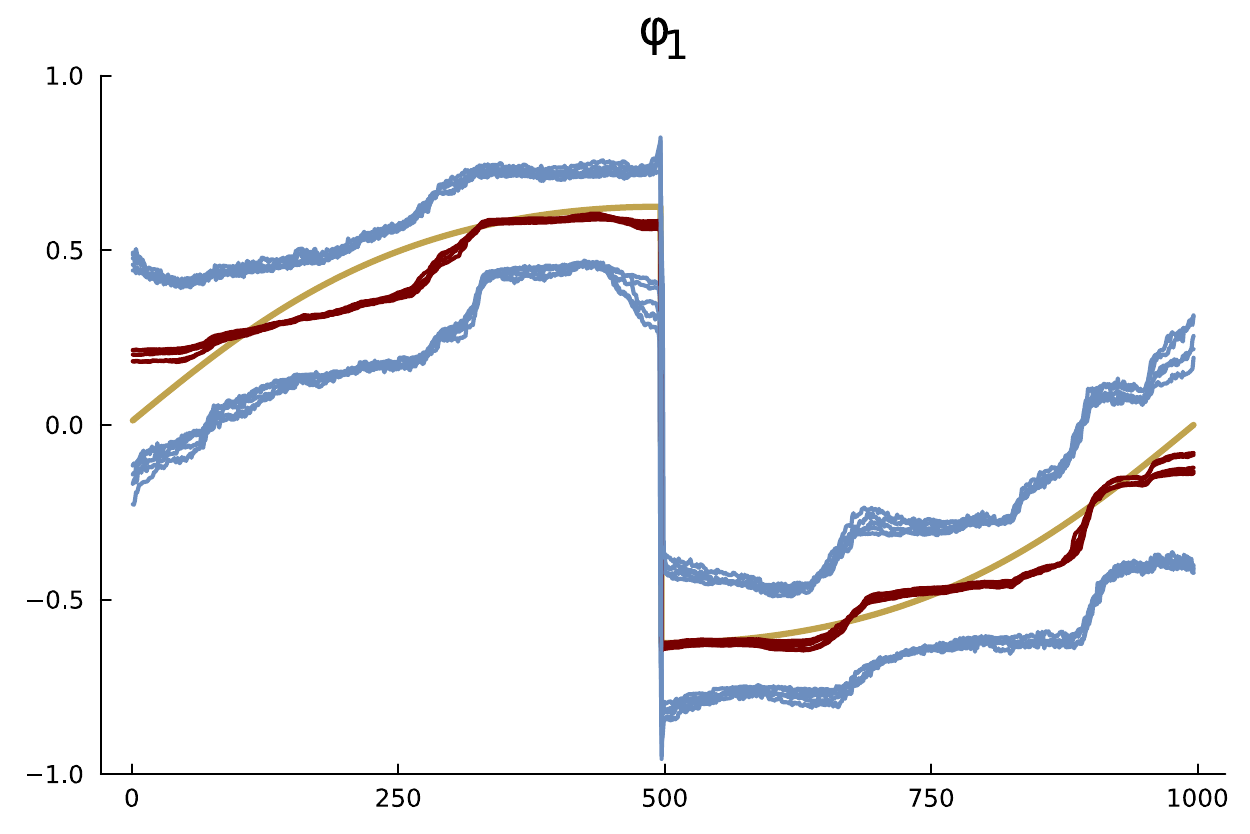}
  \includegraphics[width=0.3\textwidth]{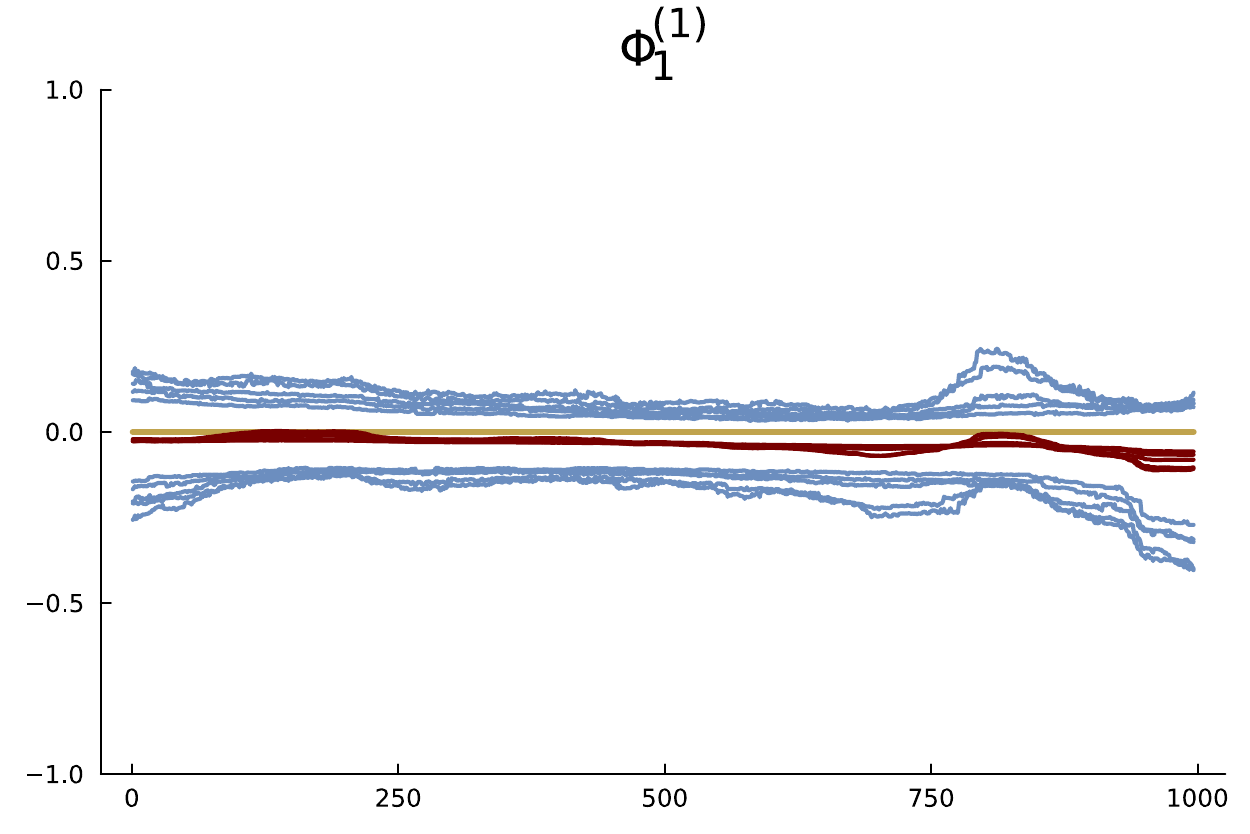}
 \includegraphics[width=0.3\textwidth]{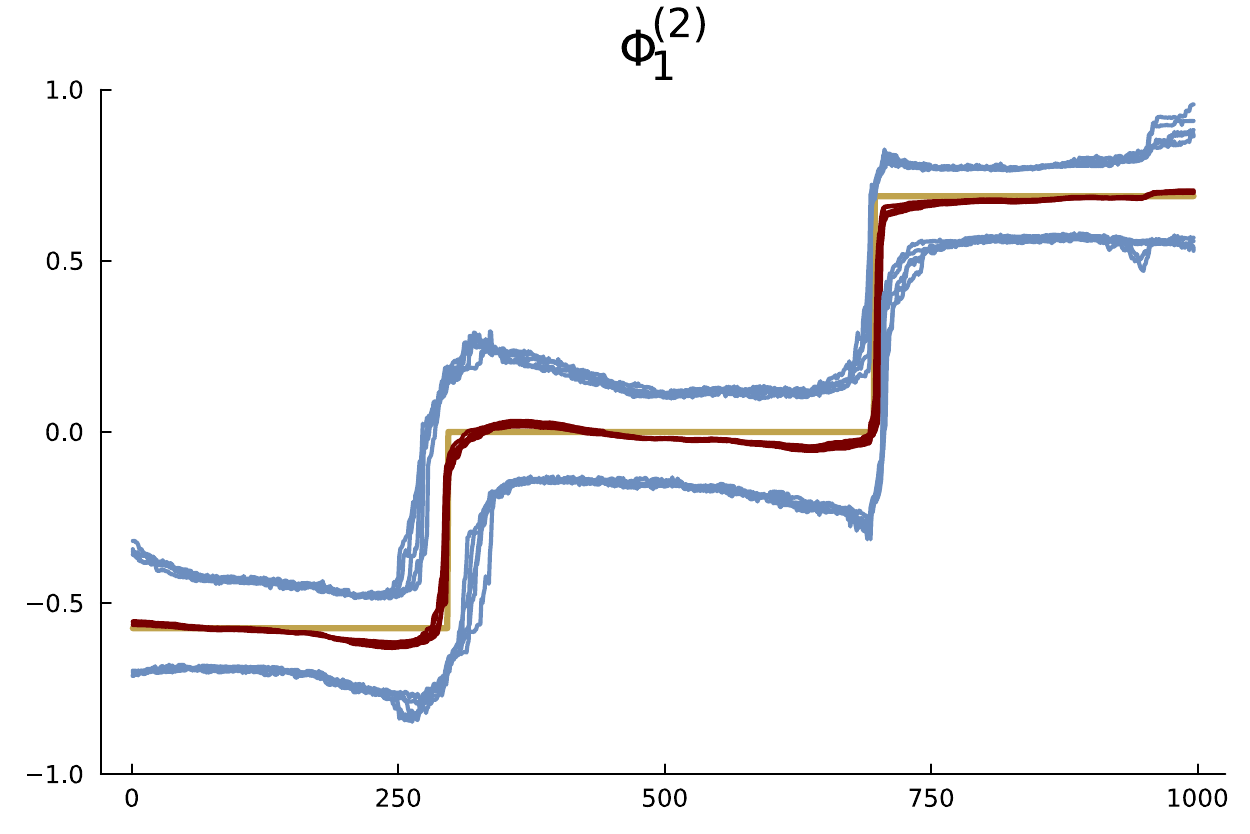}
 
\caption{Fitting TVSAR$(p=1,P=[1, 1])_{4,12}$ model to data generated from a TVSAR$(1,1)_{12}$.   MCMC convergence is assessed by re-estimating the model using three different seeds for the same dataset. The red and blue lines are posterior medians and $95\%$ HDIs over time for each of the three repeated
runs. The beige line is the true parameter evolution.} \label{fig:AR2_ev1}
\end{figure}

\begin{figure}
\label{p:redundant_4}
 \centering
  \includegraphics[width=0.3\textwidth]{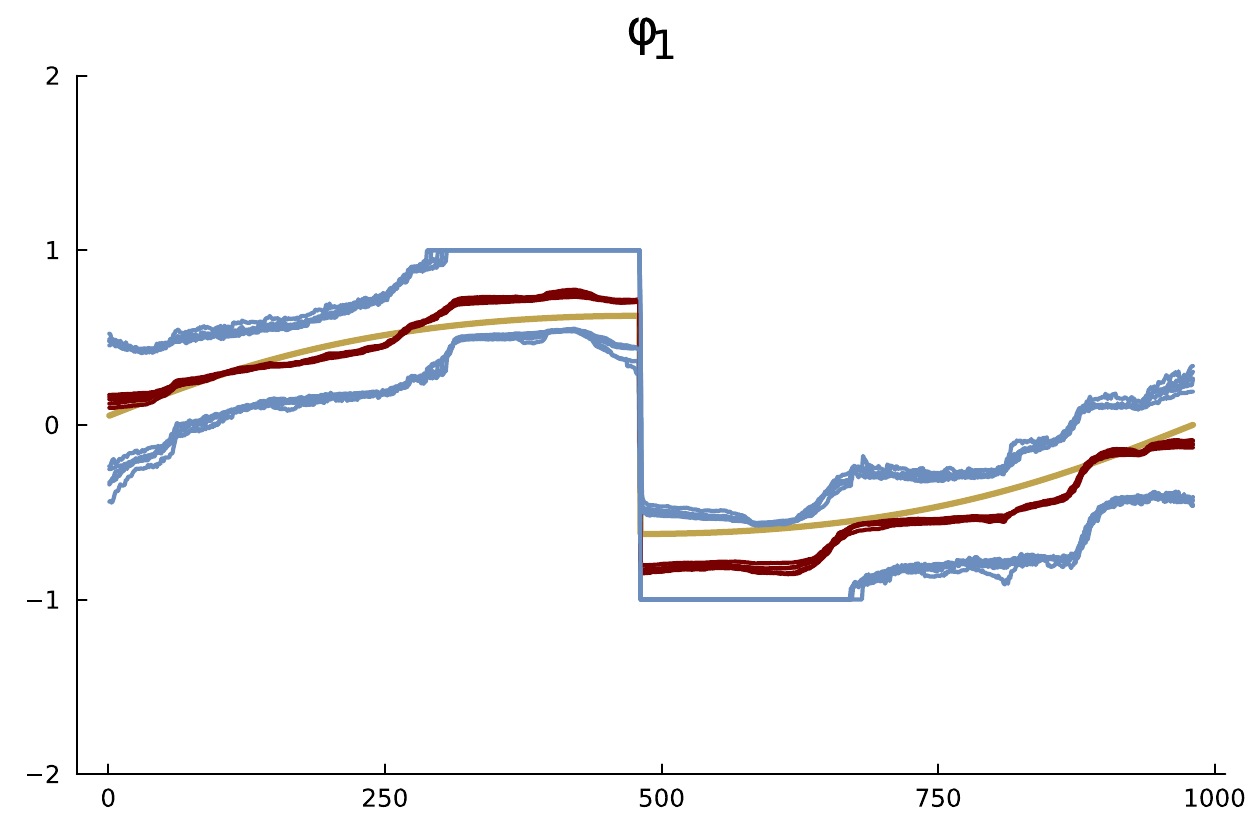}
  \includegraphics[width=0.3\textwidth]{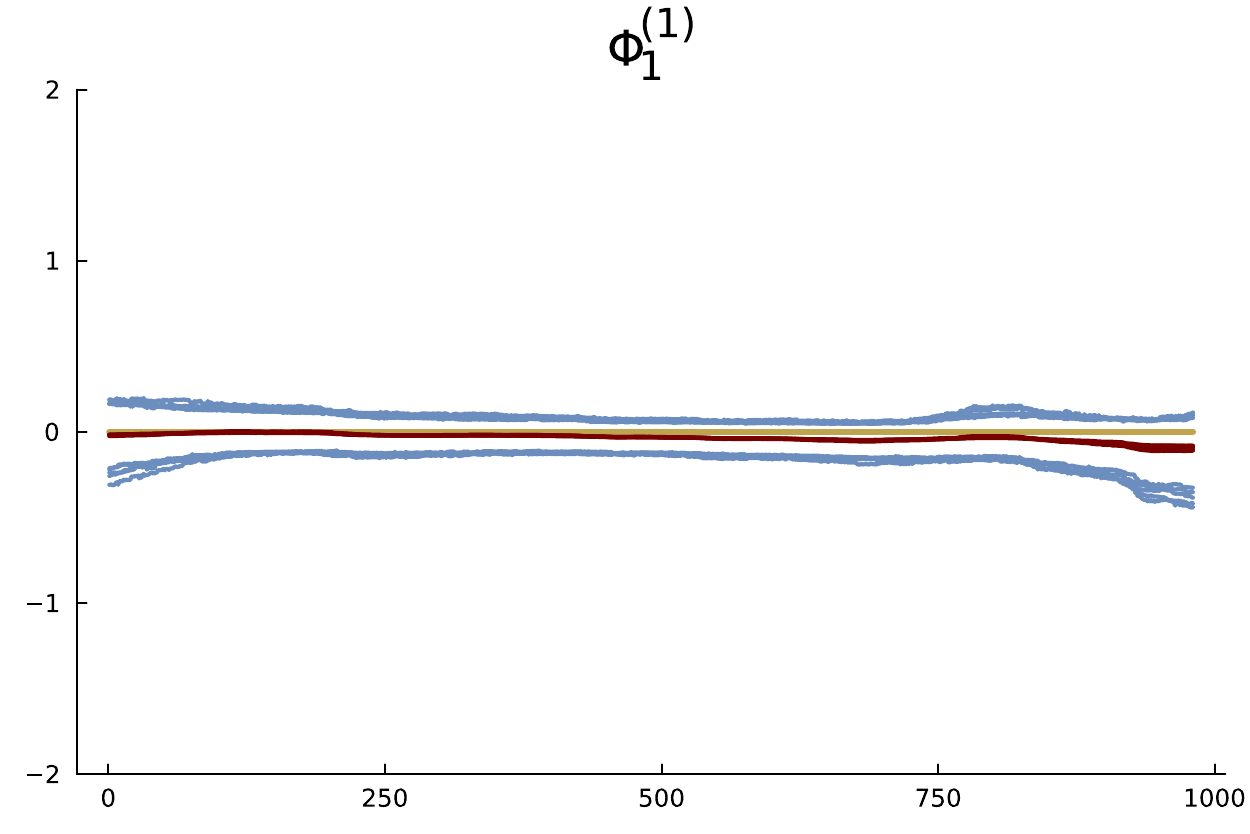}
\includegraphics[width=0.3\textwidth]{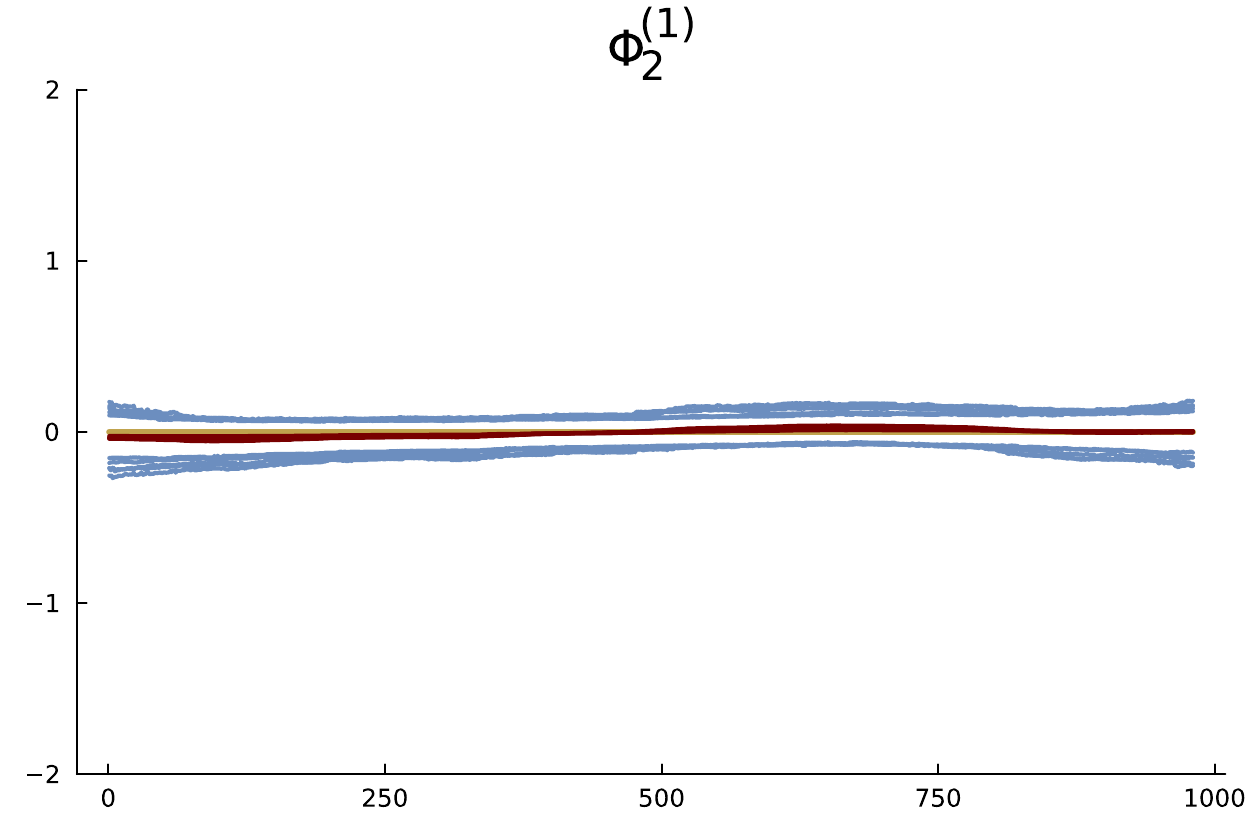}
 \includegraphics[width=0.3\textwidth]{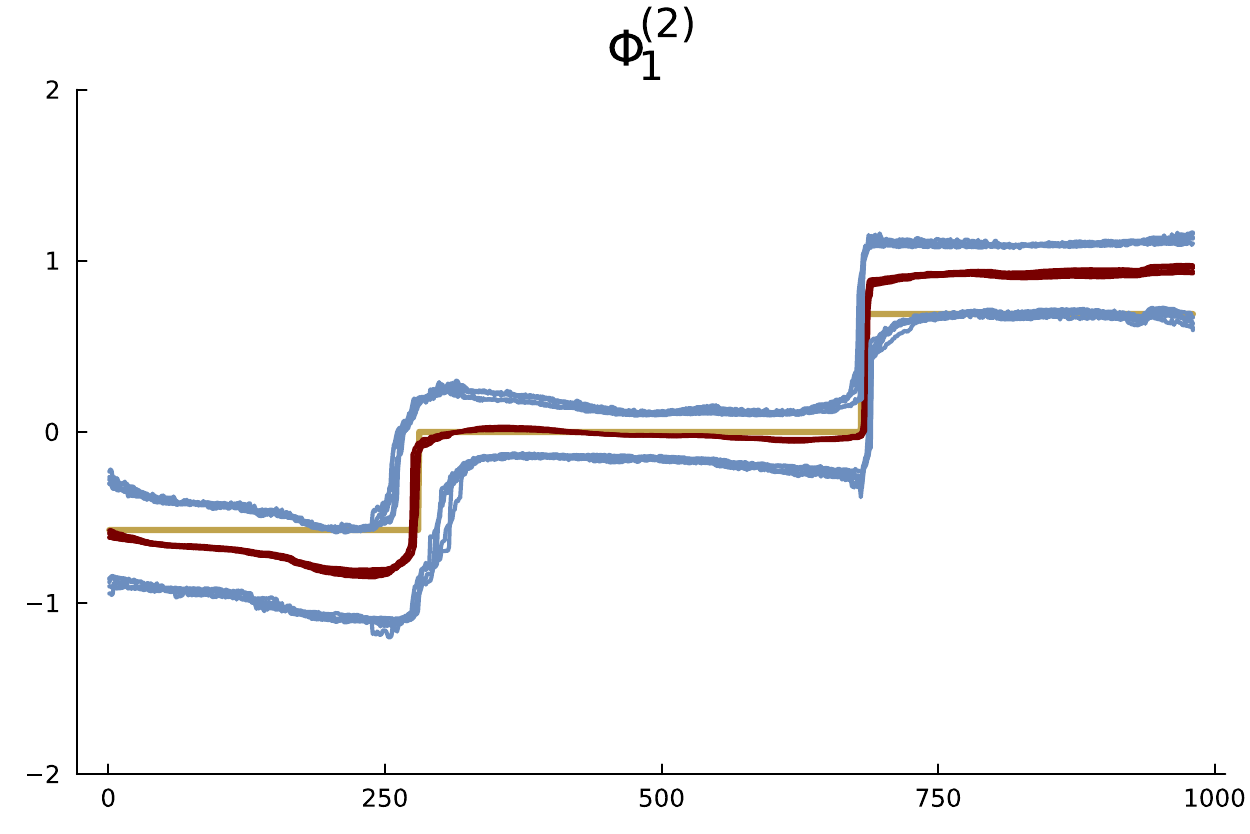}
  \includegraphics[width=0.3\textwidth]{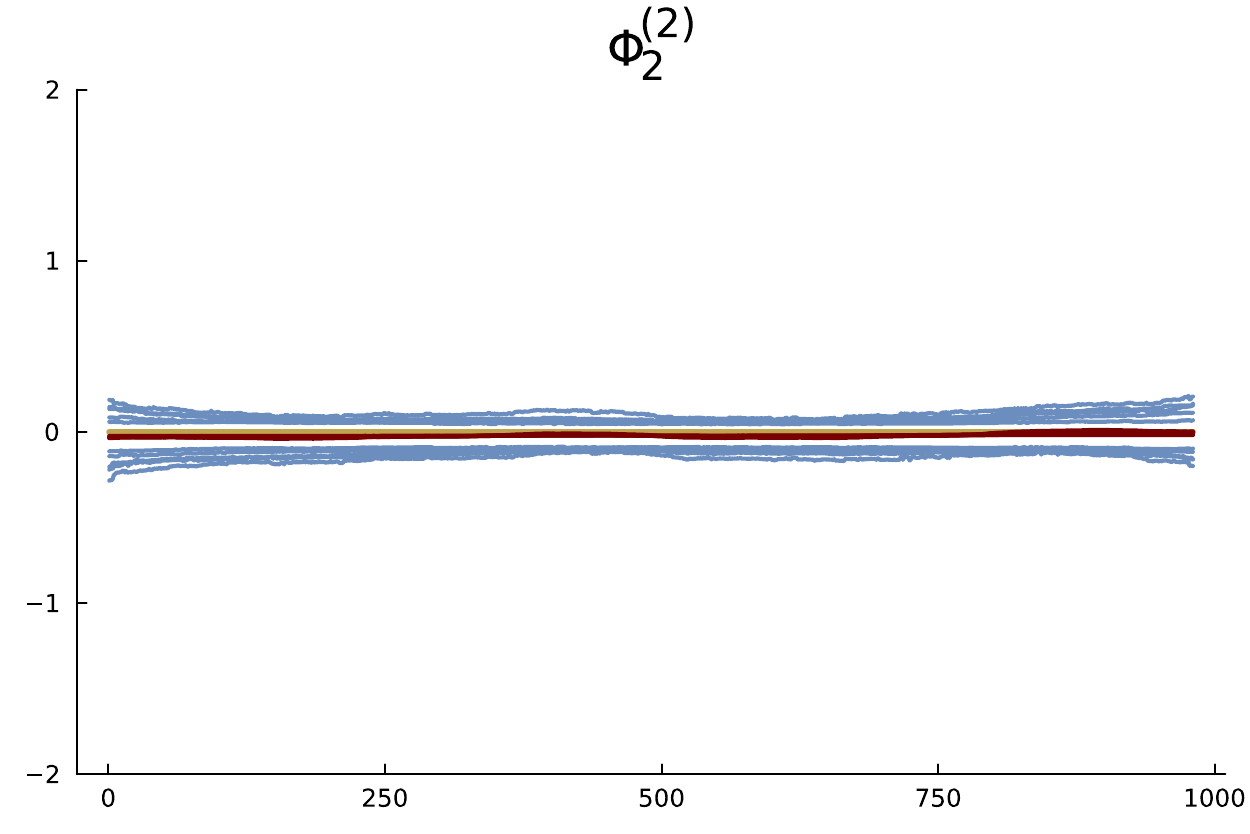}
 
\caption{Fitting TVSAR$(p=1,P=[2, 2])_{4,12}$  model to data generated from a TVSAR$(1,1)_{12}$.   MCMC convergence is assessed by re-estimating the model using three different seeds for the same dataset. The red and blue lines are posterior medians and $95\%$ HDIs over time for each of the three repeated
runs. The beige line is the true parameter evolution.} \label{fig:AR2_ev1}
\end{figure}

\begin{figure}
\label{p:redundant_5}
 \centering
  \includegraphics[width=0.45\textwidth]{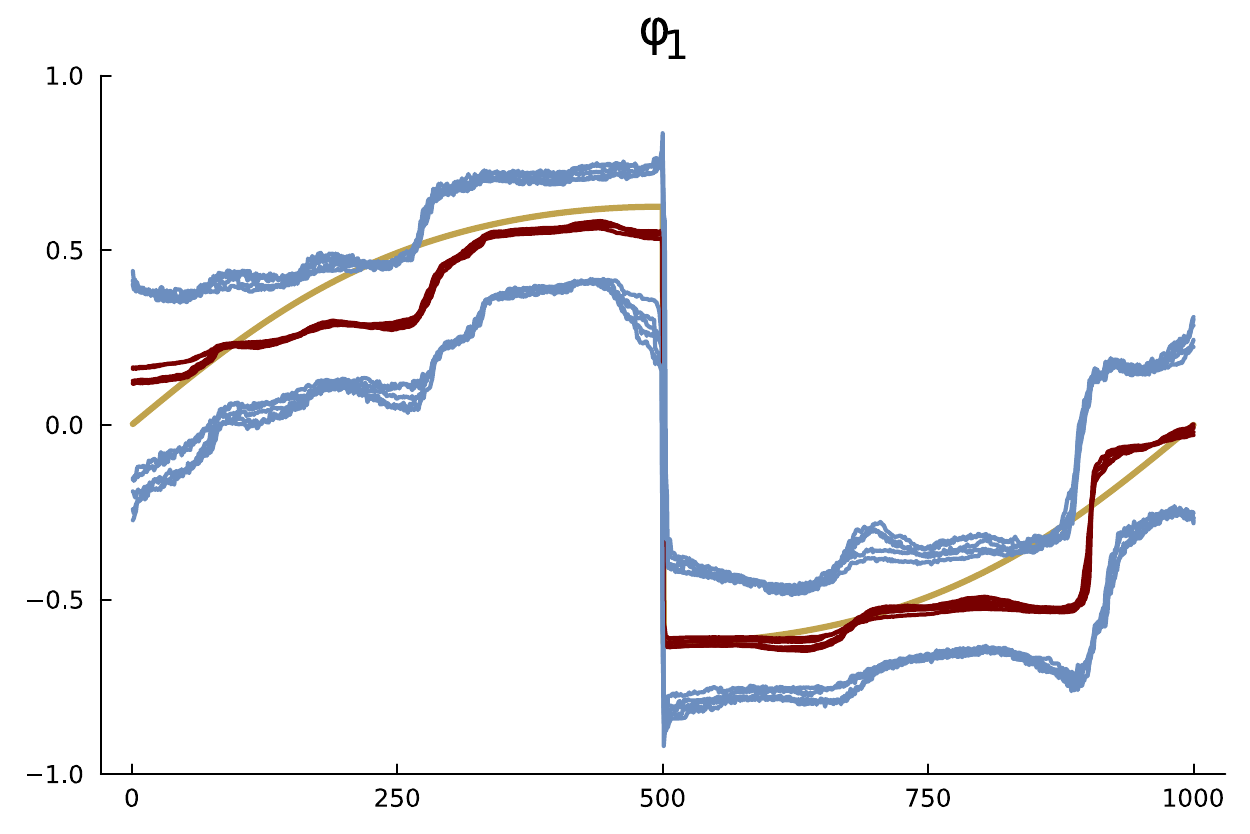}
  \includegraphics[width=0.45\textwidth]{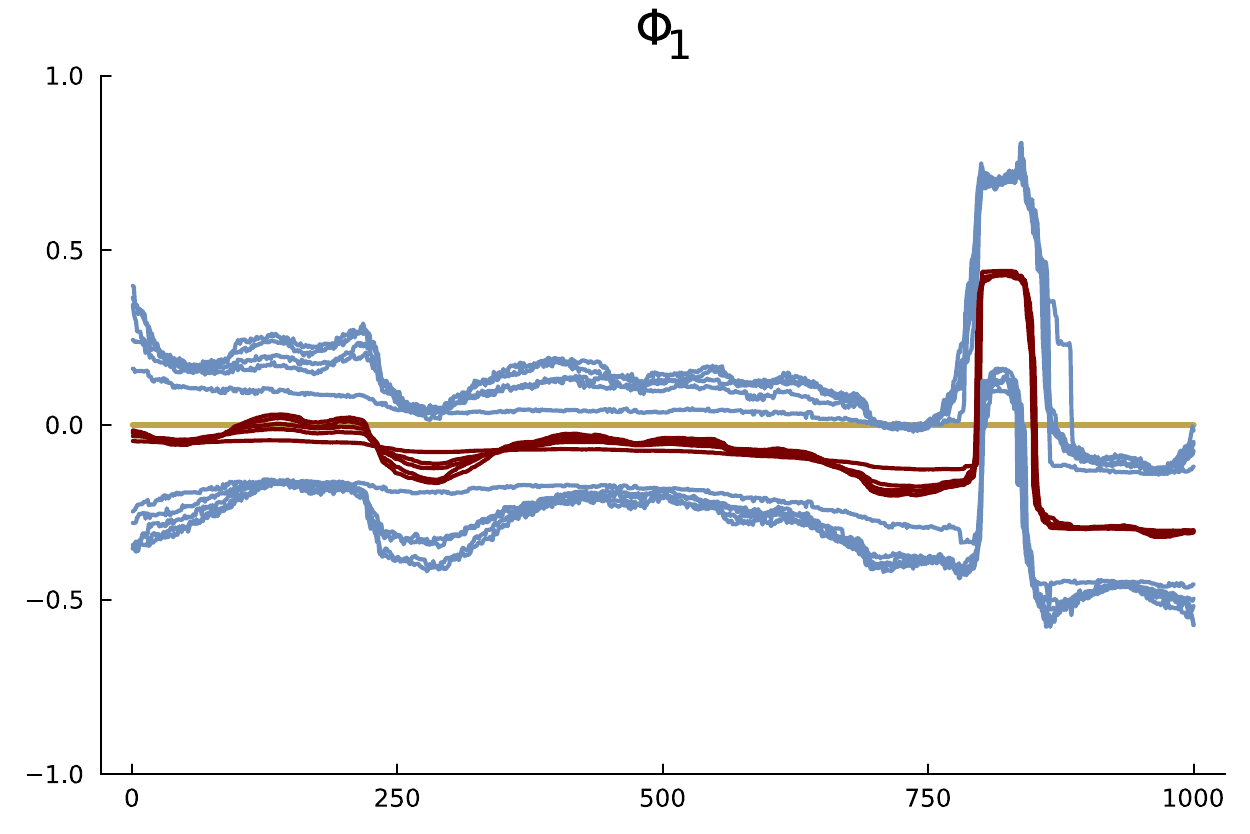}

\caption{Fitting TVSAR$(1,1)_{4}$ model to data generated from a TVSAR$(1,1)_{12}$. MCMC convergence is assessed by re-estimating the model using three different seeds for the same dataset. The red and blue lines are posterior medians and $95\%$ HDIs over time for each of the three repeated
runs. The beige line is the true parameter evolution.} \label{fig:AR2_ev1}
\end{figure}

\clearpage

\section{Additional results - US industrial production}\label{app: usip}

\begin{figure}[H]
 \centering
 \includegraphics[width=0.9\textwidth]{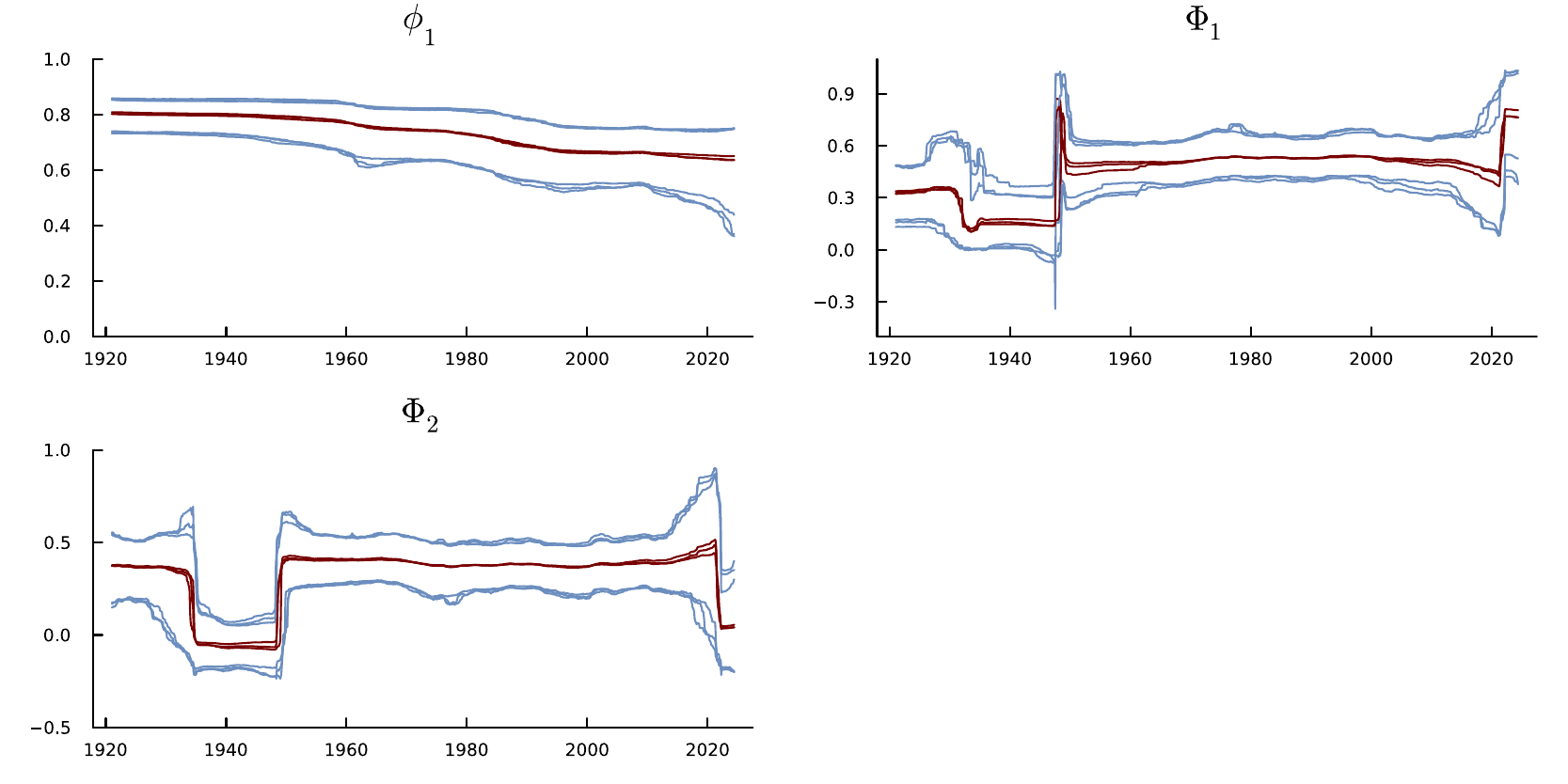}
    \caption{US industrial production. The posterior median and $95\%$ credible intervals from the SAR(1,2) model with dynamic shrinkage prior and stochastic volatility, obtained with the FFBSx algorithm using an offset of $10^{-16}$. The posterior median and $95\%$ credible intervals are shown for three runs based on different initial values drawn from the prior.}\label{fig:usip_SAR12_ffbsx_threeruns}
\end{figure}

Figure \ref{fig:usip_SAR12_ffbsx_threeruns} assesses the convergence of the FFBS-x algorithm on the US industrial production data by plotting the posterior medians and equal-tail $95\%$ credibile intervals from three runs, each with different initial values. The initial values for the global parameters are set so that $\mu_k=\mu$ and $\kappa_k = \kappa$ for all $k$, with $\mu$ and $\kappa$ drawn from their respective priors in each run of the algorithm.

\begin{figure}[H]
 \centering
 \includegraphics[width=0.9\textwidth]{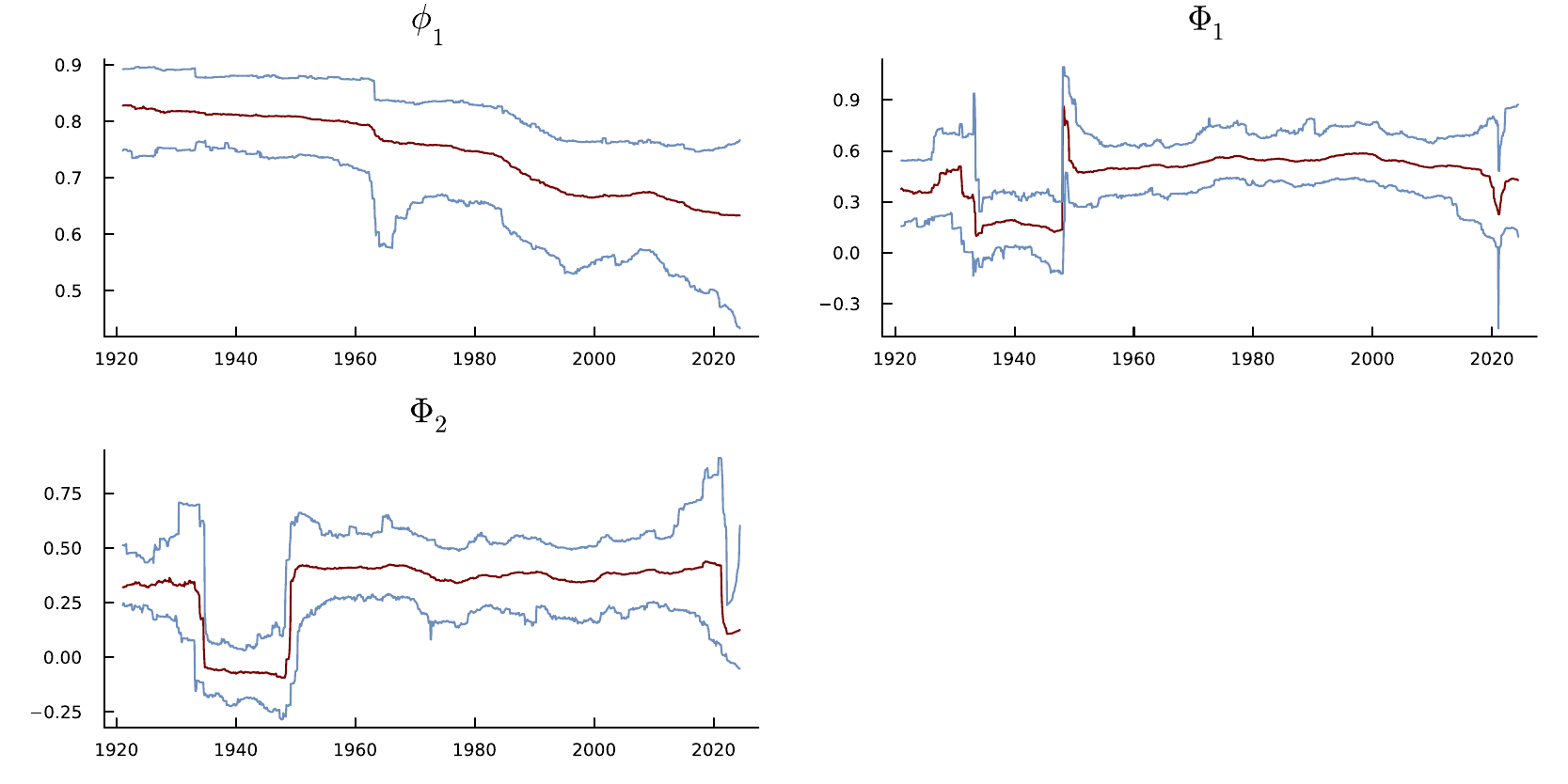}
    \caption{US industrial production. Posterior from SAR(1,2) model with stochastic volatility obtained with the PGAS algorithm with $N=100$ particles using an offset of $10^{-16}$.}\label{fig:usip_SAR12_pgas100}
\end{figure}

Figure \ref{fig:usip_SAR12_pgas100} plots the posterior median and equal-tail $95\%$ credible bands from the PGAS algorithm with 100 particles for the US industrial production data.

\begin{figure}[H]
 \centering
 \includegraphics[width=1\textwidth]{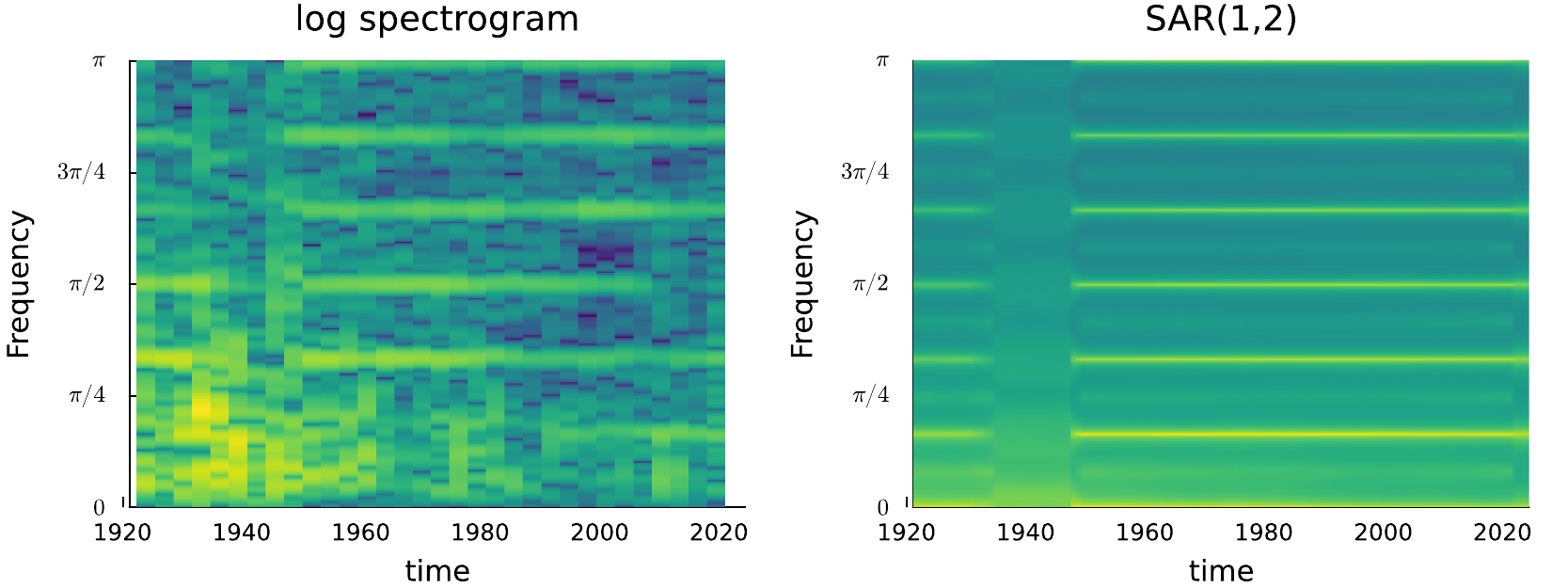}
    \caption{US industrial production. Left: Nonparametric log spectral density over time estimated by Hanning tapered periodograms over overlapping sliding windows of $120$ months, moving $36$ months between each window; note that 18 months (half the window width) are lost in the beginning and at the end of the time period. Right: posterior median time varying log spectral density from the TVSAR(1,2) model. }\label{fig:usip_SAR12_spectrograms}
\end{figure}

\begin{figure}
 \centering
 \includegraphics[width=0.8\textwidth]{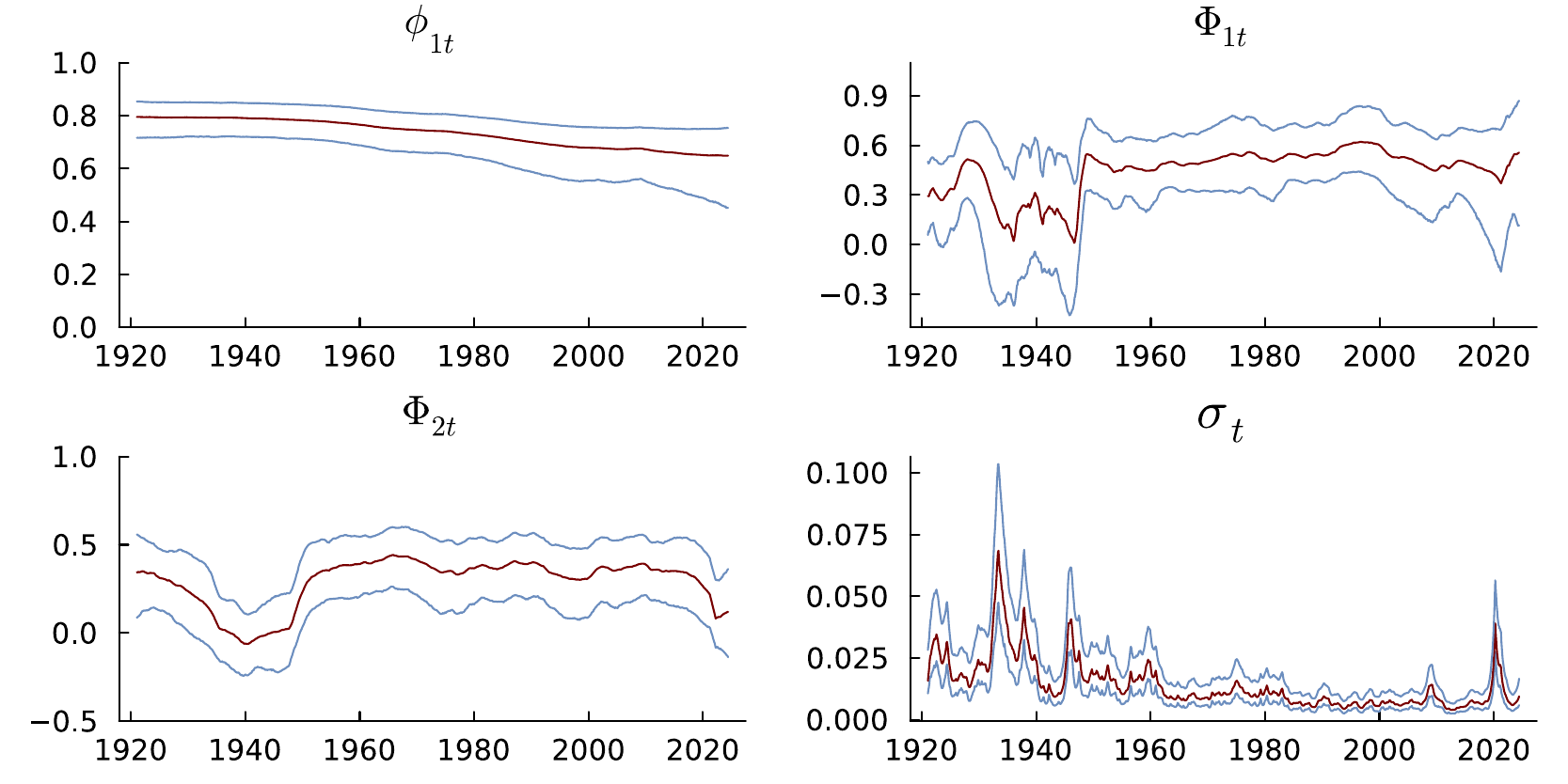}
    \caption{US industrial production. Posterior from FFBSx in the TVSAR$(1,2)_{12}$ model with homoscedastic Gaussian parameter innovations and stochastic volatility.}\label{fig:usip_SAR12_Gaussian_ffbsx}
\end{figure}

Table \ref{table:usip_times} presents the computing time of the FFBSx sampler for subsets of different lengths of the industrial production time series, and for different dimensions of the state. Both the DSP sampler \citep{kowal2019dynamic} and the Kalman-based sampling of the state \citep{thrun2002probabilistic} are known to scale linearly in the length of the time series. The computations were performed  with unoptimized Julia code where the computing time can be reduced substantially by removing type instabilities and allocating less memory.

\begin{table}[ht]
\centering
\caption{Computational times in seconds for 1,000 posterior draws using the FFBSx sampler applied to subsets of different lengths of the industrial production time series.}
\label{tab:comp_times}
\begin{tabular}{l c c c c c}
\hline
 & SAR(1,1)& SAR(2,2) & SAR(3,3) & SAR(4,4) & SAR(5,5) \\
\hline
N of state variables &  13&  26&  39&  52&  65\\
\hline
$T=300$ &  5 &  7 &  11 &  13 &  16 \\
$T=600$ &  9 &  13&   18 &   29 &  42 \\
$T=1200$ &   18 &   24 &   35 &   45 &   63 \\

\hline
\end{tabular}\label{table:usip_times}
\end{table}

\clearpage

\section{Computing the log predictive scores}\label{app: lps}

The LPS in Table \ref{table:usip_lps} are based on three separate Gibbs sampling runs. The LPS estimates are computed by running a Gibbs sampler to obtain \num{10000} posterior parameter draws after a burn-in sample of \num{3000} draws for each the $120$ observations in the test set. Using these draws, the predictive density at the test data point $y_t$ is estimated by the Monte Carlo average
\begin{equation}\label{eq:LPS_MC}
    p(y_t \vert y_{1:(t-1)}) = \iint p(y_t \vert \v x_t, \v\theta_t, \sigma_t)p(\v\theta_t, \sigma_t\vert y_{1:(t-1)})d \v\theta_t d\sigma_t \approx J^{-1}\sum_{j=1}^J p(y_t \vert \v x_t, \v\theta_t^{(j)}, \sigma_t^{(j)}),
\end{equation}
where $\v\theta_t^{(j)}$ and $\sigma_t^{(j)}$ uses draws from the joint posterior $p(\v\theta_{0:(t-1)},\v h_{0:(t-1)},\sigma_{1:(t-1)}, \v\mu,\v\kappa \vert y_{1:(t-1)})$ and the model's parameter evolution to draw $\v\theta_t$, $\v h_t$ and $\sigma_t$. The LPS estimate is then obtained by summing the log predictive density evaluations over the test periods
\begin{equation}
    \mathrm{LPS} = \sum_{t=t^\star}^T \mathrm{LPS}_t,
\end{equation}
where $\mathrm{LPS}_t = \log p(y_t \vert y_{1:(t-1)})$ and $t^\star$ is the first time point in the test dataset.

A reviewer pointed out that the estimate in \eqref{eq:LPS_MC} can be made more precise in conditionally linear Gaussian state-space models by using the Kalman filter to analytically integrate out the state trajectory (see Section A.1.2 in the supplementary material to \citet{bitto2019achieving}). A similar estimator is possible here by analytically integrating out the state using the extended Kalman filter; this will rely on the Gaussian approximation underlying the FFBSx algorithm also in the LPS computation, but the additional bias may be worthwhile from the reduction in variance. A bigger challenge with both approaches, however, is that the computational cost of LPS is substantial, unless a cluster is available to estimate the $\mathrm{LPS}_t$ for $t=t^\star,\ldots,T$ across the compared models in parallel. A more practical approach is to use an information criteria to choose the number of lags over local segments of the time series, and then fit the TVSAR with the maximal number of regular and seasonal lags over the segments. As shown in Experiment 3, the potential redundant lags will be shrunk to zero by the dynamic shrinkage process prior. We proceed with the TVSAR$(1,2)_{12}$ model in this section, but model comparison and evaluation for the TVSAR model and other state-space models with global-local shrinkage processes are important topics for future research.

\clearpage

\section{Performance near the non-stable region}\label{app: nearnonstable}

As discussed in Section \ref{sec:conclusions}, the FFBSx sampler for the TVSAR model with stability restrictions can generate a posterior distribution for the parameter evolution that deviates from the true posterior when the data generating process has long spells near the instability region. To illustrate this, we simulate a time series of length $T=500$ from a TVSAR$(1,1)_{12}$ where the seasonal polynomial is consistently close to the instability region and the regular AR polynomial is close to the non-stability instability region during certain time periods. A simple Gaussian homoscedastic parameter evolution is used to ensure that the sample from PGAS with $100$ particles accurately represents" the true posterior distribution. We fixed the static parameters to their values in the data generating process to clearly isolate the state estimation problem.

\begin{figure}[H]
 \centering
 \includegraphics[width=0.9\textwidth]{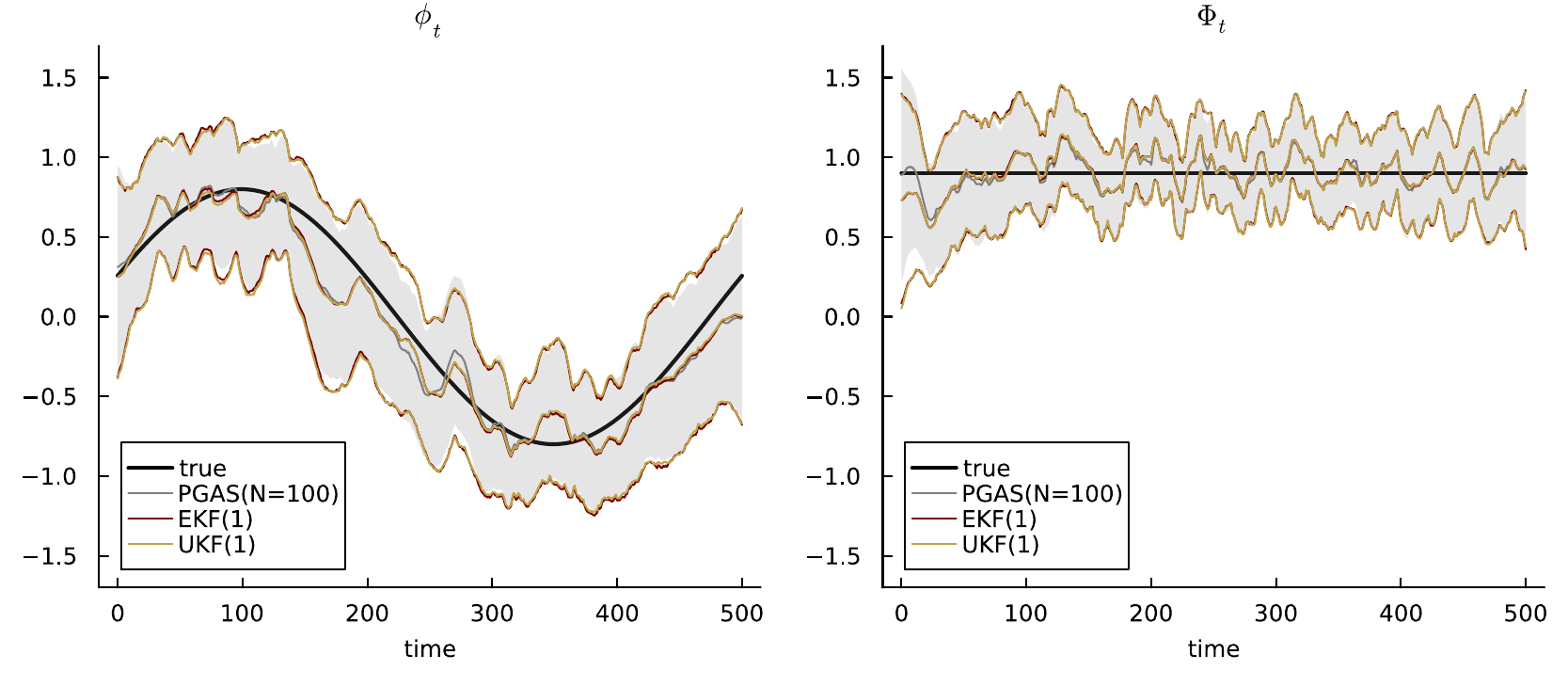}
 \includegraphics[width=0.9\textwidth]{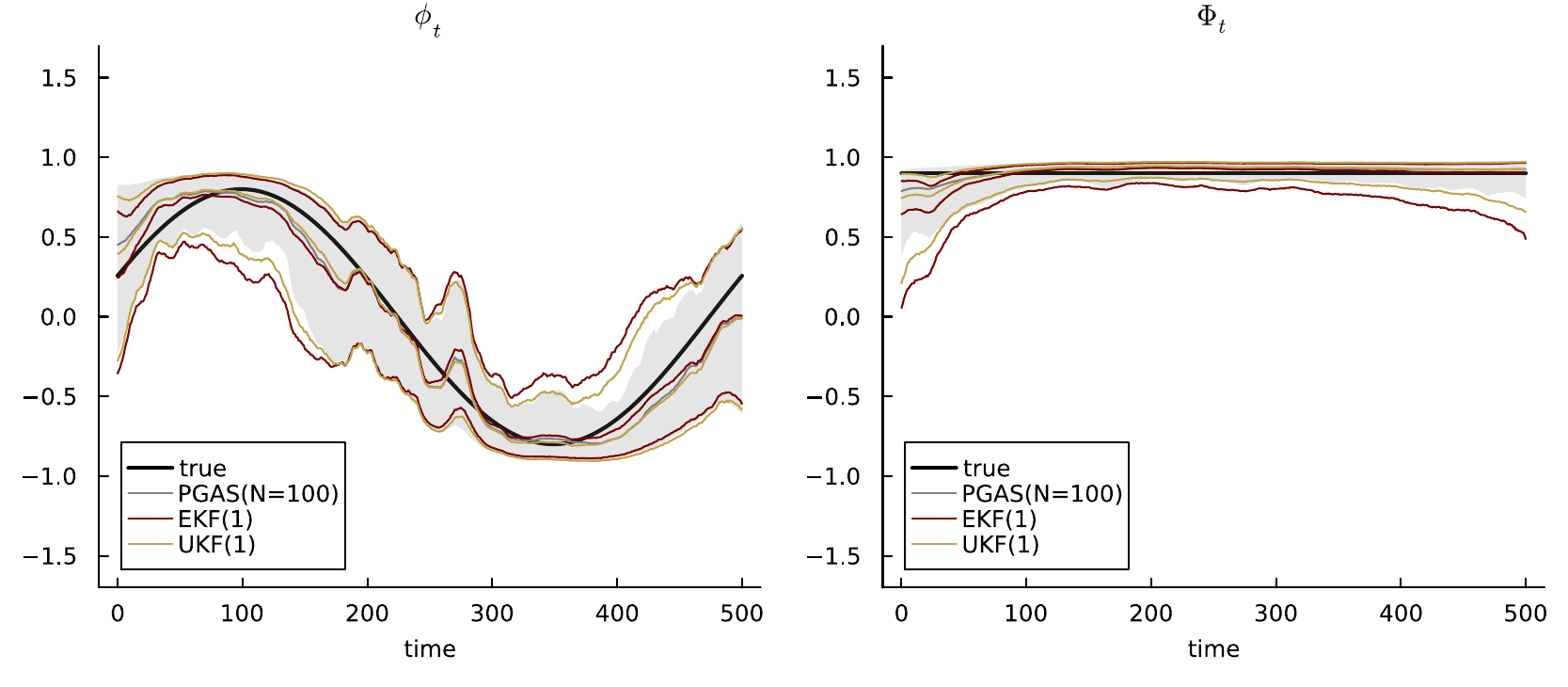}
    \caption{Fitting a TVSAR$(1,1)_{12}$ model to a time series from a near non-stable TVSAR$(1,1)_{12}$ process. Both the data generating process and the fitted model have homoscedastic Gaussian parameter innovations. In the graphs in the top panel, the parameters are \emph{not} restricted to the stable region. In the bottom panel, the parameters are restricted to the stability region. The gray line and gray shaded regions are the posterior mean and the 95\% credible intervals from the PGAS sampler with $100$ particles. The posterior from the FFBSx sampler (red lines denoted EKF) and the posterior from FFBS with the unscented Kalman filter (UKF) are also shown.}\label{fig:nearnonstable1}
\end{figure}

The top panel of Figure \ref{fig:nearnonstable1} shows the results when the fitted model has parameters that are \emph{not} restricted to the stability region, while the plots in the bottom panel are for the case with stability restrictions. The true parameter evolution (black line) for the two parameters and the inferred posterior distribution for the fitted TVSAR$(1,1)_{12}$ model are shown for several algorithms. The posterior from PGAS with $100$ particles (which was verified to be a good representation of the true posterior distribution since more particles gave very similar results) is shown as a gray line for the posterior median and the gray shaded regions for the 95\% crebilibity intervals. The posteriors from two approximate samplers are also shown: FFBSx based on the extended Kalman filter (denoted by EKF in the graphs) and FFBS based on the unscented Kalman filter (UKF, \citet{julier2004unscented}); the unscented Kalman filter uses so-called sigma points instead of linearization and can under some circumstances be seen as a second-order approximation method. 

Figure \ref{fig:nearnonstable1} (top panel) shows that when fitting models without imposing stability restrictions, all three sampler give essentially identical results, but the posterior distribution contains substantial probability mass on the non-stable processes. Note that this model is still nonlinear due to the multiplicative seasonal model. The bottom panel of Figure \ref{fig:nearnonstable1} illustrates the results when the same dataset is fitted with TVSAR$(1,1)_{12}$ with stability restrictions imposed. Here the FFBSx sampler gives a distorted posterior compared to the one from PGAS. The FFBS sampler based on the unscented Kalman filter (UKF) improves on the FFBSx. There are several other recently proposed Kalman-based approximate filters in the signal processing literature \citep{skoglund2015extended, skoglund2019iterative, garcia2015posterior} that can be used in FFBS samplers to more accurately estimate the posterior near the instability region. This is our current focus and will be reported more fully in a separate article. We are also exploring other mappings from the unrestricted parameters to the partial autocorrelations than \eqref{eq:t_transf}, which is used in our article, to improve the performance of the posterior samplers near the instability region.

\end{document}